%% file: ms.tex
\documentclass[twocolumn]{aastex63}

\newcommand{\teff}{T_{\rm eff}}
\newcommand{\ebv}{E(B\, -\, V)}
\newcommand{\dmn}{(m\, -\, M)_0}
\newcommand{\kms}{\rm km~s^{-1}}
\newcommand{\vphi}{v_\phi}

\shorttitle{A Blueprint for the Milky Way's Stellar Populations.\ IV}
\shortauthors{An et al.}

\submitjournal{the AAS Journals}

\begin{document}

\title{A Blueprint for the Milky Way's Stellar Populations.\ IV.\ A String of Pearls -- the Galactic Starburst Sequence}

\author{Deokkeun An}
\affiliation{Department of Science Education, Ewha Womans University, 52 Ewhayeodae-gil, Seodaemun-gu, Seoul 03760, Republic of Korea; deokkeun@ewha.ac.kr}

\author{Timothy C.\ Beers}
\affiliation{Department of Physics and Astronomy and JINA Center for the Evolution of the Elements, University of Notre Dame, Notre Dame, IN 46556, USA}

\author{Young Sun Lee}
\affiliation{Department of Physics and Astronomy and JINA Center for the Evolution of the Elements, University of Notre Dame, Notre Dame, IN 46556, USA}
\affiliation{Department of Astronomy and Space Science, Chungnam National University, Daejeon 34134, Republic of Korea}

\author{Thomas Masseron}
\affiliation{Instituto de Astrof\'{i}sica de Canarias (IAC), E-38200 La Laguna, Tenerife, Spain}
\affiliation{Departamento de Astrof\'{i}sica, Universidad de La Laguna (ULL), E-38206 La Laguna, Tenerife, Spain}

\begin{abstract}

We continue our series of papers on phase-space distributions of stars in the Milky Way based on photometrically derived metallicities and Gaia astrometry, with a focus on the halo-disk interface in the local volume. To exploit various photometric databases, we develop a method of empirically calibrating synthetic stellar spectra based on a comparison with observations of stellar sequences and individual stars in SDSS, SMSS, and PS1, overcoming band-specific corrections employed in our previous work. In addition, photometric zero-point corrections are derived to provide an internally consistent photometric system with a spatially uniform metallicity zero point. Using our phase-space diagrams, we find a remarkably narrow sequence in the rotational velocity ($\vphi$) versus metallicity ([Fe/H]) space for a sample of high proper-motion stars ($>25$~mas~yr$^{-1}$), which runs along Gaia Sausage/Enceladus (GSE) and the Splash sub-structures, and is linked to the disk, spanning nearly $2$~dex in [Fe/H]. Notably, a rapid increase of $\vphi$ from a nearly zero net rotation to $\sim180\ \kms$ in a narrow metallicity interval ($-0.6 \la {\rm [Fe/H]} \la -0.4$) suggests that some of these stars emerged quickly on a short gas-depletion time scale. Through measurements of a scale height and length, we argue that these stars are distinct from those heated dynamically by mergers. This chain of high proper-motion stars provides additional support for recent findings that suggest a starburst occurred when the young Milky Way encountered the gas-rich GSE progenitor, which eventually led to the settling of metal-enriched gas onto the disk.

\end{abstract}


\section{Introduction}

Photometric survey databases are useful resources for studying stellar populations and structures of the Milky Way Galaxy. The size of spectroscopic samples has grown rapidly in recent years, but photometric surveys still cover a significantly larger volume of space and thereby can provide the least-biased sample of Galactic stars. Multi-band observations are particularly useful, since they can be used to constrain fundamental stellar parameters, such as effective temperature ($\teff$) and metallicity ([Fe/H]), with sufficient accuracy. Specifically, the overall shape of a spectral energy distribution as traced by multi-band photometry depends on $\teff$, and ultra-violet excess provides information about a star’s metallicity. When these data are combined with all-sky, high-precision astrometry from Gaia, they can provide rich information on chemical and kinematical properties of stars, as we demonstrated in a series of papers \citep[][hereafter Paper~I, II, and III, respectively]{paper1,paper2,paper3}.

To obtain a clear view of Galactic stellar populations, one needs to establish an accurate relationship between photometry and fundamental stellar parameters on an empirical basis or by using theoretical predictions. To take advantage of each method, we adopted a hybrid approach in previous papers in this series to calibrate theoretical isochrones of the main sequence using observations of well-studied Galactic globular and open clusters. The models were taken from YREC \citep{sills:00}, and were combined with MARCS \citep{gustafsson:08} synthetic spectra, in order to convert $\teff$ and luminosities into photometric colors and magnitudes. Differences of the models from observations typically amount to a few hundredths of a magnitude for warm ($\teff > 5000$~K) stars, but they become as large as a few tenths of a magnitude for cooler stars. The model offsets are also a function of metallicity, in that more metal-rich cluster sequences tend to exhibit larger color deviations.

The observed offsets from the models are systematic in nature and cannot be simply reconciled by adjusting input cluster parameters, which implies that they may originate from errors in the input physics and physical parameters. To overcome these difficulties, we took the observed offsets as empirical correction functions that one needs to apply to our specific choice of theoretical stellar models. When models are used with empirical corrections, we obtain distances from SDSS photometry that are consistent with Gaia parallaxes, and our photometrically derived metallicities ([Fe/H]) are also in overall agreement with spectroscopic measurements in SDSS (Paper~I).

While we developed a method of deriving photometric metallicities from SDSS, this set of color-$\teff$ corrections is only valid for observations taken in the SDSS $ugriz$ filter set. Other photometric surveys also adopt filter sets similar to that of SDSS, but their transmission curves are not exactly the same as each other, leading to non-negligible differences in magnitudes. In this sense, direct calibration of synthetic spectra can serve as an alternative way of establishing such relations in various filter passbands. Importantly, it enables us to combine various photometric survey databases and produce chemo-kinematical phase-space maps over the entire celestial sphere in an internally consistent manner.

One of the goals in this study is to generalize our empirical-correction procedure, and construct a set of corrected synthetic spectra in order to generate magnitudes in any given filter set with high confidence. An essential requirement to achieve this goal is to finely sample flux for a set of calibration stars over a wide range of wavelength and stellar parameters using multi-band photometry, which has become practical in the era of massive photometric surveys. The basic idea of calibrating model fluxes of theoretical stellar spectra, as opposed to making corrections on individual color indices in the models, was introduced by \citet{lejeune:97,lejeune:98}. However, the current work is based on a significantly larger set of photometric, spectroscopic, and astrometric data, which were unavailable then.

The other goal of this work is to probe a multi-dimensional data cube of Galactic stars, constructed based on the revised metallicity estimates. In addition to chemical information from photometry, we exploit kinematic data from Gaia, as in the previous papers of this series. The majority of main-sequence stars in our sample are too faint to have radial velocity measurements. However, along the great circle perpendicular to the direction of disk rotation ($l=0\arcdeg$ and $180\arcdeg$), rotational velocities ($\vphi$) in the cylindrical coordinate in the rest frame of the Galaxy do not depend on radial velocity. This enables to derive $\vphi$ by utilizing the proper motions of stars near the Galactic prime meridian and construct phase-space diagrams in $\vphi$ and [Fe/H], which can subsequently be used to characterize kinematical and chemical properties of individual populations. 

Because our method relies on calibration of theoretical models for main-sequence stars, our current approach is effectively limited to a local volume ($d < 6$~kpc). Nonetheless, in our previous papers, we demonstrated the usefulness of such data by providing an unbiased, global perspective on local stellar populations, including Gaia Sausage/Enceladus \citep[GSE;][]{belokurov:18,helmi:18} and the Splash \citep{belokurov:20}. In particular, we applied Gaussian mixture models in Paper~III to isolate individual stellar populations, and evaluated their fractional contributions as a function of distance from the Galactic plane ($Z$) and Galactocentric distance ($R_{\rm GC}$). In this work, we present evidence for yet another stellar population, which appears to have been formed during a starburst episode in the early history of the Galaxy, possibly driven by the GSE merger.

This paper is organized into two parts, which describe each of the above two subjects: the revised calibration of theoretical models (\S~\ref{sec:calib}) and its application to photometric databases (\S~\ref{sec:phase}). We summarize our results in \S~\ref{sec:summary}.

\section{Spectrum-based Empirical Corrections on Theoretical Models}\label{sec:calib}

Stellar metallicities presented in this study are computed using a set of theoretical stellar isochrones with revised empirical corrections. As described below, our new color-$\teff$-[Fe/H] relations, which convert theoretically predicted quantities ($\teff$ and luminosity) into observables (colors and magnitudes), hinge on both stellar sequences and field stars with spectroscopic metallicity estimates; see \citet{an:09,an:13}, Papers~I and II, and references therein, for more information on our previous model corrections. The same set of base theoretical models \citep{sills:00,gustafsson:08} with identical model parameters, including an age-metallicity relation and $\alpha$-element abundance mixtures, are adopted in this work. In contrast to the models adopted in the previous papers of this series, this newer version of calibration utilizes photometry in various filter passbands over a wide range of wavelength, and thereby enables fine-tuning of synthetic stellar spectra.

\subsection{Calibration Samples}\label{sec:sample}

\input{tab1.tex}
\input{tab2.tex}

As summarized in Table~\ref{tab:tab1}, we adopt both cluster sequences and a set of individual stars with spectroscopic metallicity estimates for the calibration of models. As shown in the first two columns, a total $19$ passbands from various photometric surveys are utilized in this study: $ugriz$ photometry from the Sloan Digital Sky Survey (SDSS) DR14, $uvgriz$ from the SkyMapper Sky Survey (SMSS) DR2 \citep{onken:19}, $grizy$ from the Pan-STARRS1 surveys \citep[PS1,][]{chambers:16}, and $BV$ from the AAVSO Photometric All-Sky Survey (APASS) DR10 \citep{henden:18}. The $gri$ photometry in APASS is not used owing to large photometric zero-point offsets \citep[see also][]{tonry:18}. We add to the list standard-star photometry in $BVI_C$ constructed by P.\ Stetson \citep[see][]{stetson:00}.\footnote{https://www.canfar.net/storage/list/STETSON/Standards.} Some of these databases use the same notation for their filter passbands ($ugriz$), but their response functions are not identical. Below we make a distinction between these filters by specifying the survey names.

Table~\ref{tab:tab2} lists the stellar sequences adopted in this study. As in our previous exercise, we employ a set of well-studied Galactic globular and open clusters (M15, M92, M13, M3, M5, M67, and NGC~6791) over a wide range of metallicity ($-2.4 \leq {\rm [Fe/H]} \leq 0.4$). We use fiducial sequences from \citet{an:08}, which were derived from the SDSS imaging data. Zero-point corrections \citep{an:13} are applied to tie \citeauthor{an:08} cluster photometry to SDSS DR14 \citep{dr14}. Fiducial sequences in PS1 for the same set of clusters are taken from \citet{bernard:14}. For Stetson's photometry, we only take data in $BVI_C$, since the $U$-band and $R$-band are less well-defined than the others. At the time of this writing, cluster photometry in the SMSS and APASS filter systems is not yet available, but we plan to improve the calibration of synthetic spectra by incorporating such data whenever they become available.

In addition to clusters, we employ the Gaia double sequence, which appears on a color-magnitude diagram from stars with large proper motions \citep{gaiahrd}. As demonstrated in Paper~II, each of the sequences represents two dominant populations in the local halo -- GSE and Splash -- and has [Fe/H]$\approx-1.3$ and $-0.4$, respectively \citep[see also][]{sahlholdt:19}. This dynamically defined group of stars provides a powerful constraint on the shape of a sequence, bridging the gap between globular and open clusters at intermediate metallicities. We follow the procedure developed in Paper~II to extract individual sequences from the Gaia double sequence (see Appendix~\ref{sec:cteff}). In short, this technique relies on metallicity-sensitive $u$-band photometry to separate the two chemically distinct populations. For SDSS and SMSS, we use their $u$-band data. For PS1 and APASS, we use SDSS $u$. Additional cuts on kinematics further help to isolate each of the populations. As in our previous work, we compute $\vphi$ using Gaia's proper motions and parallaxes, but without radial velocity measurements, within $\pm30\arcdeg$ along the Galactic prime meridian. To construct a clean sequence, we use objects with good astrometry, having less than $20\%$ uncertainty in parallax and $30\%$ uncertainty in proper motion. We also apply cuts on $\ebv < 0.1$ and $|b| > 20\arcdeg$. As in Paper~II, we impose $-150 < \vphi \leq -50\ \kms$ and $120 < \vphi \leq 150\ \kms$ on the sample to derive the blue (metal-poor) and red (metal-rich) main sequences, respectively, for which we compute weighted median colors in bins of $0.5$~mag in $M_r$ ($\Delta M_r=0.2$~mag in $u\, -\, r$ CMDs).

For individual calibration stars with spectroscopic metallicities, we utilize the Sloan Extension for Galactic Understanding and Exploration \citep[SEGUE,][]{yanny:09,rockosi:22} and the Galactic Archaeology with HERMES (GALAH) survey \citep{buder:21}, as they are among the largest and most uniform spectroscopic data sets in the Northern and Southern Hemisphere, respectively. Specifically, we adopt metallicity estimates from a rerun of the updated SEGUE Stellar Parameter Pipeline \citep[SSPP;][]{lee:08a,lee:08b}, performed by one of the coauthors (Y.\ S.\ Lee).  We apply cuts based on a signal-to-noise ratio (SNR) of the spectra ($>30$), uncertainty in $\teff<200$~K, and $\log{g}>3.5$ to select main-sequence stars with high-quality parameter estimates. For the GALAH sample, we also require that a stellar parameter quality flag ({\tt flag\_sp}) and an overall iron abundance quality flag ({\tt flag\_fe\_h}) are not set.

The advantage of the SEGUE and GALAH samples is corroborated by the availability of photometric data in various passbands (such as SMSS $uv$; see Table~\ref{tab:tab1}). All SEGUE stars are covered by SDSS and PS1 imaging surveys, but can only be matched to objects in SMSS near the equatorial region. Likewise, the majority of GALAH stars have good matches to SMSS photometry, but only a small fraction of its survey area overlaps with SDSS. Primary stellar sources ({\tt type} $=6$) in the SDSS are kept, with a set of minimal quality flags in the $r$-band measurements to ensure that sources do not have issues such as de-blending, interpolation, and saturation. Similarly, primary detections in PS1 from its stacked imaging catalog are taken. We select point-like sources by imposing a maximum $0.05$~mag difference in $i$-band photometry between a point-spread function and Kron magnitudes. For SMSS, we apply cuts on a number of photometric quality flags to only retain good photometric measurements: {\tt class\_star} $>0.9$, {\tt flags} $<3$, {\tt nch\_max} $=1$, {\tt prox} $>7.5$, {\tt ngood\_min} $>1$, and {\tt nimaflags} $=0$ in each passband.

Photometric catalog objects are matched with Gaia Early Data Release 3 \citep[EDR3,][]{edr3} and Data Release 3 \citep[DR3,][]{gdr3} using a $1\arcsec$ search radius. Zero-point corrections on parallax by the Gaia team \citep{lindegren:21} are adopted. In the following calibration, objects with good parallaxes ($\sigma_\pi/\pi < 0.2$) are used; about $68\ \%$ of them have parallaxes within $2\%$ from those inferred based on Bayes' theorem \citep{bailerjones:21}. The foreground reddening values in \citet{schlegel:98} are adopted, except for clusters, along with extinction coefficients at $R_V=3.1$ in \citet{schlafly:11} for SDSS, PS1, and Johnson-Cousins (those listed as `Landolt') bands. For SMSS, we adopt values in \citet{wolf:18}. These coefficients assume $14\%$ reduction in the original $\ebv$ in \citet{schlegel:98}.

\subsection{Spectrum-based Corrections}

\subsubsection{Scope}

In this study, we aim to provide a set of isochrones with empirically calibrated synthetic spectra over a wide range of $\teff$ and [Fe/H], which in turn can be employed to derive such quantities in other stars. Since we restrict our analysis to main-sequence stars, $\teff$ can be directly mapped onto mass, luminosity, and surface gravity ($\log{g}$) of a star in the isochrones; we take $\teff$ as an independent variable in the following model comparisons. We compute theoretical flux ratios in various filter passbands as a function of wavelength, $\teff$, and [Fe/H], and attribute any deviation from the calibration samples to systematic errors in the models.

For this purpose,  we employ YREC isochrones \citep{sills:00} and synthetic spectra generated using MARCS model atmospheres \citep{gustafsson:08}; see \citet{an:09} for more information on the construction of the MARCS model library. We adopt the same age-metallicity and [Fe/H]-[$\alpha$/Fe] relations for Galactic stars as in our previous papers of this series \citep[see also][]{an:13}: ([Fe/H], [$\alpha$/Fe]) $=\{(-3.0,+0.4)$, $(-2.0,+0.3)$, $(-1.0,+0.3)$, $(-0.5,+0.2)$, $(-0.3,0.0)$, $(+0.4,0.0)\}$ and ([Fe/H], age) $=\{(-3.0$, 13 Gyr), ($-1.2$, 13 Gyr), ($-0.3$, 4 Gyr), ($+0.4$, 4 Gyr)\}, with a linear interpolation in this metallicity grid. Inhomogeneous $\alpha$-element abundance ratios in the Milky Way have a net effect of changing the overall metallicity of a star, but its impact is only mild; $\Delta {\rm [Fe/H]} \sim \pm0.2$ for $\Delta {\rm [\alpha/Fe]} \sim \pm0.2$ \citep[e.g.,][]{kim:02}. The effect of age is minimized in this study by restricting our sample to low-mass main-sequence stars.

We derive synthetic stellar colors in SDSS $ugriz$ from the MARCS library using filter-response curves on the project webpage.\footnote{https://www.sdss.org/instruments/camera/\#Filters} References for other filter transmissions include \citet[][PS1 $grizy$]{tonry:12}, \citet[][SMSS $uvgriz$]{bessell:11}, and \citet[][Johnson-Cousins $BVI_C$]{bessell:12}. APASS $BV$ are taken from the SVO filter profile service.\footnote{http://svo2.cab.inta-csic.es/svo/theory/fps3} When deriving a Vega magnitude, we use a Vega model from the HST CALSPEC library \citep{bohlin:14}, with the suggested flux re-scaling in \citet{riello:21}. Effective wavelengths ($\lambda_{\rm eff}$) of the filter passbands are computed for each model, as they have a mild dependence on the underlying stellar spectrum (primarily on $\teff$). All flux ratios are referenced to the SDSS or SMSS $r$-band among different filter passbands, because of the large amount of flux collected in this passband, and its relatively weak metallicity sensitivity in the $\teff$ regime considered in this study ($4000 < \teff < 7000$~K). Its bolometric corrections are also less prone to systematic errors.

For a given [Fe/H], differences in colors between observational data and models are computed as a function of absolute magnitude in the $r$-band ($M_r$). We use a $M_r$-$\teff$ relation of the isochrones to infer $\teff$ from $M_r$. For the calibration samples, we determine $\teff$ photometrically using stellar isochrones, instead of taking spectroscopic $\teff$, because individual spectroscopic $\teff$ measurements are not typically available for stars in clusters, and it is well known that there exists a $\teff$ scale difference of a few hundred kelvin between photometric and spectroscopic approaches \citep[e.g.,][]{pinsonneault:04}. This significantly reduces systematic differences in color-$\teff$ relations from our heterogeneous data sets, and makes our correction procedures internally more consistent. The infrared flux method (IRFM) is another useful way of deriving $\teff$ from photometry, but our approach has an advantage of making a specific prediction on $\teff$ and $\log{g}$ of a star, both of which are necessary for generating precise synthetic model colors.

\begin{figure}
\centering
\epsscale{1.1}
\plotone{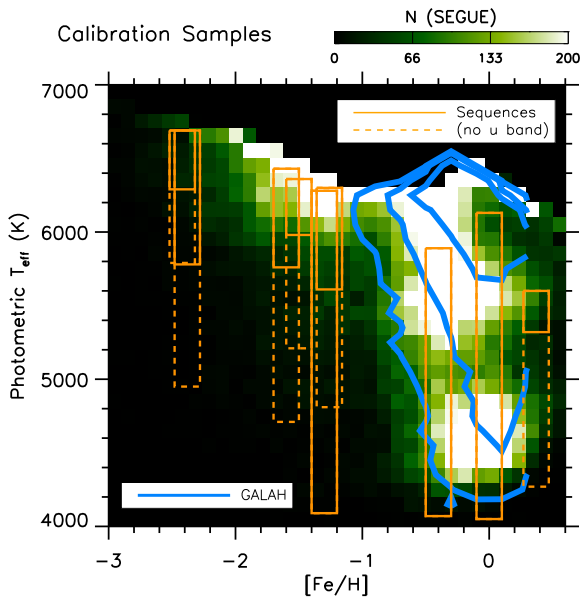}
\caption{The $\teff$-[Fe/H] space covered by the calibration sample. The orange solid rectangles exhibit main sequences of clusters and the Gaia double sequence with $u$-band measurements in SDSS, while more extended sequences with deeper $gr$ photometry are shown by orange-dashed lines. The number density of SEGUE and GALAH stars is shown by the green 2-D histogram and blue contours, respectively. Only those having SNR $>30$, $\log{g}>3.5$, and photometric $\teff$ errors less than $200$~K are retained for these spectroscopic samples.}
\label{fig:sample}
\end{figure}

Figure~\ref{fig:sample} displays a $\teff$-[Fe/H] space covered by the stellar sequences and spectroscopic targets in our sample, where the $\teff$'s are determined from our isochrones. Stellar sequences in SDSS are shown by orange boxes with a width of $\pm0.1$~dex in [Fe/H]. The dotted lines represent the same sequences, but without valid $u$-band measurements, demonstrating the necessity of deeper $u$-band photometry. Stellar sequences have discrete metallicities, while the spectroscopic sample (the green 2-D histogram for SEGUE and contours for GALAH targets) fills up the remaining space. Our spectroscopic sample is heavily biased toward more metal-rich stars, with a significantly lower number in the metal-poor regime ([Fe/H] $ < -1$). The upper right corner is not covered by both samples, owing to the increased metal content and relatively old ages of the stars.


\subsubsection{Model Comparisons}

For each star in the spectroscopic sample, an isochrone is generated by interpolating the model grid at the star's metallicity ([Fe/H]), and differences in flux are computed in various filter passbands, as done for the stellar sequences. However, to evenly sample cluster sequences and the spectroscopic targets, we bin each of the SEGUE and GALAH samples in [Fe/H] and compute {\it mean} flux offsets as a function of $\teff$ (see Appendix~\ref{sec:cteff}). In accordance with the metallicity of the observed stellar sequences (Table~\ref{tab:tab2}), the central metallicities are set to [Fe/H] $=\{-2.4$, $-1.6$, $-1.3$, $-0.4$, $0.0$, $+0.3\}$. To match this binning and keep the average metallicity of a subset of stars as close as possible to the central metallicity values, the spectroscopic sample is divided into [Fe/H] $=\{(-2.9,-2.1)$, $(-1.8,-1.4)$, $(-1.5,-1.1)$, $(-0.5,-0.3)$, $(-0.1,+0.1)$, $(+0.2,+0.4)\}$. A large width ($0.8$~dex) is set in the lowest metallicity bin to compensate for the small number of stars in the sample. Because the flux difference from our model changes mildly with metallicity, our adopted bin sizes have little impact on the following calibration. Nonetheless, mean flux offsets from GALAH are taken only at [Fe/H] $=\{-0.4$, $0.0$, $+0.3\}$ due to the lack of metal-poor stars in the sample.

\begin{figure*}
\centering
\includegraphics[scale=0.36]{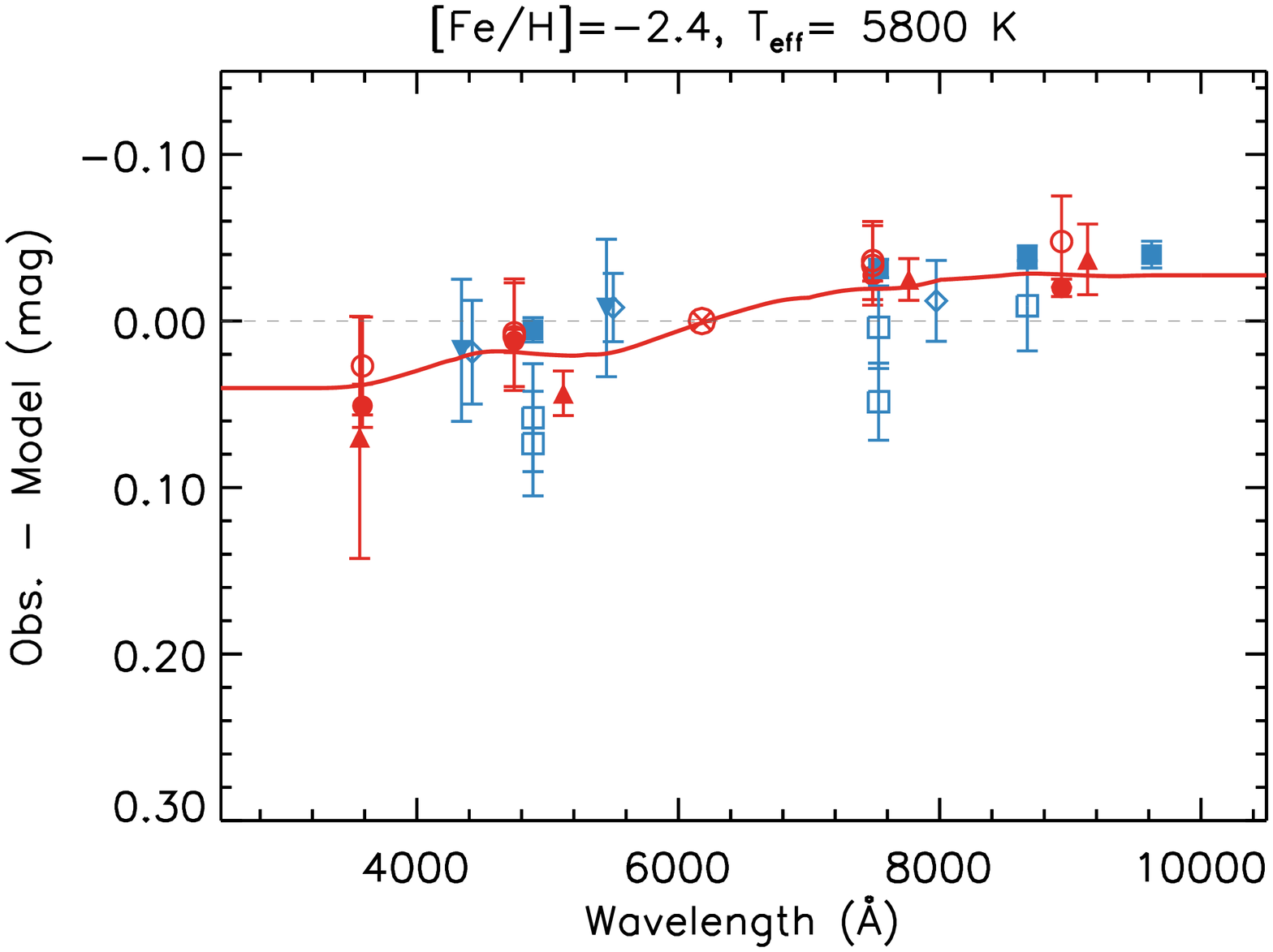}
\includegraphics[scale=0.36]{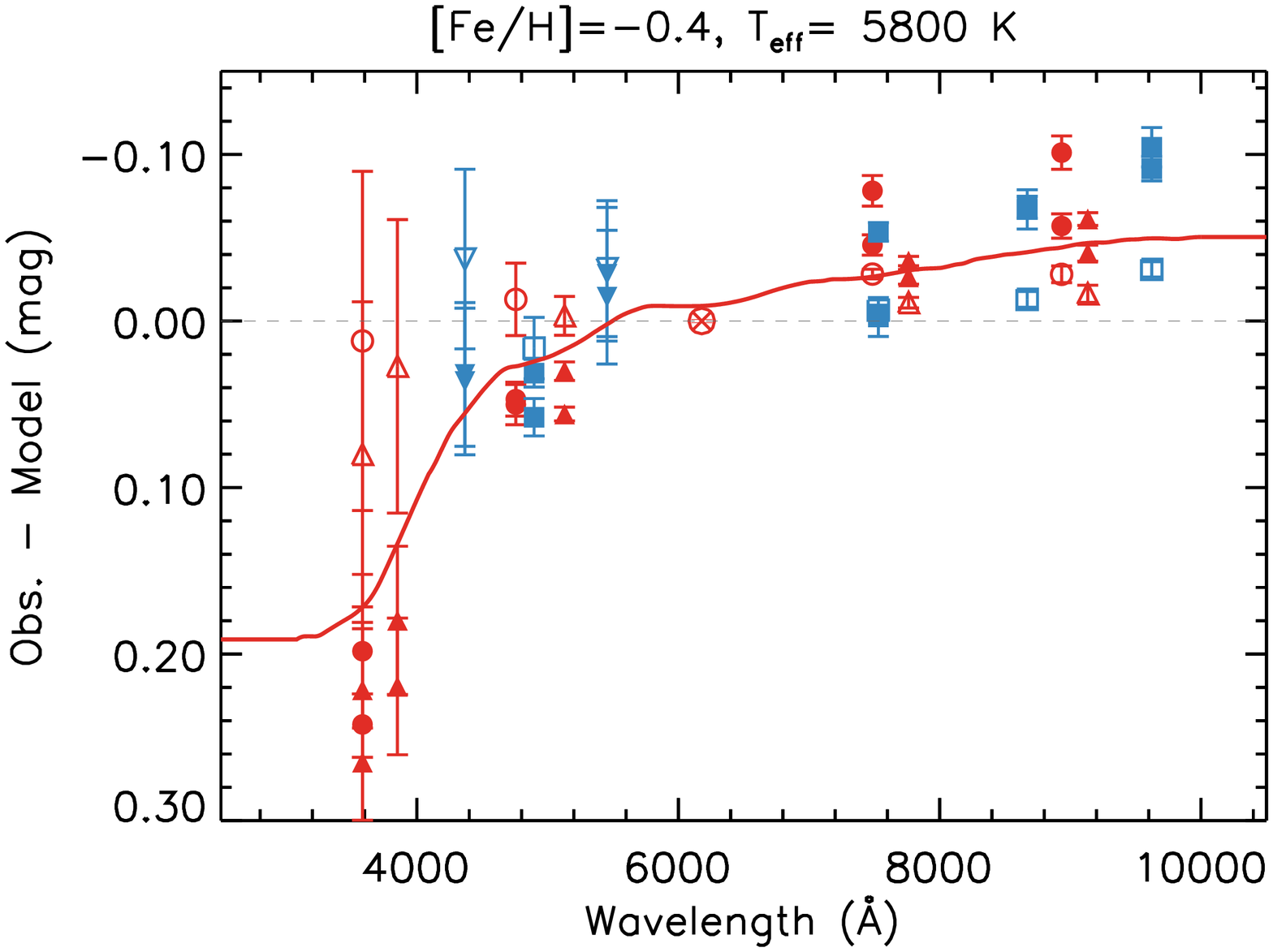}
\includegraphics[scale=0.36]{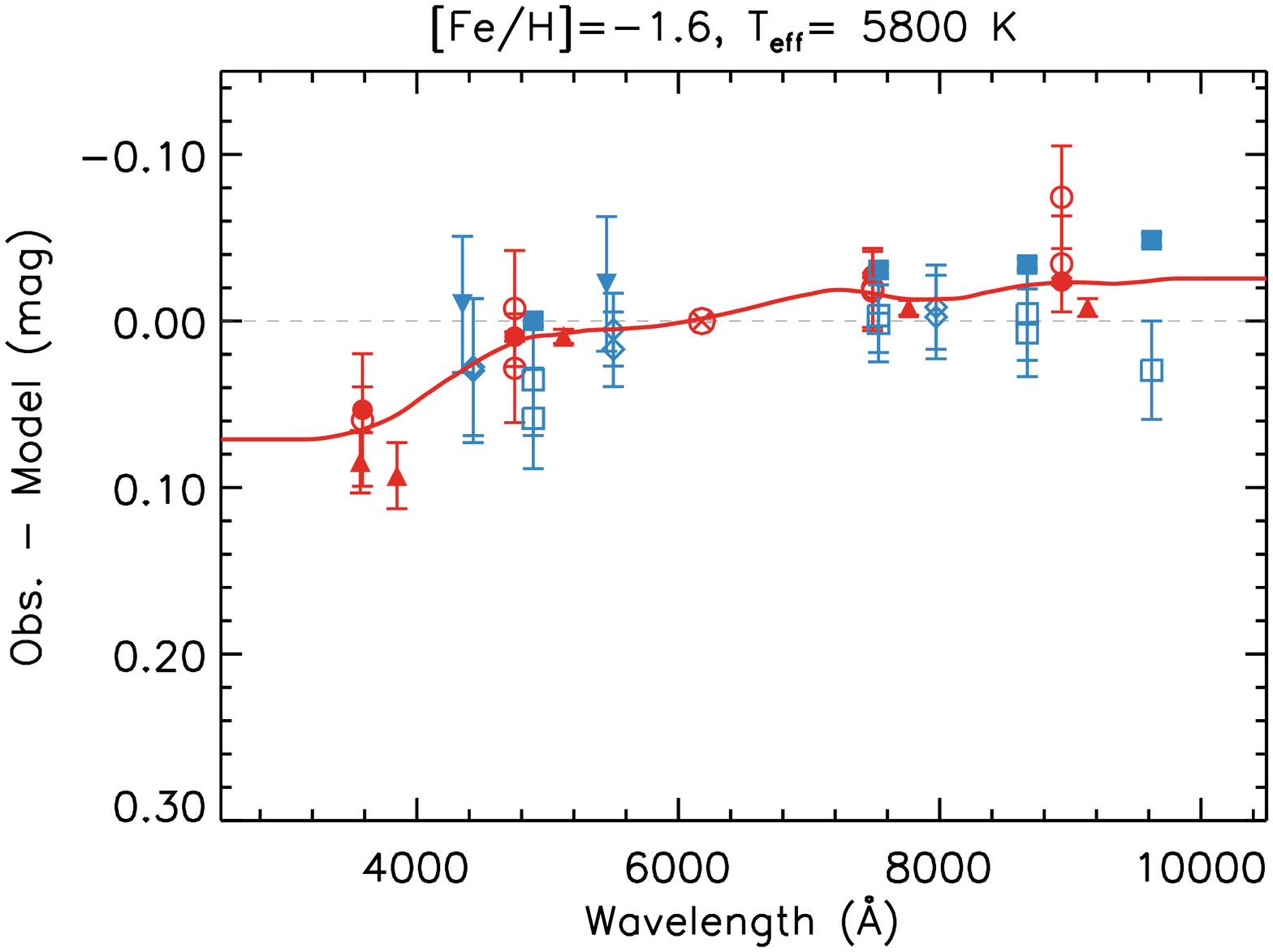}
\includegraphics[scale=0.36]{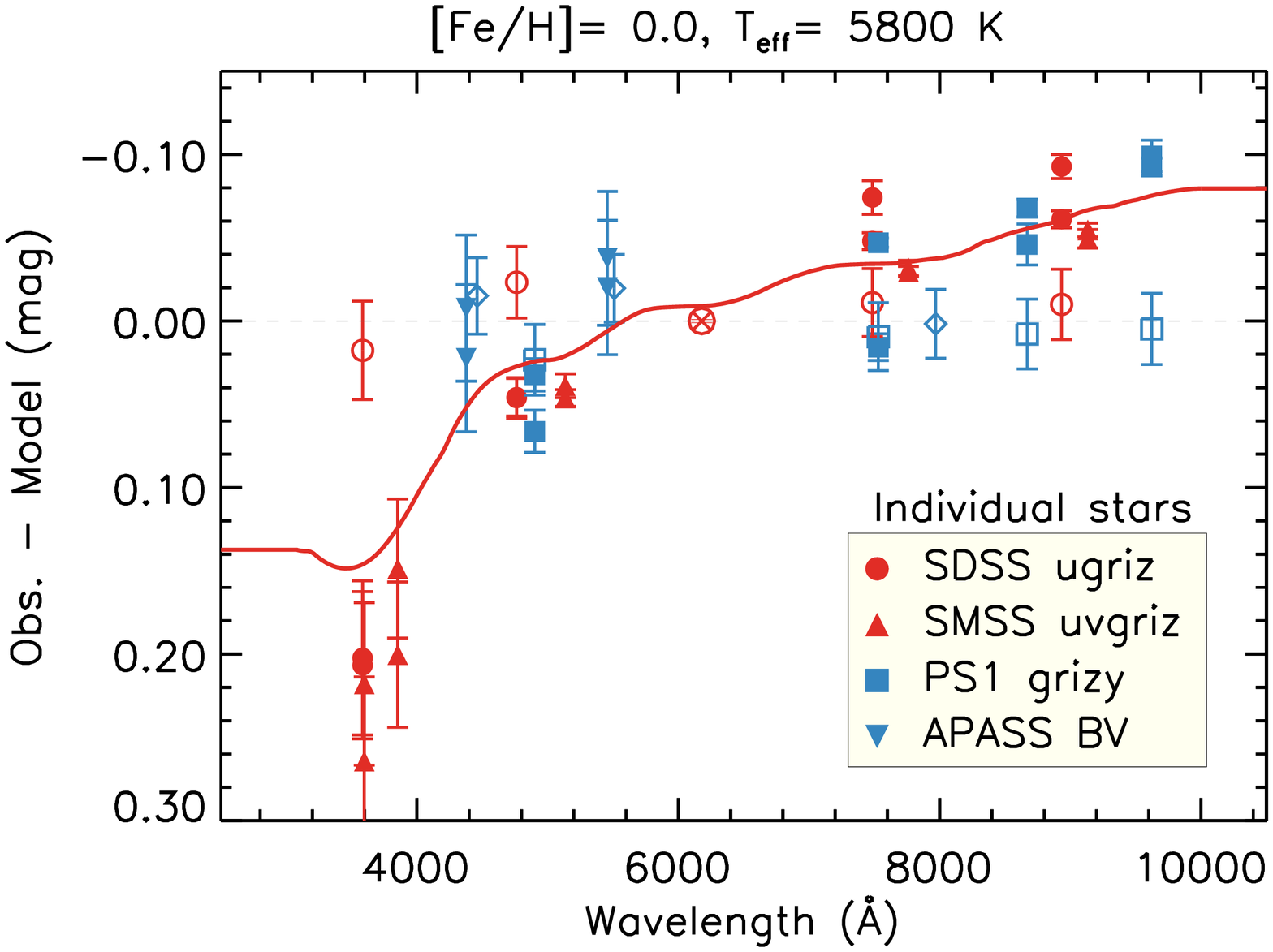}
\includegraphics[scale=0.36]{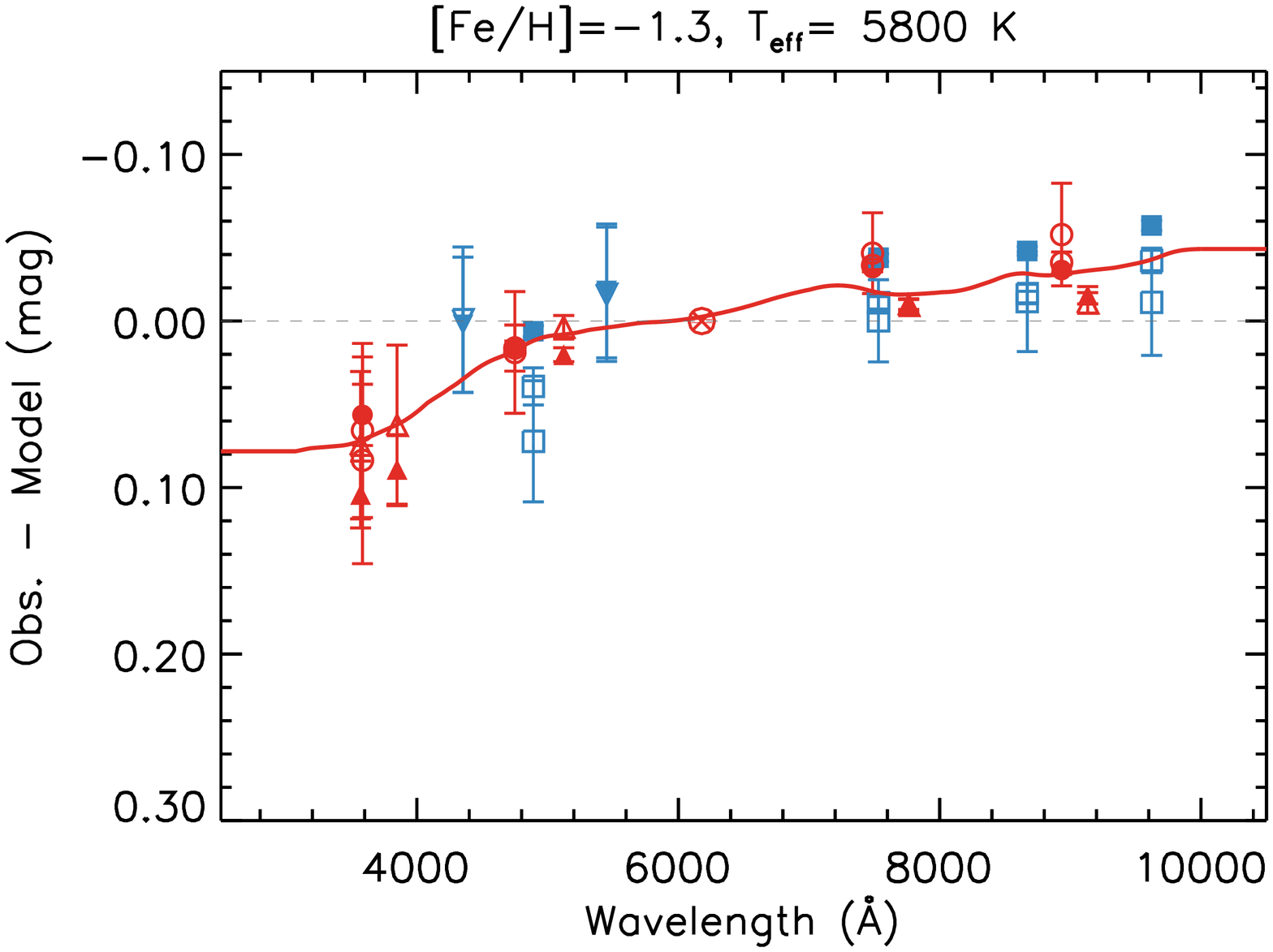}
\includegraphics[scale=0.36]{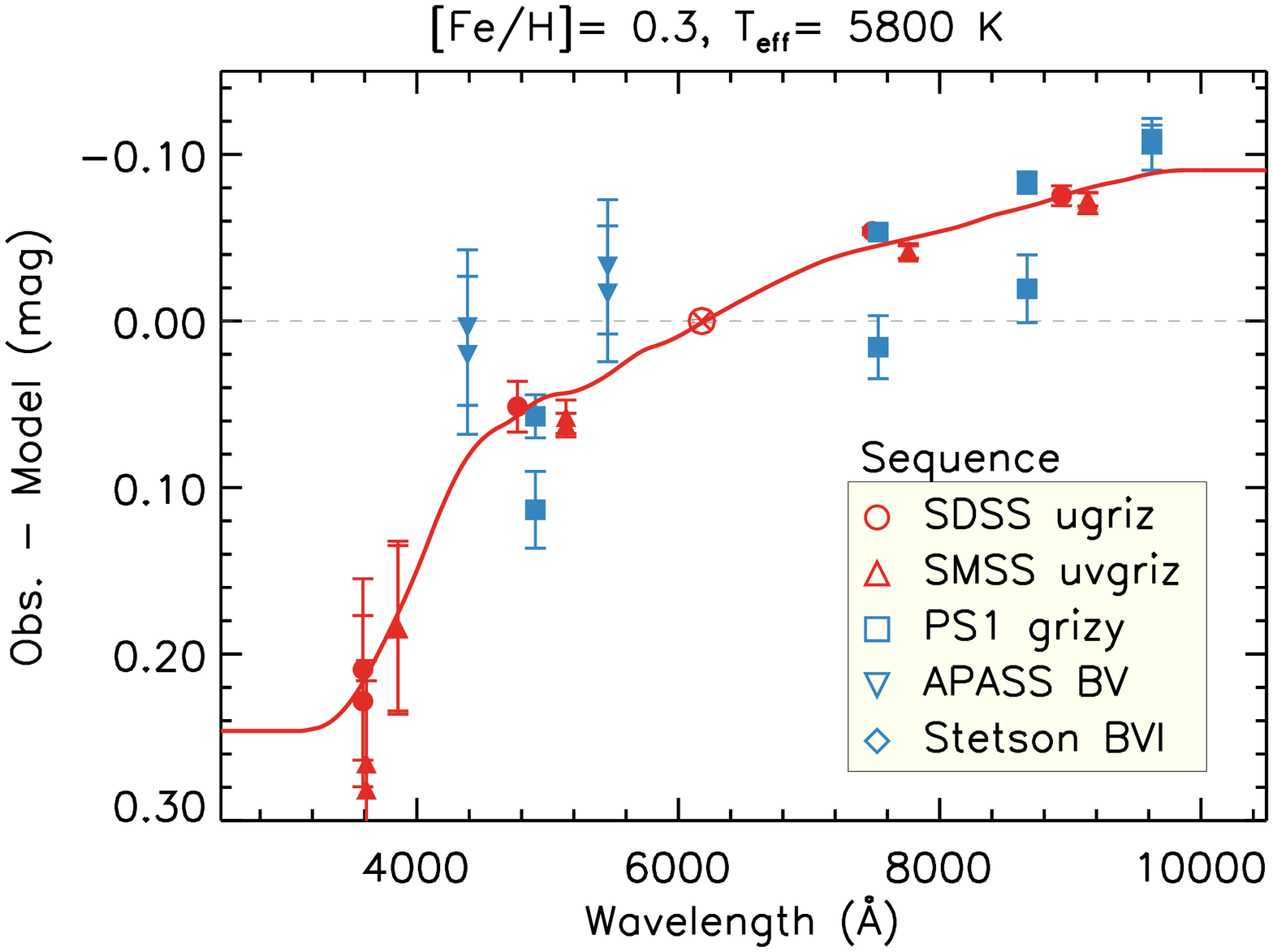}
\caption{Magnitude differences between observed data and original isochrones as a function of $\lambda_{\rm eff}$ of filter passbands. A $\teff=5800$~K case is shown above to display the overall quality of a model comparison with stellar sequences (globular, open clusters, and Gaia's double sequences; open symbols) and spectroscopic measurements  of individual stars from SEGUE and GALAH (filled symbols). Comparisons are made in six metallicity bins at [Fe/H] $=-2.4$, $-1.6$, $-1.3$, $-0.4$, $0.0$, and $0.3$, respectively (from top left to bottom right panels). The red solid line represents moving averages of the difference as a function of $\lambda_{\rm eff}$ (see text).}
\label{fig:cteff}
\end{figure*}

\begin{figure*}
\centering
\includegraphics[scale=0.36]{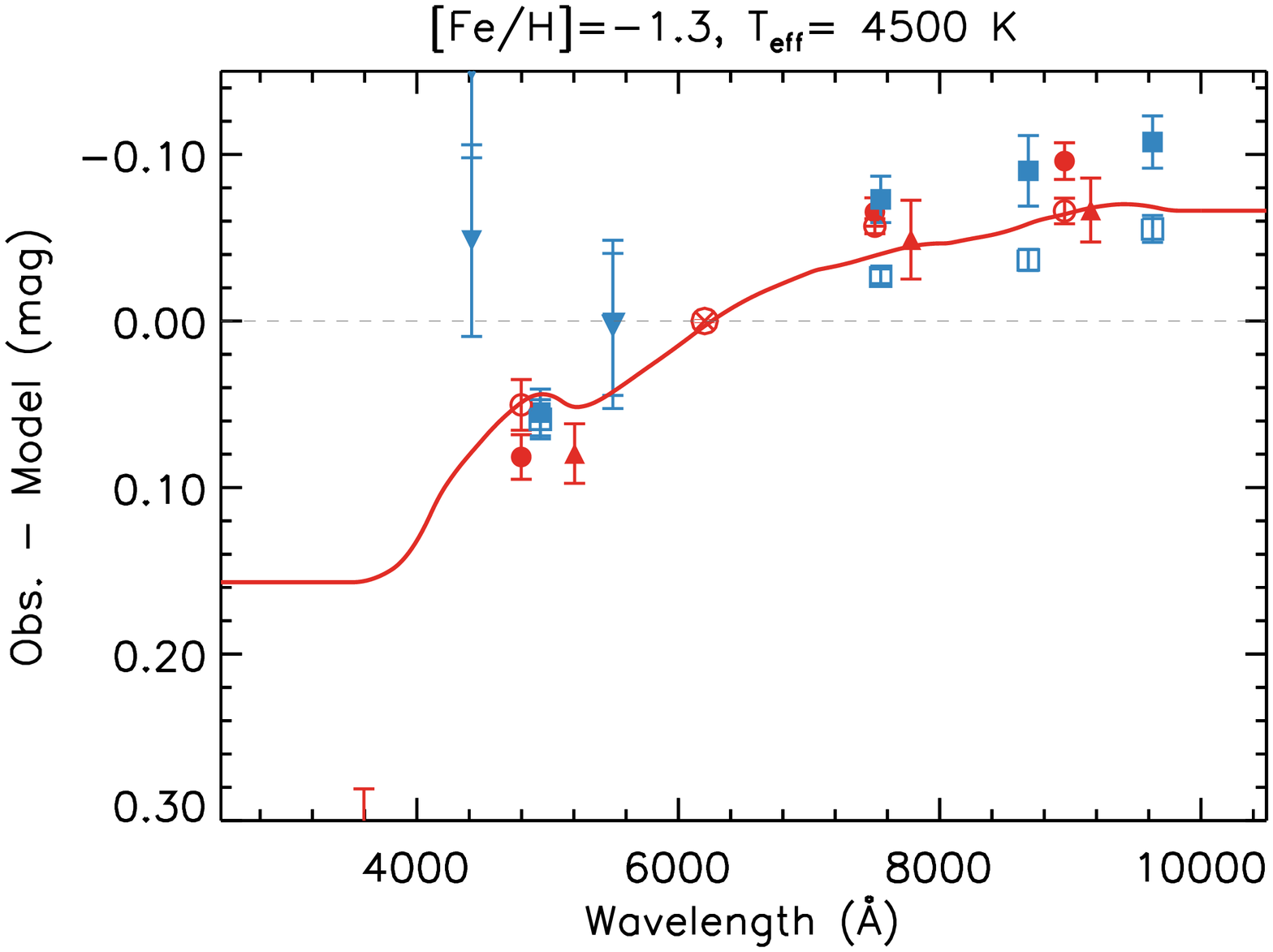}
\includegraphics[scale=0.36]{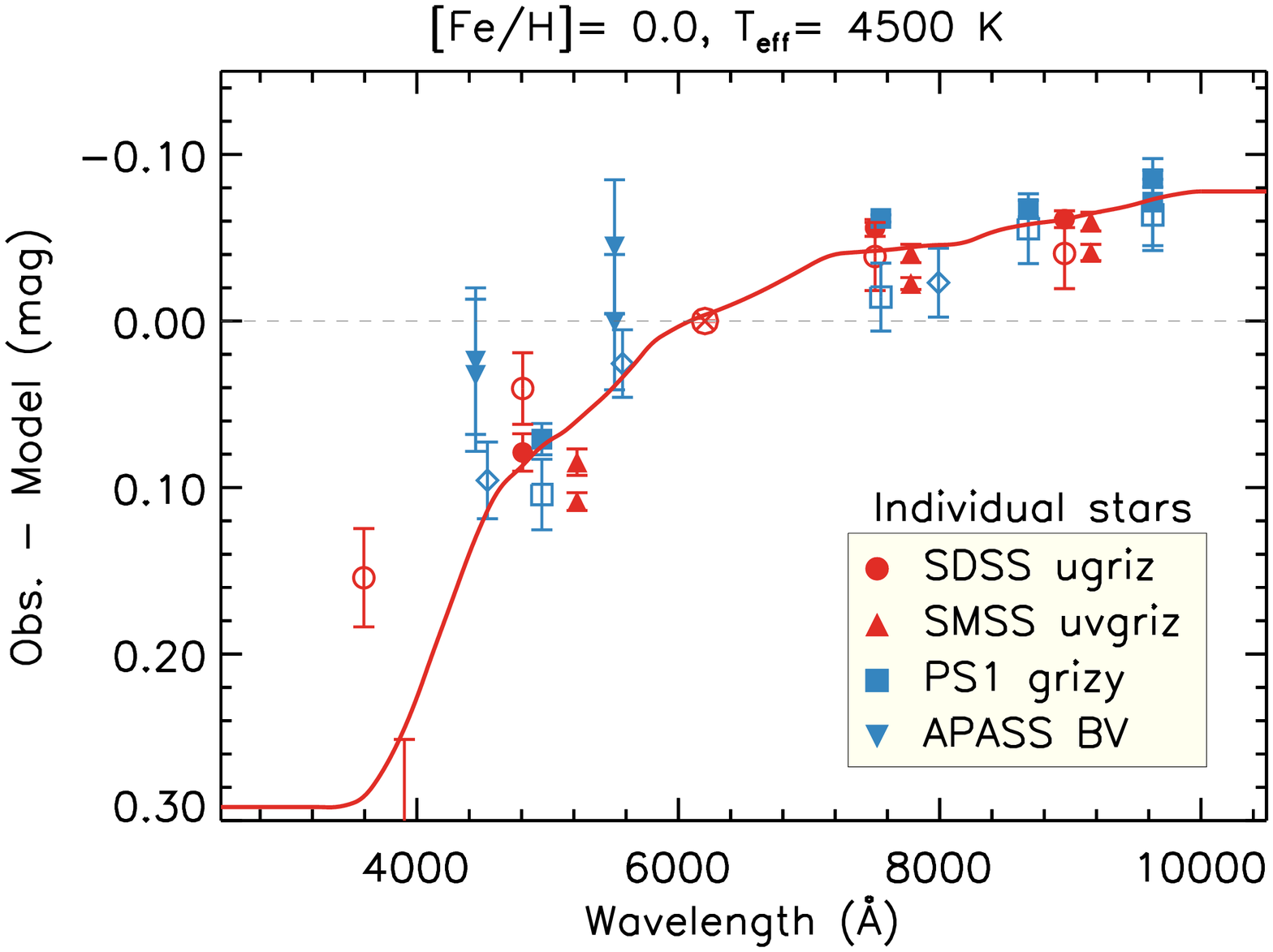}
\includegraphics[scale=0.36]{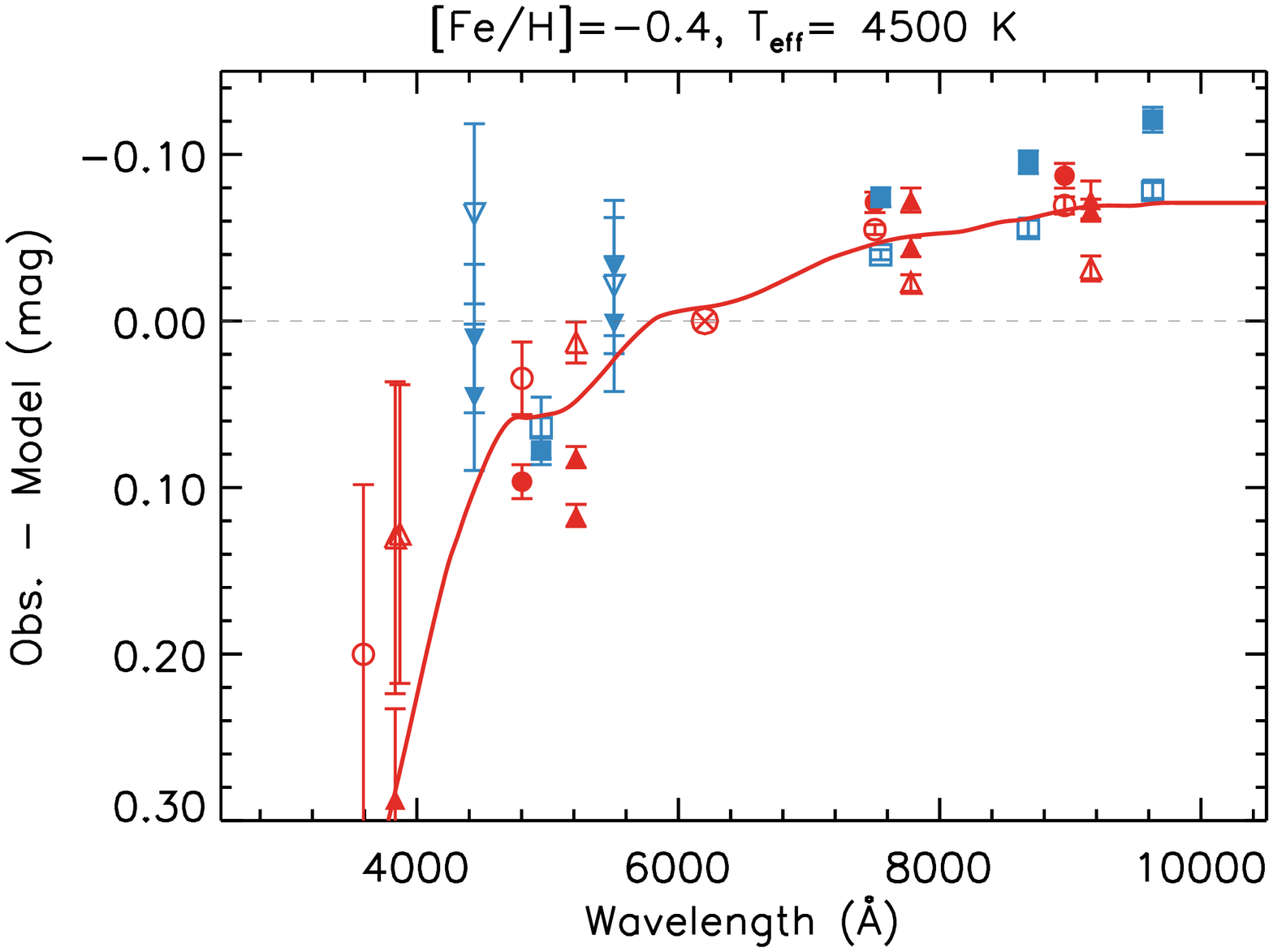}
\includegraphics[scale=0.36]{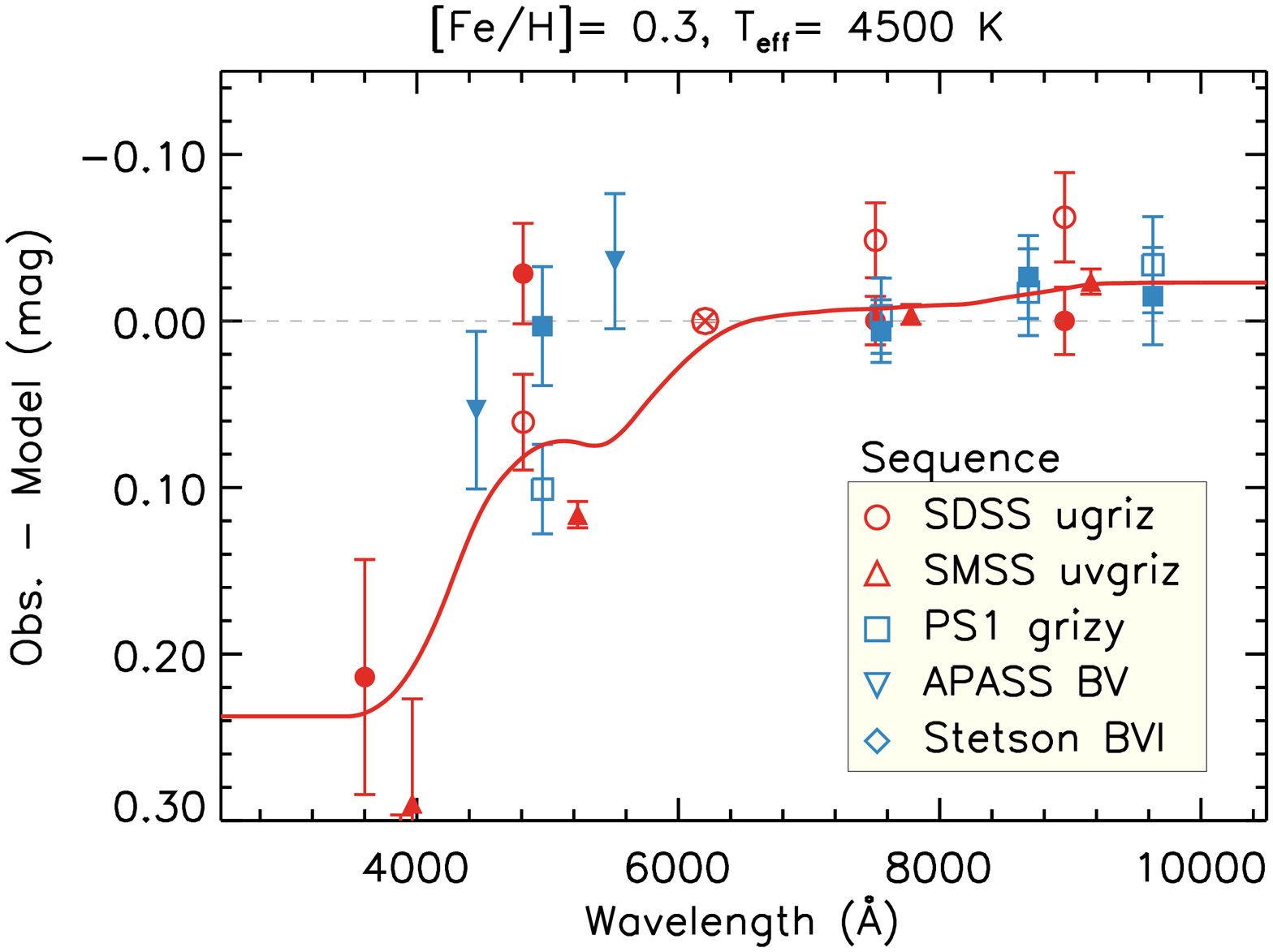}
\caption{Same as in Figure~\ref{fig:cteff}, but at $\teff=4500$~K. Model comparisons in the two lowest metallicity bins are not shown owing to the lack of observational data.}
\label{fig:cteff2}
\end{figure*}

Figures~\ref{fig:cteff} and \ref{fig:cteff2} show magnitude differences between models and observations, as a function of $\lambda_{\rm eff}$, at two selected $\teff$ ($5800$~K and $4500$~K, respectively). In each panel, differences are shown for the sequences and an ensemble of spectroscopic targets in open and filled symbols, respectively; different symbols are used to indicate references for photometry. All flux differences are registered to the SDSS $r$-band, or the SMSS $r$-band for the GALAH sample, owing to a relatively small number of cross-matches with SDSS objects in the Southern sky. The differences from the model at these two $r$ passbands are nearly the same, and are defined to be zero (i.e., their bolometric corrections are assumed to be correct).\footnote{Similarly, because cluster fiducial sequences in PS1 \citep{bernard:14} are not directly tied to SDSS photometry, color indices are registered using the PS1 $r$-band, instead of the fiducial SDSS $r$-band. Nonetheless, SDSS $r$-band and PS1 $r$-band have nearly the same magnitude offsets from models (see Appendix~\ref{sec:cteff}), so switching between the two passbands has a negligible impact on the model comparison.}

Error bars for the stellar sequences in Figures~\ref{fig:cteff} and \ref{fig:cteff2} represent propagated uncertainties from photometry and input parameters ([Fe/H], $(m\, -\, M)_0$, $\ebv$, and age; Table~\ref{tab:tab2}). For the fiducial sequences in SDSS and PS1, constant uncertainties of $0.02$~mag are assumed in the color indices. Similarly, a $2\%$ error is adopted for the mean colors of Stetson's $BVI_C$ cluster sequences as a conservative limit. Photometry of Gaia's double sequence is collected across a large area on the sky, and therefore an observed scatter is taken as uncertainties in the mean colors, since it represents a sum of random and systematic zero-point errors, unless propagated photometric uncertainties are larger. A comparison of APASS photometry to Stetson's standard photometry for the sample clusters (except NGC~6791) reveals an rms dispersion of $0.04$~mag in $BV$. Thus, it is added in quadrature to a total error budget for the APASS-based double sequence.

For the spectroscopic sample, the error bars in Figures~\ref{fig:cteff} and \ref{fig:cteff2} also indicate the quadratic sum of random and systematic uncertainties. The random component includes uncertainties in photometry, spectroscopic [Fe/H], and Gaia parallax. For the systematic uncertainty, a $0.1$~dex in [Fe/H] is assumed to take into account a scale difference between our models and the spectroscopic determinations. A $20\%$ uncertainty in age is adopted for all stars. Flux differences from these systematic uncertainties are typically less than $0.01$~mag in $gizyBV$, but are as high as $0.05$~mag in $uv$ at high metallicities. In APASS, $0.04$~mag uncertainty in photometry is further incorporated into the final uncertainty (see above).

In Figures~\ref{fig:cteff} and \ref{fig:cteff2}, there are systematic differences between the two classes of samples. The stellar sequences tend to show smaller flux deviations from the models than the spectroscopic sample; for example, at [Fe/H] $\geq-0.4$, our result indicates that the spectroscopic sample is fainter at $\lambda<4000$~\AA, but brighter at $\lambda>7000$~\AA\ than the sequences. Indeed, the cluster sequences in PS1 $grizy$ \citep{bernard:14} exhibit the smallest differences overall. The observed discrepancy between stellar sequences and spectroscopic samples can be caused by inconsistent metallicity scales. However, other sources of errors, such as adopted ages, may also contribute to the observed offset, although the absolute model deviation changes monotonically with age, without modifying the observed wavelength-dependent offsets.

In Figures~\ref{fig:cteff} and \ref{fig:cteff2}, the red line indicates an average model deviation as a function of $\lambda_{\rm eff}$. Mean magnitude differences between models and observational data are computed by linearly interpolating values at three adjacent filter $\lambda_{\rm eff}$. They are smoothed by applying a boxcar average with a width of $1000$\ \AA, which is comparable to the FWHM of a broad filter passband. Average differences indicate that our models greatly over-estimate flux below $5000$\ \AA, by up to $20\%$, while they under-estimate flux at longer wavelengths, by $10\%$ at the most. The $BV$ photometry from APASS (open and filled downward triangles) exhibits consistently larger fluxes than the other calibration sample by $0.05$--$0.1$~mag, but the differences from the mean line are within our estimated $1\sigma$ uncertainties. We suspect all-sky photometric zero-point errors (at roughly $4\%$ levels) as a likely source of the systematics (see above).

\subsubsection{Construction of a Correction Cube}

\begin{figure*}
\centering
\includegraphics[scale=0.36]{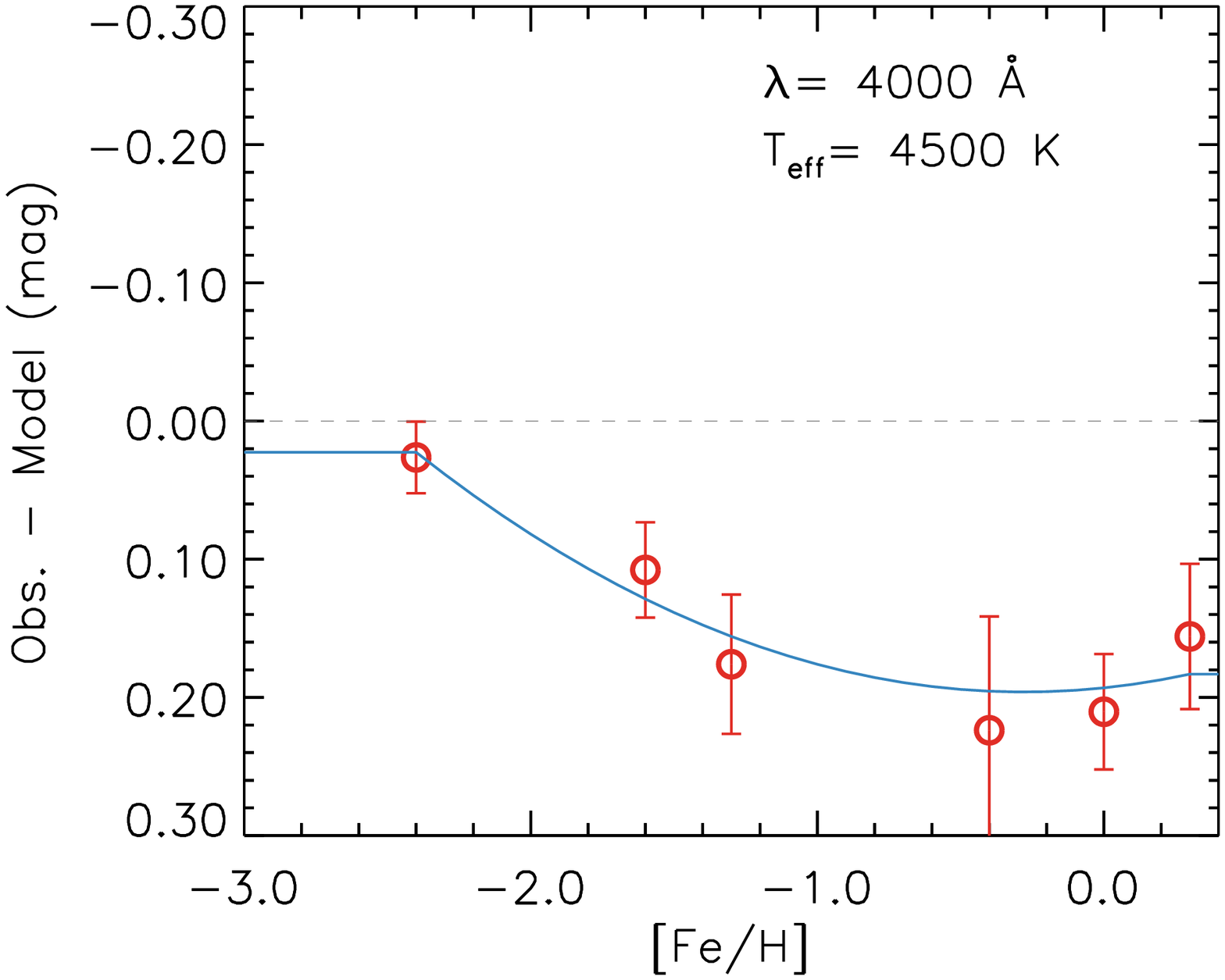}
\includegraphics[scale=0.36]{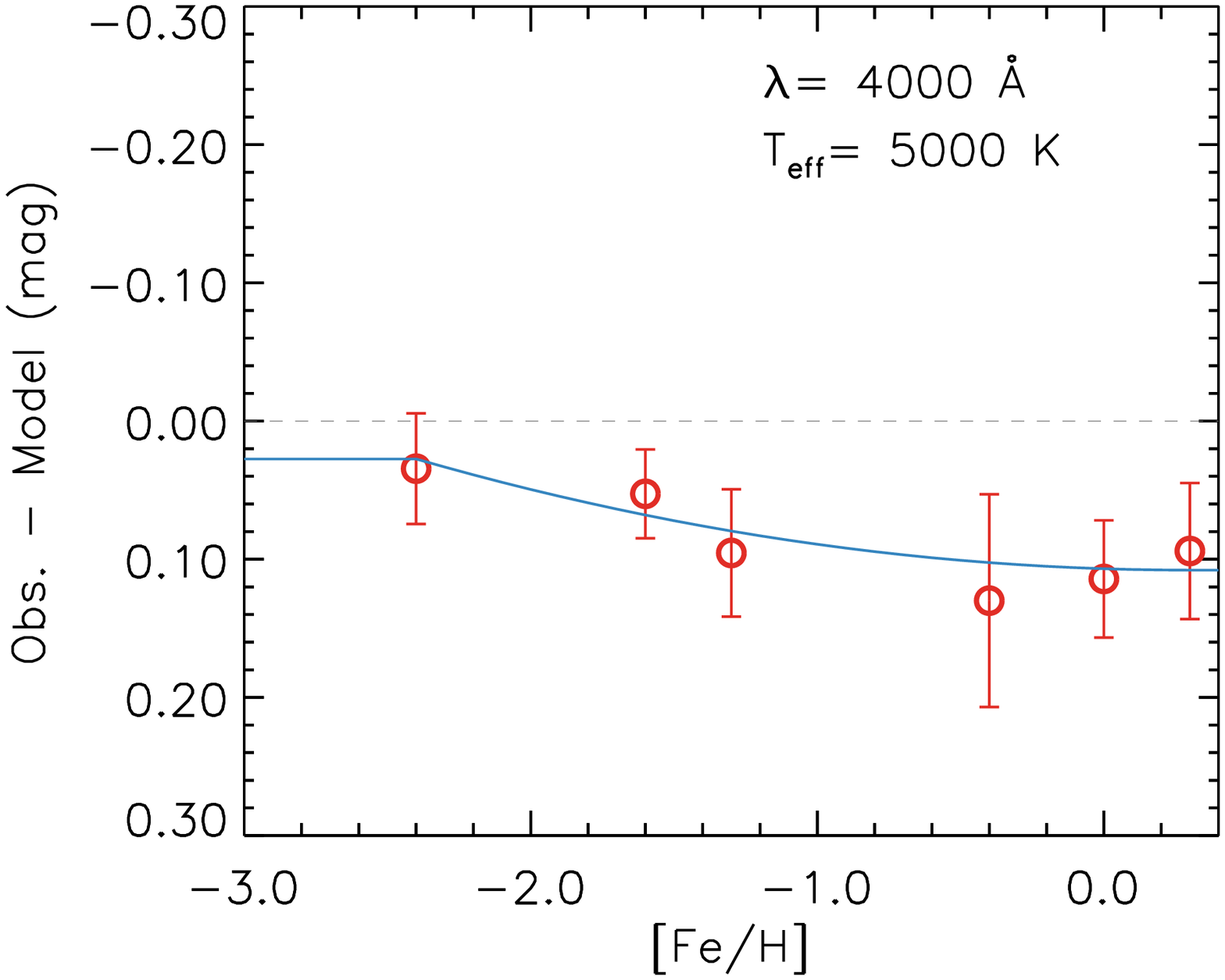}
\includegraphics[scale=0.36]{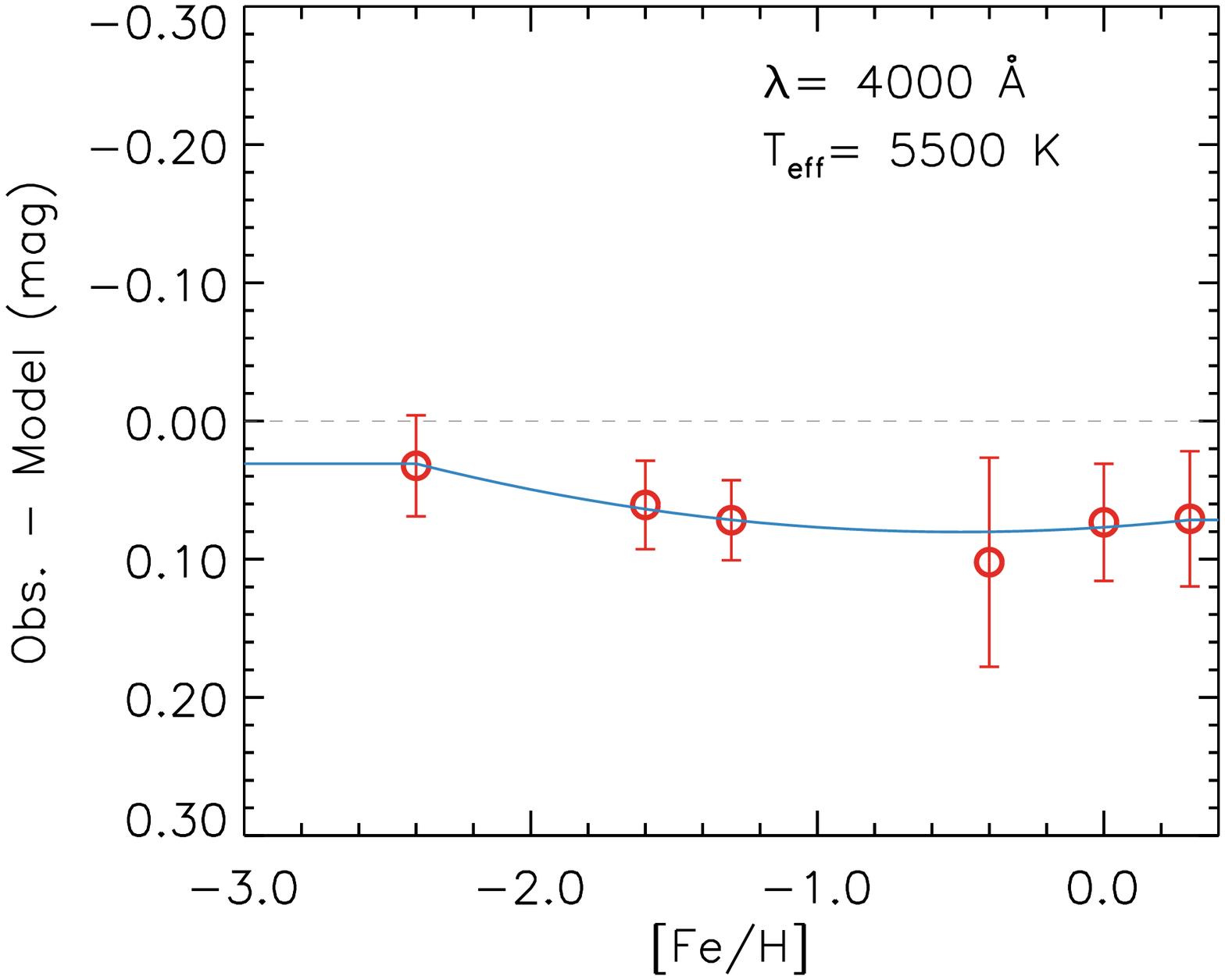}
\includegraphics[scale=0.36]{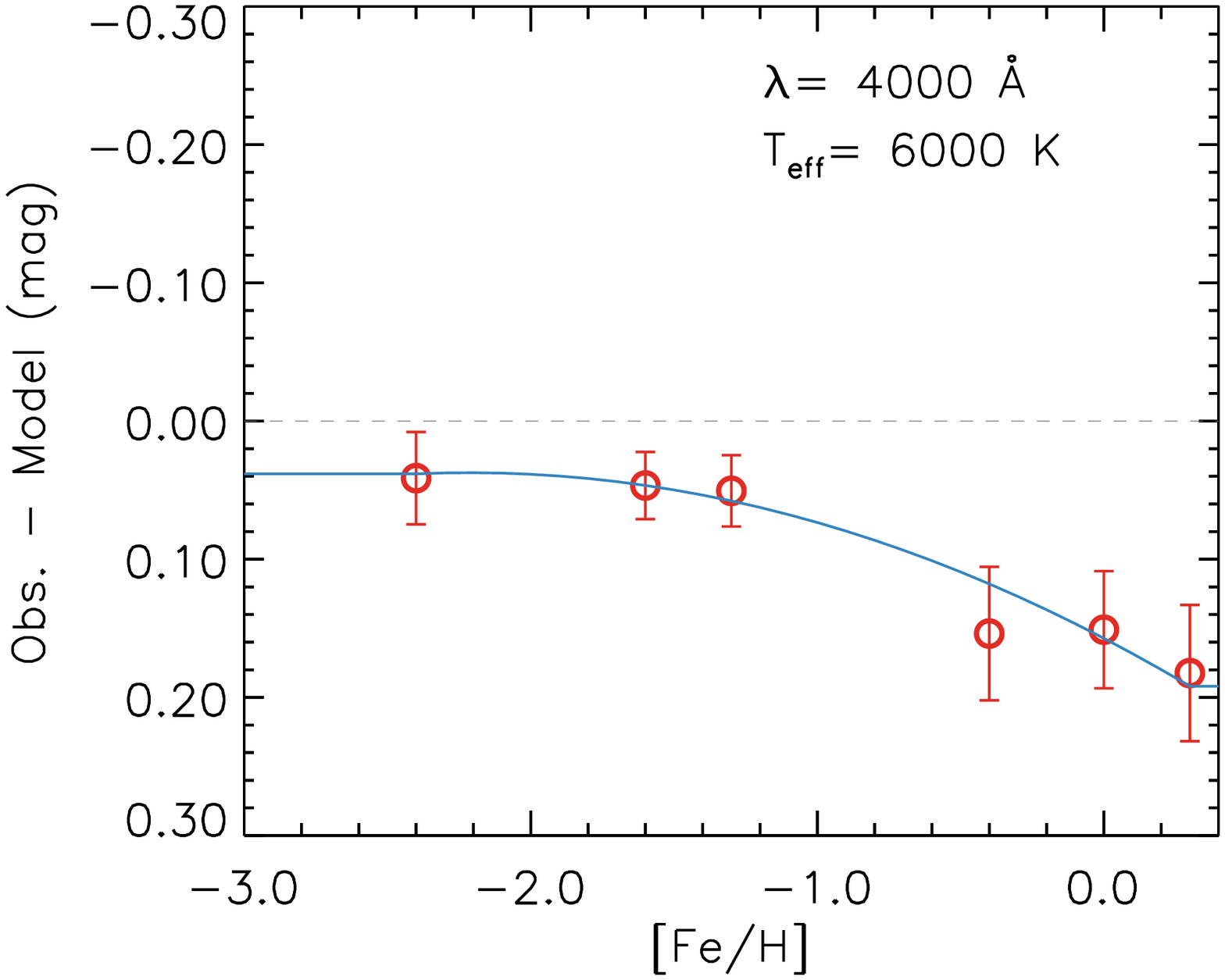}
\caption{Mean magnitude differences at $\lambda=4000$~\AA, as a function of [Fe/H], at selected $\teff$. Each point represents one of the metallicity groups in this work. The blue line represents the best-fitting $3$rd-order polynomial. Constant offsets are assumed beyond the metallicity range covered by the data.}
\label{fig:cteff3}
\end{figure*}

\begin{figure*}
\centering
\includegraphics[scale=0.32]{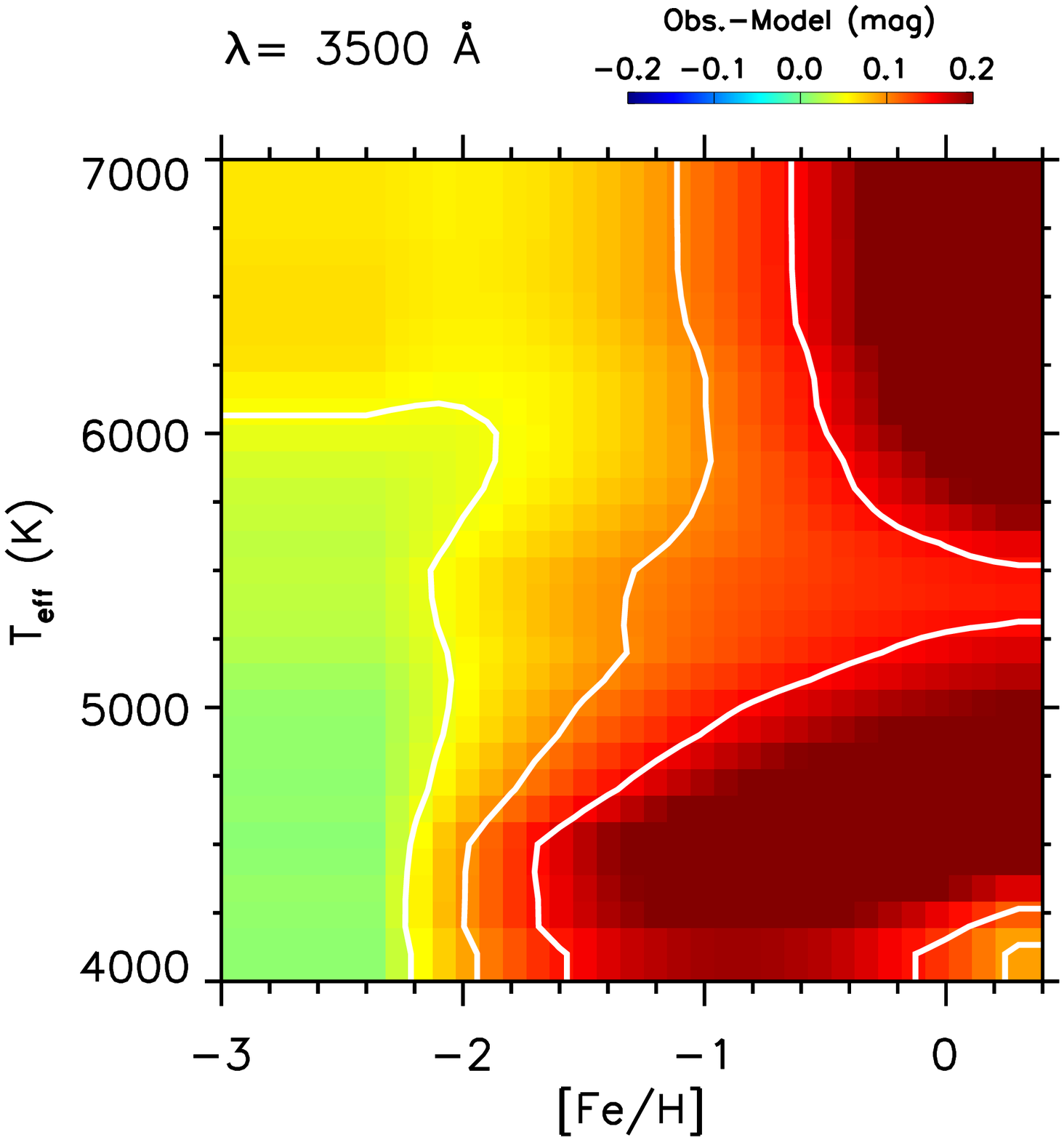}
\includegraphics[scale=0.32]{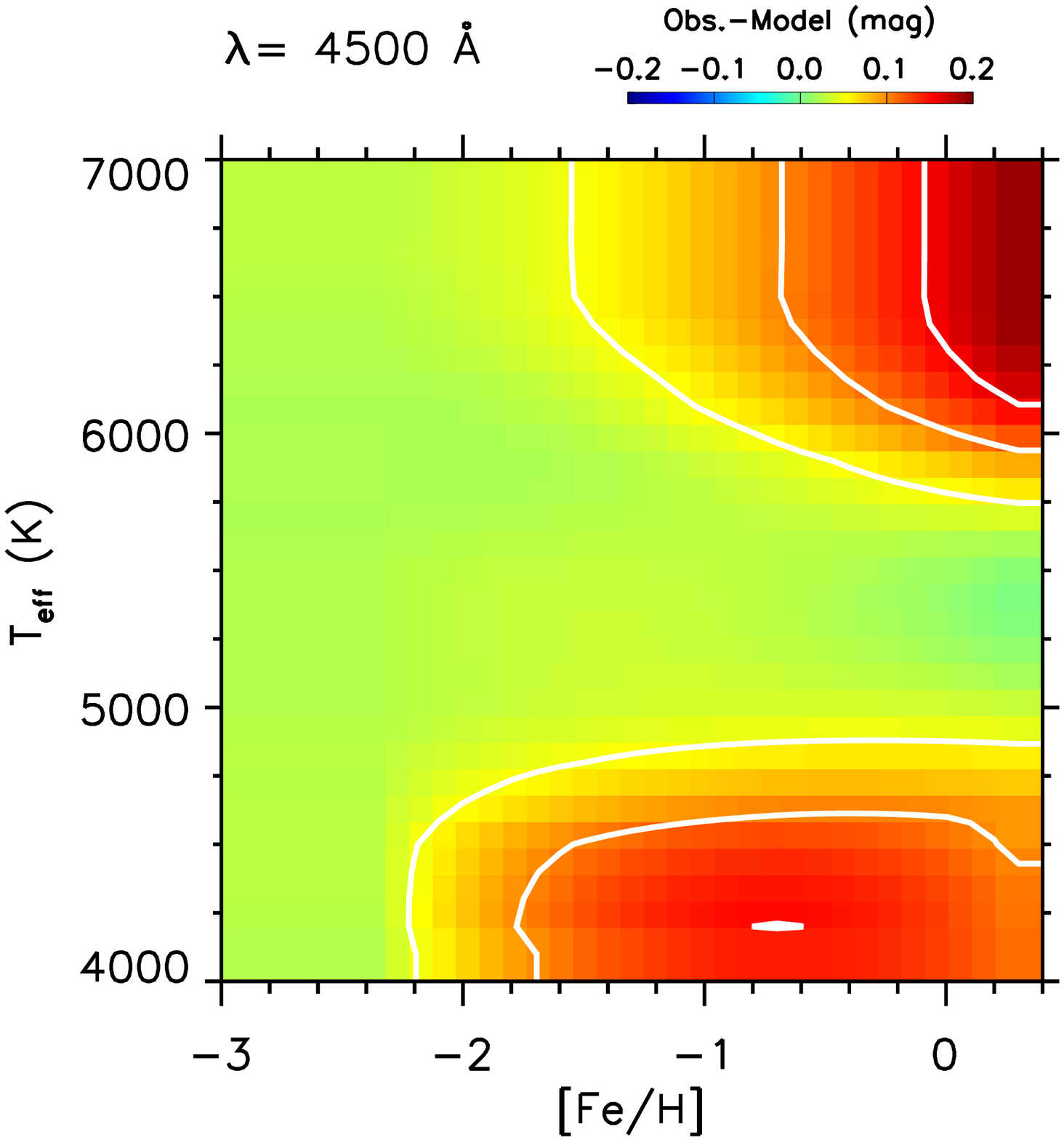}
\includegraphics[scale=0.32]{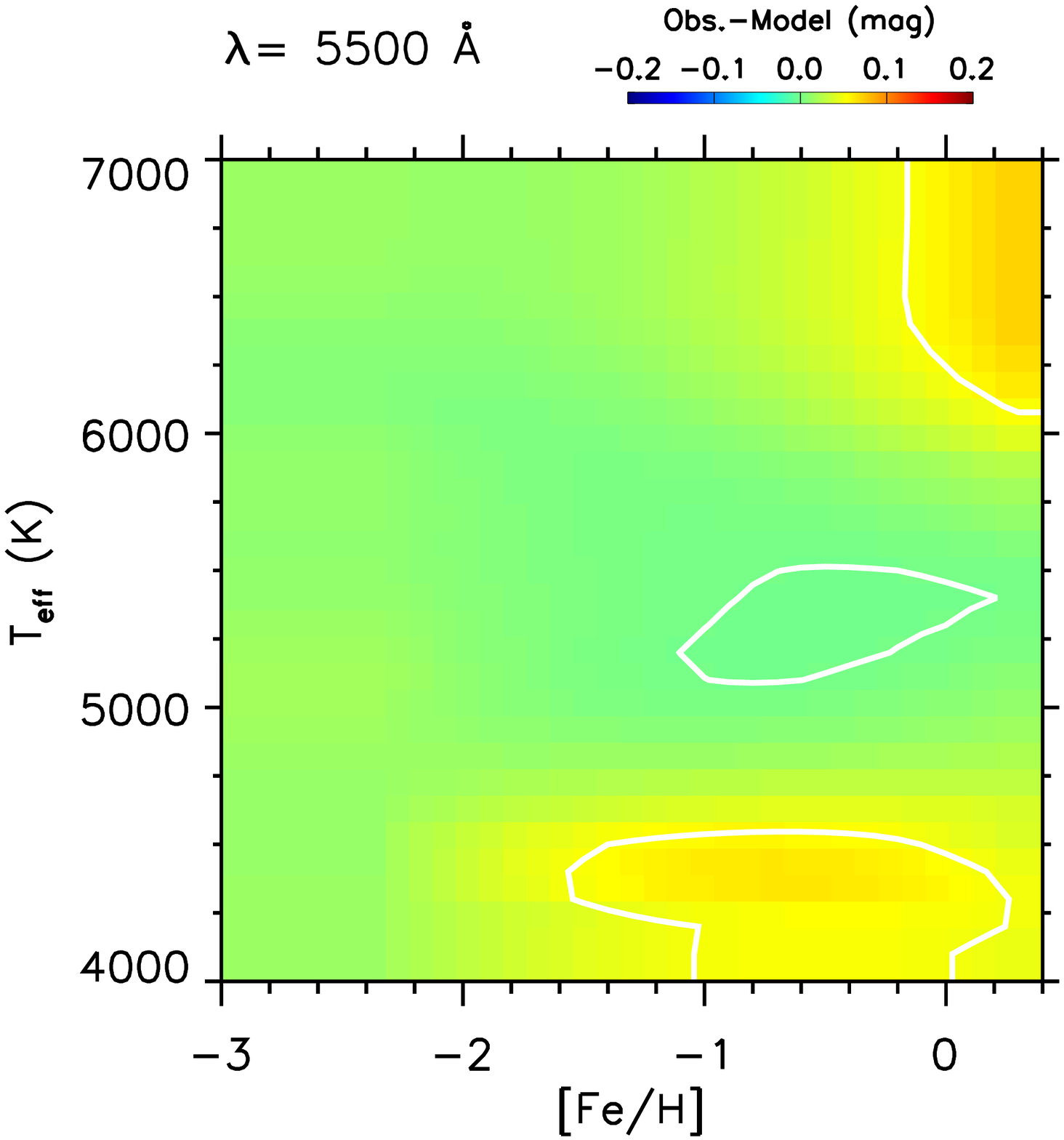}
\includegraphics[scale=0.32]{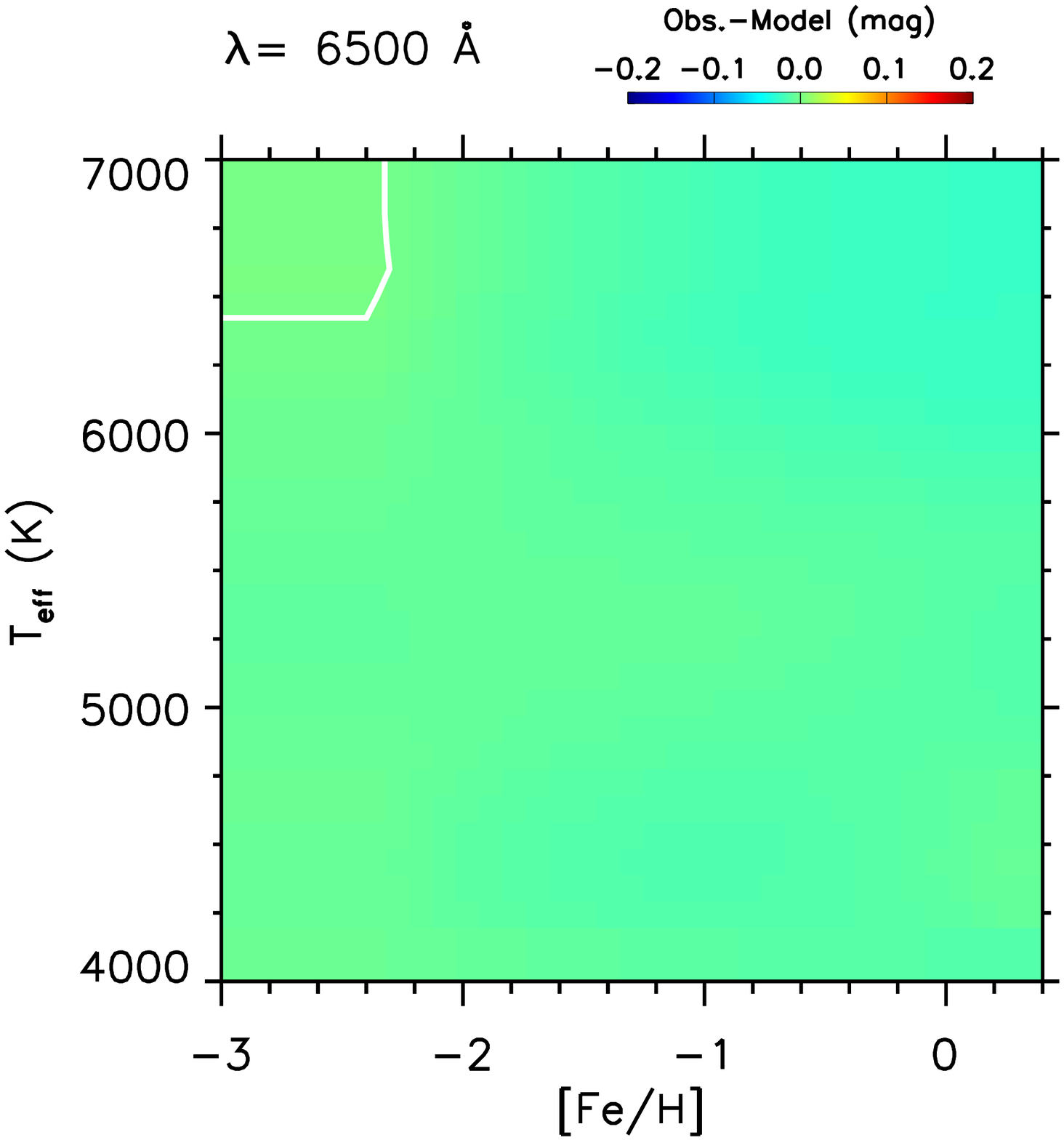}
\includegraphics[scale=0.32]{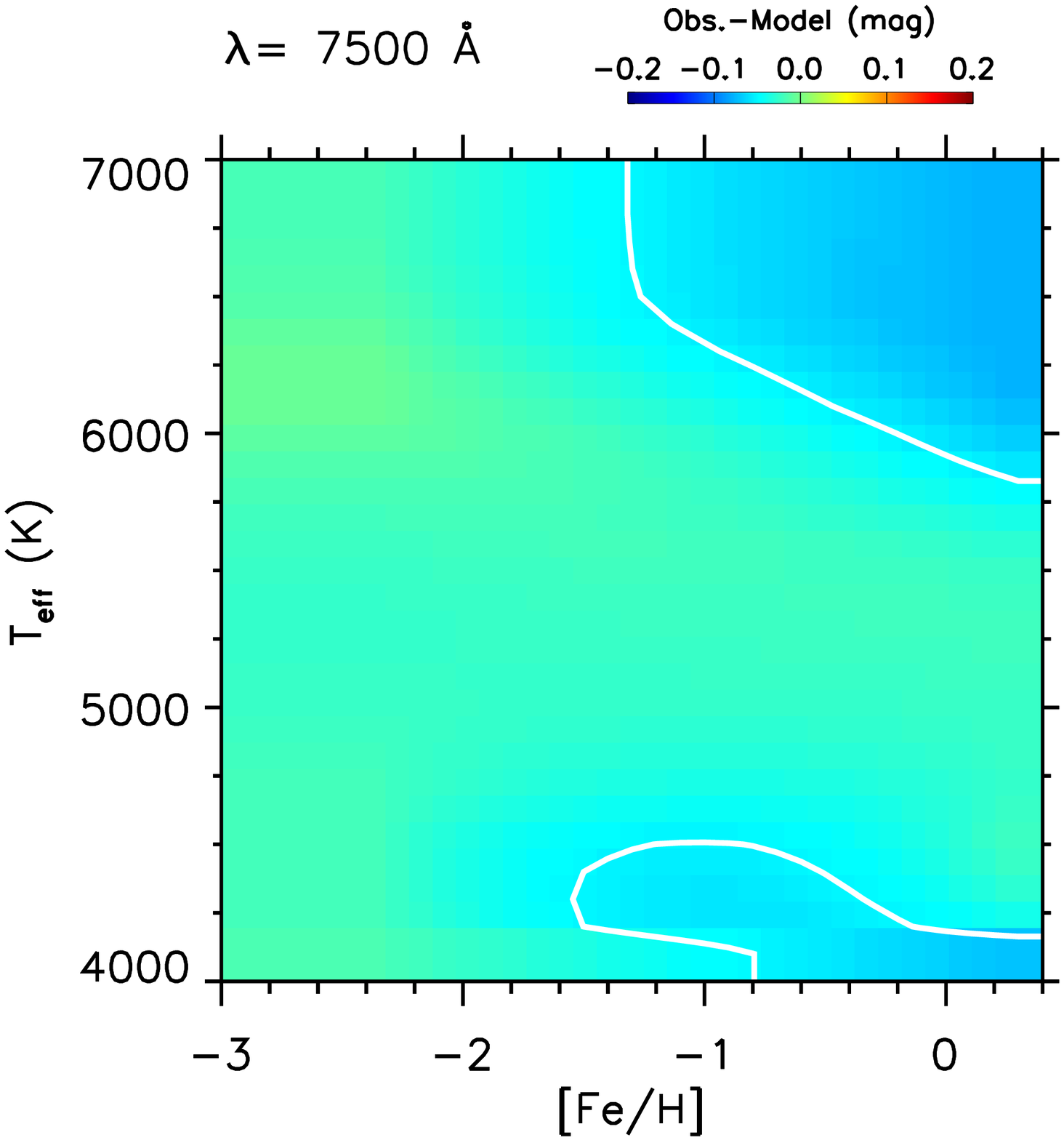}
\includegraphics[scale=0.32]{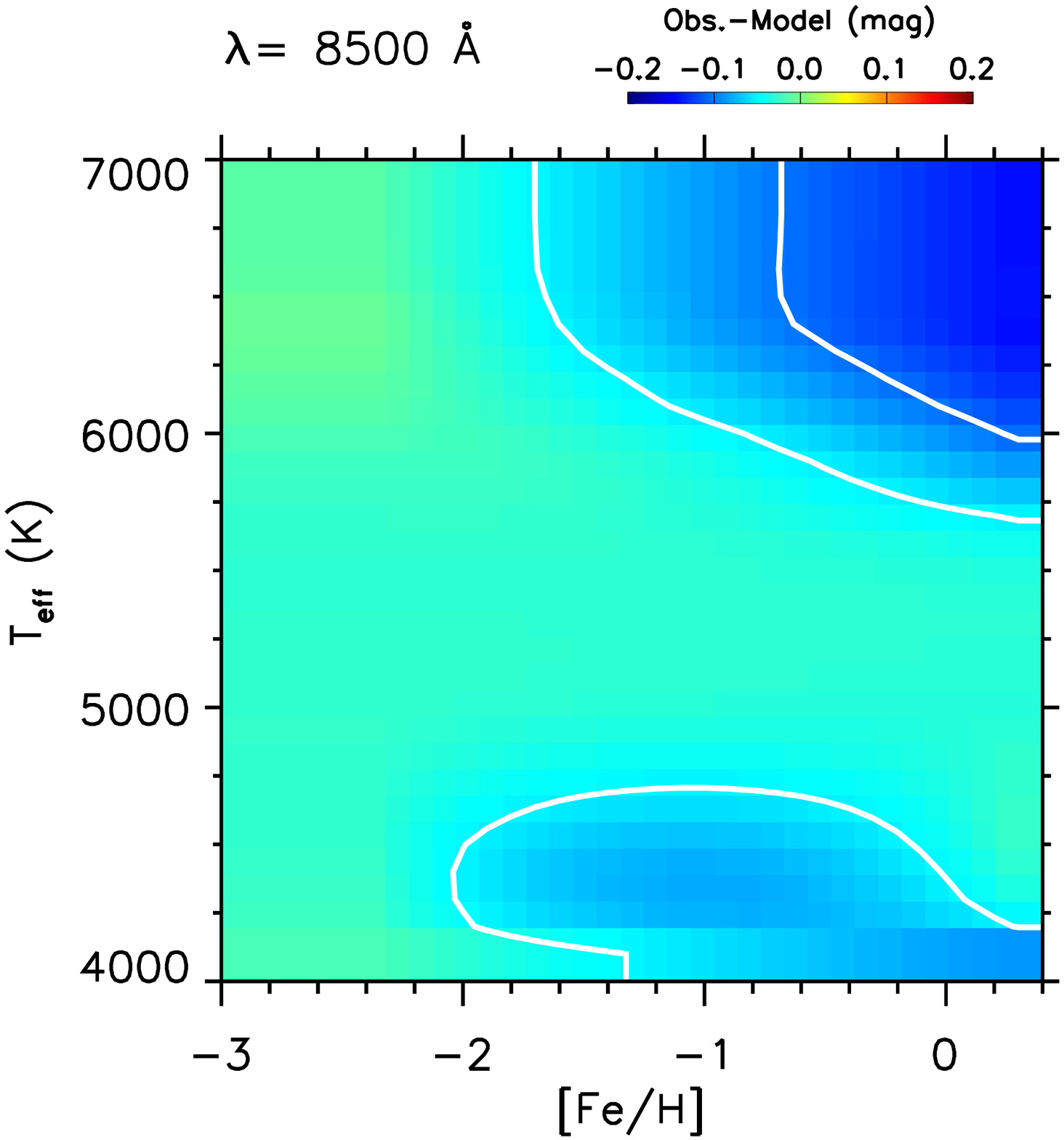}
\caption{Mean magnitude offsets of the original models from observational data in $\teff$ versus [Fe/H] at selected wavelengths.}
\label{fig:final}
\end{figure*}

Figure~\ref{fig:cteff3} displays model differences as a function of [Fe/H]. Each data point represents one of the metallicity groups in this work. As shown by the blue lines, a $3^{\rm rd}$-order polynomial function is used to depict the observed trend. Beyond the metallicity range covered by the sample, a constant offset is assumed in the model deviation. Figure~\ref{fig:final} displays slices of these mean model deviations at some selected wavelengths. Red colors indicate that models over-predict the flux, while the blue colors show regions with under-predicted flux. In this way, we construct a three-dimensional data cube of model deviations as a function of $\teff$, [Fe/H], and wavelength.

\begin{figure}
\epsscale{1.9}
\plottwo{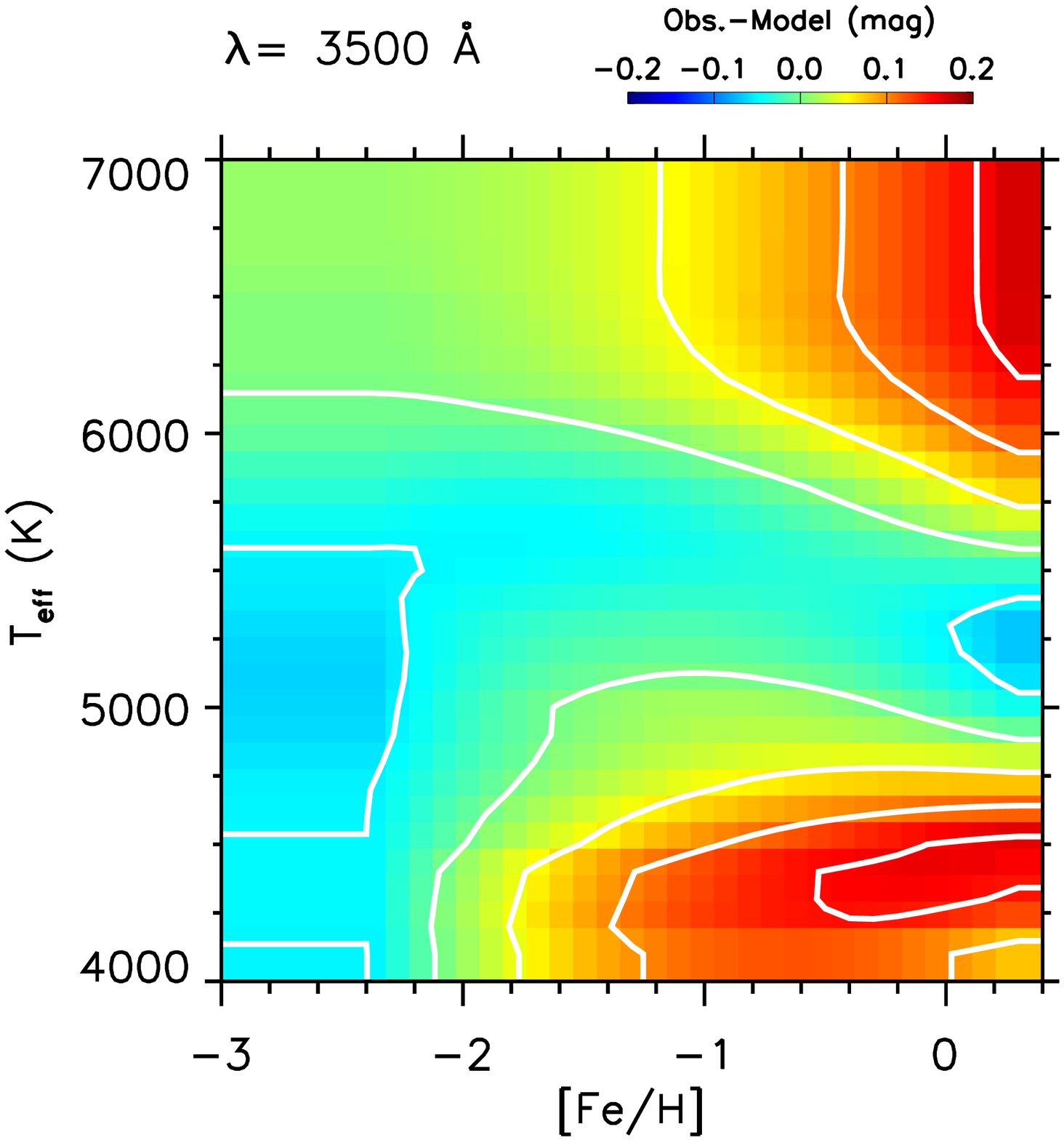}{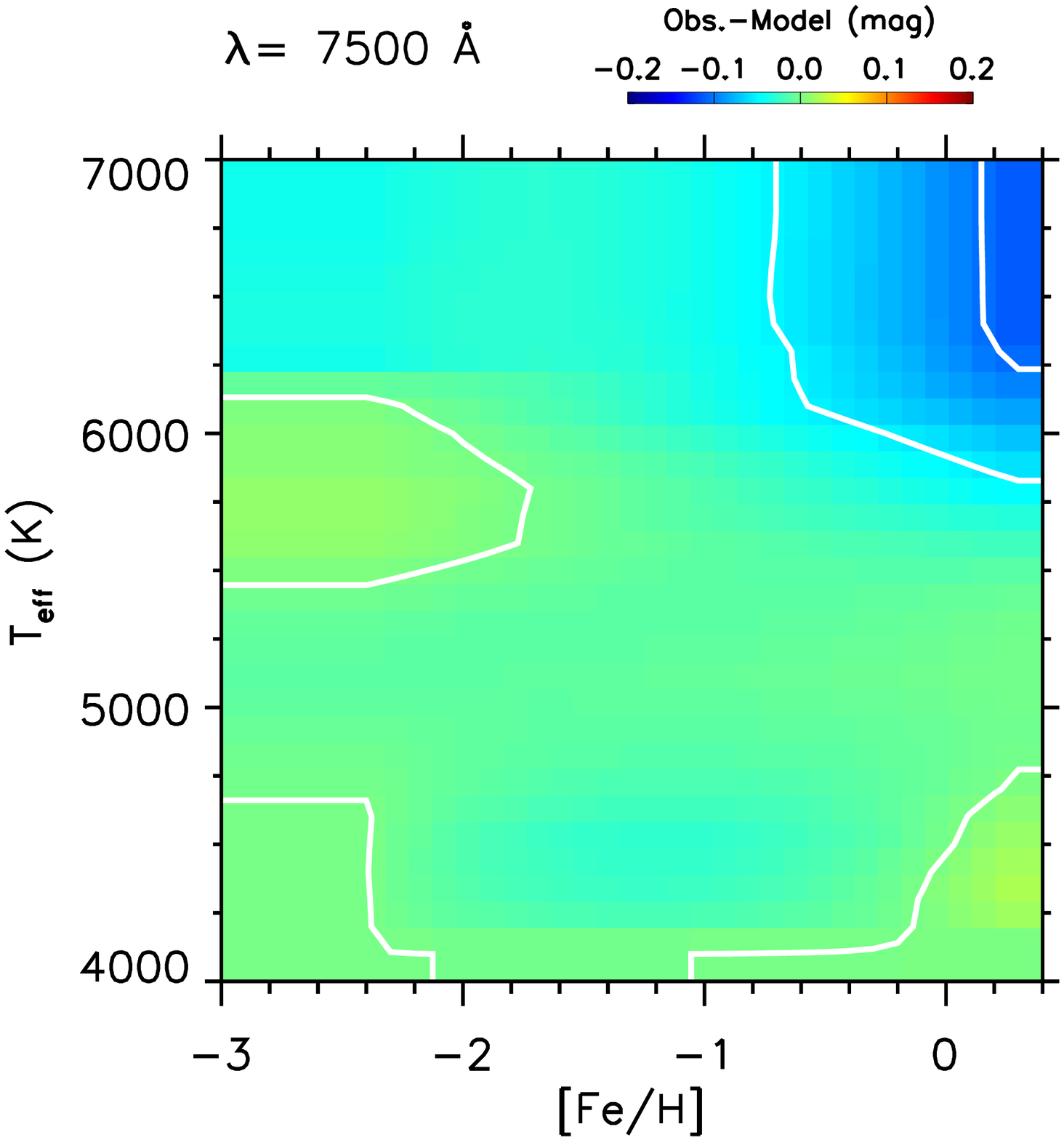}
\caption{Same as in Figure~\ref{fig:final}, but assuming $400$~K cooler $\teff$ in the YREC isochrones.}
\label{fig:alt}
\end{figure}

To first order, the model deviations change monotonically with wavelength: models are brighter than observations at shorter wavelengths, while the sign reverses at longer wavelengths. This may suggest an offset in the $\teff$ scale of the isochrones as a major source of the systematic mismatch. However, the required amount of offset must be very large, by about $400$~K, even for warm stars (Figure~\ref{fig:alt}); it is even larger for cooler stars. Even if there are systematic differences between different approaches of determining $\teff$, such as the IRFM and spectroscopic determination from excitation/ionization balance, this is beyond the accepted range of errors in the models. Therefore, it seems that the observed offsets originate from a combination of various sources of systematic errors, such as incorrect input physics or under-estimated line absorption in the models. Boundary conditions in stellar-interior models may also be incompatible with the atmosphere models in this study. On the observational side, an inconsistent metallicity scale, incorrect assumptions on elemental-abundance ratios, or errors in the assumed age could be responsible for the systematic $\teff$ offsets.

The model differences are highly non-linear in the [Fe/H] versus $\teff$ space, and no simple function can be adopted to remedy the problem. For this reason, we take the observed flux offsets in Figure~\ref{fig:final} as a correction matrix for our choice of stellar isochrones and synthetic spectra. More specifically, we employ a semi-empirical approach to correct synthetic spectra based on observations, while keeping stellar-interior models intact. Unlike in our previous work, the data cube in Figure~\ref{fig:final} provides a continuous function of magnitude correction in wavelength. Thus, the corrected synthetic spectra can be applied to any filter sets in the wavelength range covered by our calibration sample.

\subsubsection{Comparison with Previous Calibration}\label{sec:comparison}

In comparison to purely theoretical models, the net result of our empirical correction is redder colors or a higher photometric $\teff$, due to over-estimation of the model flux at shorter wavelengths and under-estimation at longer wavelengths. Apart from this fundamental change in the models, the revised calibration also differs from our earlier versions of the empirical corrections. The biggest change is the inclusion of individual spectroscopic targets in the sample, which inevitably modifies the metallicity scale of the models. As shown earlier in Paper~I, isochrones calibrated using fiducial clusters produce photometric metallicities that are in agreement with spectroscopic estimates in SEGUE within $\Delta {\rm [Fe/H]}\sim0.1$~dex at [Fe/H] $>-1.5$, but the difference amounts to $\sim0.4$~dex at [Fe/H] $=-2$, in the sense of a lower metallicity from our cluster-based approach. Such a difference is a direct consequence of a systematic offset in the metallicity scale between the cluster sequences and the SEGUE stars.

\begin{figure}
\epsscale{1.15}
\plotone{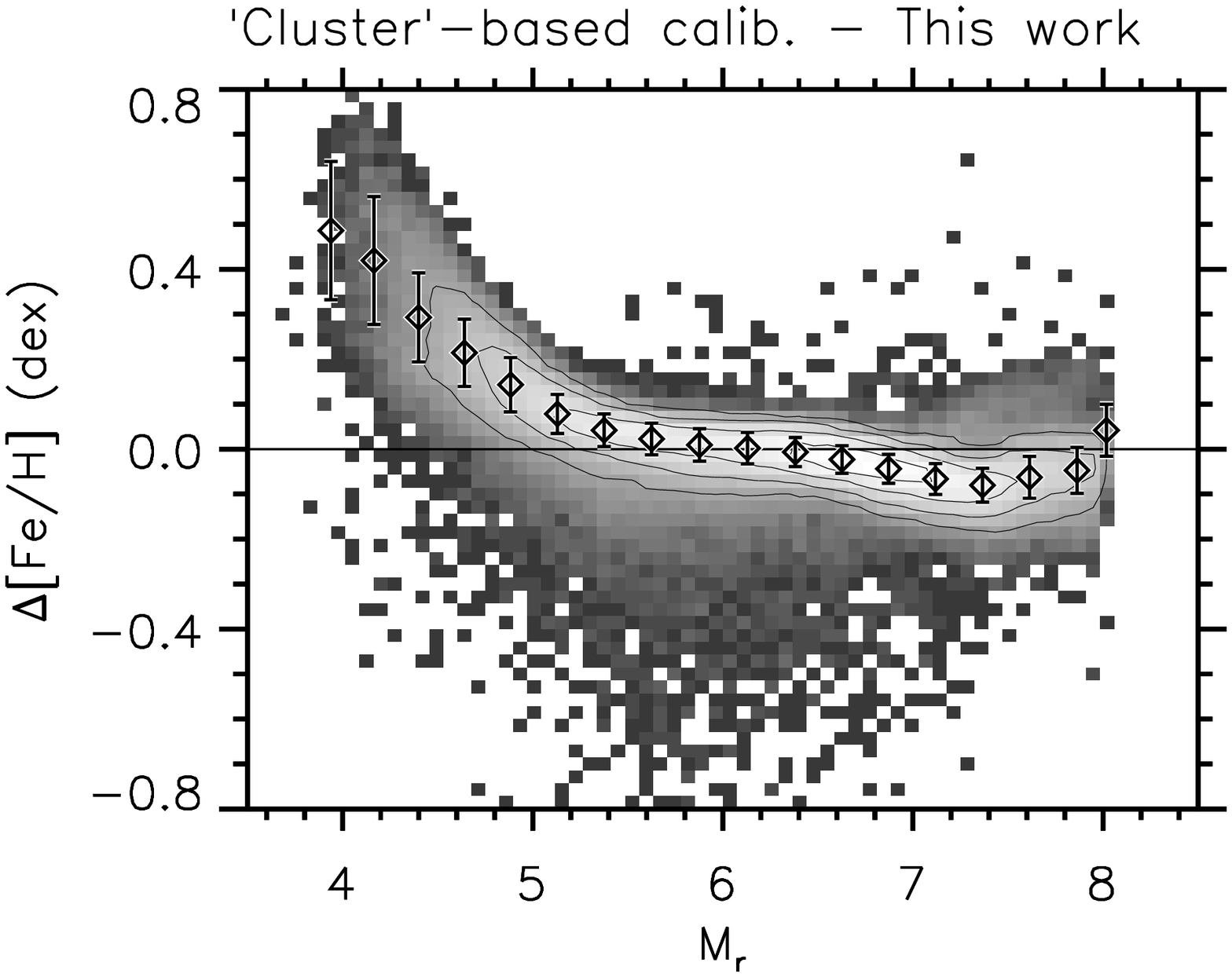}
\plotone{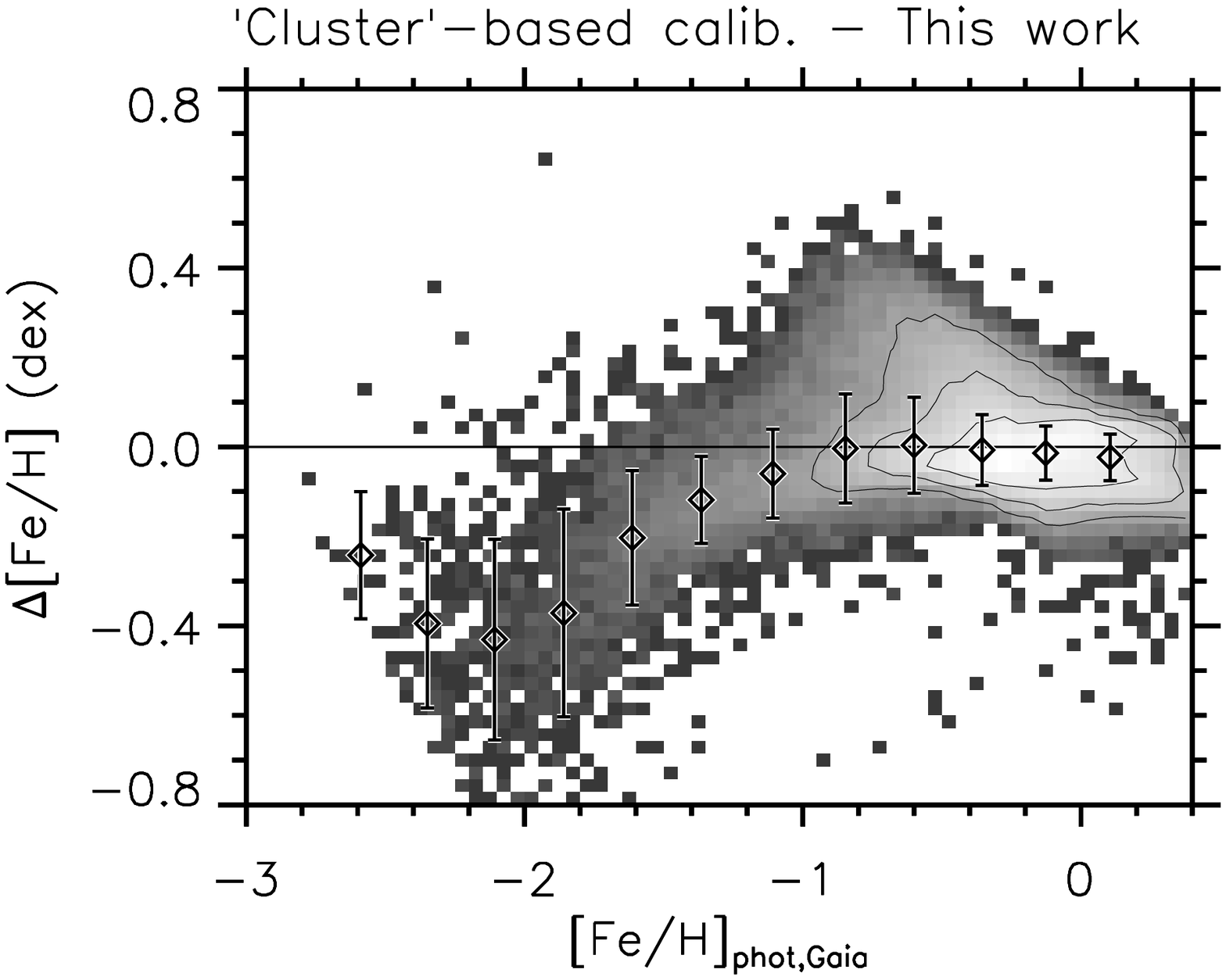}
\caption{Comparison between photometric metallicity estimates using calibrated models in Paper~III (cluster-based calibration) and this work (combined from both cluster sequences and spectroscopic samples). A logarithmic number density is shown with arbitrary contour levels. The open diamonds show a weighted median difference in bins of $\Delta M_r=0.25$ and $\Delta {\rm [Fe/H]}=0.25$ dex, respectively, with error bars indicating a standard deviation of the differences. Only those having $\sigma_\pi/\pi < 0.1$ are included in the above comparisons; the data in the bottom panel are further restricted to $4.5 < M_r < 7.5$. Other constraints are the same as in the main analysis of this work.}
\label{fig:comp}
\end{figure}

Figure~\ref{fig:comp} compares photometric metallicities from Paper~III with those in this work, based on the revised calibration. The comparisons are shown for the solutions based on Gaia parallaxes using SDSS and PS1 photometry. Only stars having parallax uncertainties less than $10\%$ are included. Other constraints are the same as for the main sample, as described in the next section, except in the top panel, where the metallicity difference is displayed over the full range of $M_r$. The large deviations for bright stars are evident, which are caused by the lack of hot stars in the calibration sample (Figure~\ref{fig:sample}). In our subsequent analysis, including the bottom panel of Figure~\ref{fig:comp}, we adopt $4.5 < M_r < 7.5$ to avoid regions with potentially large calibration errors.

The weighted median difference in Figure~\ref{fig:comp} indicates that the two calibration versions agree at high metallicity ([Fe/H] $>-1$), but that photometric metallicities from this work become larger for metal-poor stars, amounting to $0.35$~dex at [Fe/H] $=-2$. Because the GALAH sample is confined to metal-rich stars in our calibration, the systematic trend highlights an inconsistent metallicity scale between the cluster sequences and SEGUE stars. The SSPP estimates have been checked thoroughly using clusters and high-resolution spectroscopic abundance determinations, and the overall agreement is impressive \citep{rockosi:22}. Nonetheless, such comparisons were performed mostly using giants and main-sequence turn-off stars in the metal-poor regime, due to the lack of metal-poor main-sequence dwarfs in the SEGUE sample with a sufficiently high SNR. Therefore, the difference may originate from an internally inconsistent metallicity scale between dwarfs and giants in the SEGUE sample.

In summary, because neither metallicity scale is preferred over the other, the revised isochrones obtained in the current experiment should be taken as an alternative to our earlier cluster-based calibration. More precisely, a metallicity distribution of metal-poor stars in this study is hinged on an intermediate metallicity scale between SEGUE and the cluster-based work, set by their relative weights to the final calibration sample. The sense is that metallicity estimates in this study are systematically higher than those in our previous work at [Fe/H] $<-1$.

\begin{figure}
\epsscale{1.15}
\plotone{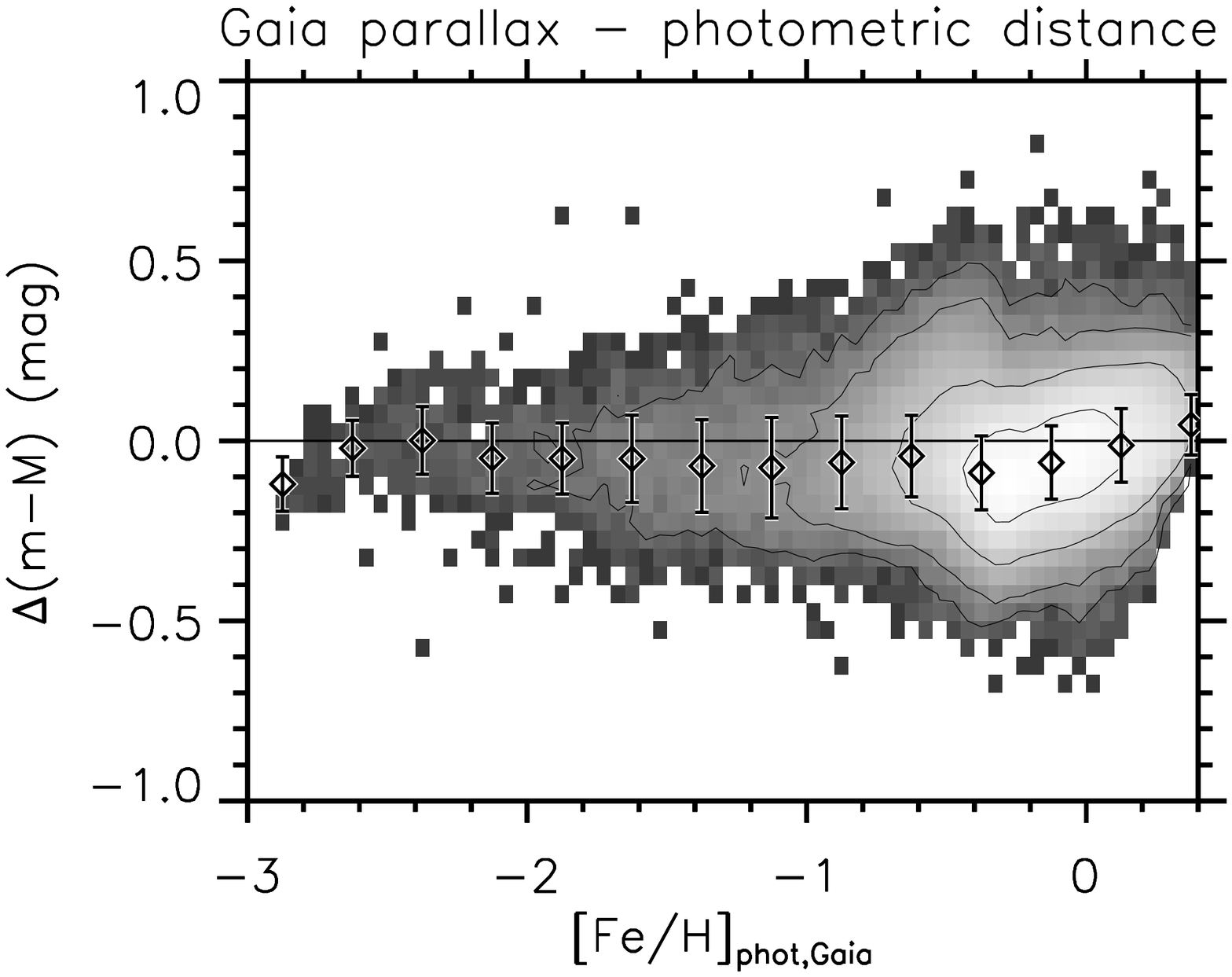}
\caption{Comparison of distance moduli estimated using calibrated models in this study with Gaia EDR3 parallaxes. The sample cuts used in the bottom panel of Figure~\ref{fig:comp} are applied, including a requirement of $\sigma_\pi/\pi < 0.1$. The plot shows a logarithmic number density with arbitrary contour levels. The open diamonds represent the weighted median difference in bins of $\Delta {\rm [Fe/H]}=0.25$ dex, with error bars indicating the standard deviation of the differences.}
\label{fig:comp2}
\end{figure}

Figure~\ref{fig:comp2} shows a comparison between our distance estimates based on SDSS (combined with PS1) and Gaia parallaxes. In this case, we use distances that are determined jointly with metallicity, without relying on Gaia parallaxes. The same set of data as in the bottom panel of Figure~\ref{fig:comp} is used, but a more stringent test is performed by restricting a comparison to those having a reduced $\chi^2$ fit of the models equal to unity. We note that the majority of unresolved photometric binaries should have been rejected in Figure~\ref{fig:comp2} (and the bottom panel of Figure~\ref{fig:comp}), as we display photometric metallicity estimates at Gaia parallaxes on the abscissa. This is because unresolved binaries are systematically brighter than single stars, and therefore show poor model fits to the observed fluxes, if a distance is fixed at a true value in the metallicity estimates \citep[see also][]{an:13}. Our models produce a local distance scale that differs by at most $5\%$ from Gaia EDR3 parallaxes. Error bars represent a standard deviation of the differences in bins of ${\rm [Fe/H]}$, but the errors in the `mean' differences are very small ($<1\%$). In fact, the above good agreement with Gaia distances is not unexpected, since our calibration relies on Gaia parallaxes for nearby spectroscopic samples. We can expect a similar level of agreement with \citet{bailerjones:21}, as their Bayesian distance estimates closely align with those in the Gaia catalog when the parallax accuracy is less than $10\%$.

Figure~\ref{fig:comp2} shows that the weighted standard deviation of the differences in distance is approximately $\sigma [\dmn] = 0.10$--$0.14$~mag across the range of metallicities displayed. However, the quadrature sum of uncertainties from both methods is estimated to be in the range of $0.2$--$0.3$~mag. This suggests that our distance measurement uncertainties may be overestimated by a factor of approximately three. The comparison with distance estimates based on SMSS (combined with PS1) also shows a similar level of discrepancy. One possible explanation for this discrepancy is that our model fitting does not account for correlations between photometric uncertainties in SDSS passbands.

In addition, we compare our photometric metallicity estimates with metallicities from the GALAH sample. For stars with [Fe/H] $>-1$, the weighted standard deviation of the difference in metallicity is $\sigma {\rm ([Fe/H])} \approx 0.15$~dex, when photometric metallicities are estimated using SDSS or SMSS photometry without Gaia parallax priors. On the other hand, the expected value from propagation of uncertainty measurements is nearly $0.25$~dex, indicating that our estimated uncertainties of photometric metallicities are overestimated by a factor of about two. However, when Gaia parallax priors are used in the computation of photometric metallicity, the difference between the estimated standard deviation and the propagated value is marginal; our measurement uncertainties are overestimated by only up to approximately $25\%$, depending on the stellar metallicity.

\section{Chemo-Kinematical Properties of the Local Halo}\label{sec:phase}

In this section, we apply our newly calibrated set of models to large photometric catalogs, and provide new insights for Galactic stellar populations in the local volume. In addition to chemical information from photometry, we exploit kinematic data from Gaia to generate phase-space maps at various distances from the Galactic plane and Galactocentric distances. As in our previous papers, we restrict our analysis to a strip within $\pm30\arcdeg$ from the Galactic prime meridian ($l=0\arcdeg$ and $180\arcdeg$), where a conversion from transverse motions into $\vphi$ is reliable (see Paper~III for more details). Below, we first inspect metallicity distributions and phase-space diagrams to validate our new calibration (\S~\ref{sec:validation}), and present distributions of scale heights and lengths for each group of stars in bins of $\vphi$ and [Fe/H] (\S~\ref{sec:scale}). Based on phase-space diagrams of high proper motion stars, we demonstrate that our data reveal yet another stellar population formed during a period of Galactic starburst activity (\S~\ref{sec:starburst}).

\subsection{Validation of Photometric Metallicity Estimates}\label{sec:validation}

In the following applications, we use Gaia EDR3 as a main source catalog, since $\vphi$ in the rest frame of the Galaxy is a major ingredient of our phase-space diagrams. We combine Gaia EDR3 astrometric data with photometry in SDSS, SMSS, and PS1 using a $1\arcsec$ match radius. As there is a little overlap between SDSS and SMSS, two sets of catalogs --- SDSS $\cap$ Gaia and SMSS $\cap$ Gaia, respectively --- are created, in addition to a master photometric catalog from all survey data [(SDSS $\cup$ SMSS) $\cap$ Gaia]. PS1 $grizy$ photometry is added to each data set, when there exist SDSS or SMSS photometry, to better constrain the stellar parameters.

To estimate metallicities for individual stars in each catalog set, we employ the calibrated models and conduct a grid search. Gaia parallaxes degrade rapidly beyond $\sim2.5$ kpc from the Sun, whereas distance uncertainties exhibit a more gradual increase when distances are derived photometrically (see Appendix~\ref{sec:error}). Therefore, we calculate photometric metallicities under two conditions: with or without priors from the Gaia parallax. The former approach enables us to obtain photometric metallicities based on the best available parallax data, but is restricted to nearby stars. In contrast, the purely photometric approach determines both distance and metallicity simultaneously, and covers a larger volume of space. To assess the uncertainty in metallicity, we determine the $\Delta \chi^2 = 2.3$ boundary for two degrees of freedom when Gaia parallaxes are used in the parameter estimation ($\Delta \chi^2 = 3.53$ when using a purely photometric approach). In cases where we utilize Gaia priors, we determine the difference in the metallicity estimation from the $\pm1\sigma$ uncertainty in parallax and add it to the uncertainty in quadrature.

We have established certain criteria in our analysis to ensure the quality and reliability of our results. First, we only consider sources that have been detected in at least five photometric passbands in SDSS or SMSS, which guarantees that sources are observed in at least $u$ or $v$. Additionally, we require that solutions have a reasonable fit to the model, as indicated by a reduced $\chi^2$ of the best-fit model being less than 3, provided they are within the range of $4.5 < M_r < 7.5$. To ensure accuracy in our photometric estimates, we exclude low-latitude regions with $|b| < 20\arcdeg$ and some areas where cumulative extinction exceeds $\ebv = 0.1$. Moreover, we set a minimum safeguard by setting a maximum allowable uncertainty in metallicity of less than $1.5$ dex. By implementing these selection criteria, we aim to minimize the potential for systematic errors and ensure that our results are of high quality and accuracy.

\begin{figure*}
  \centering
  \gridline{\fig{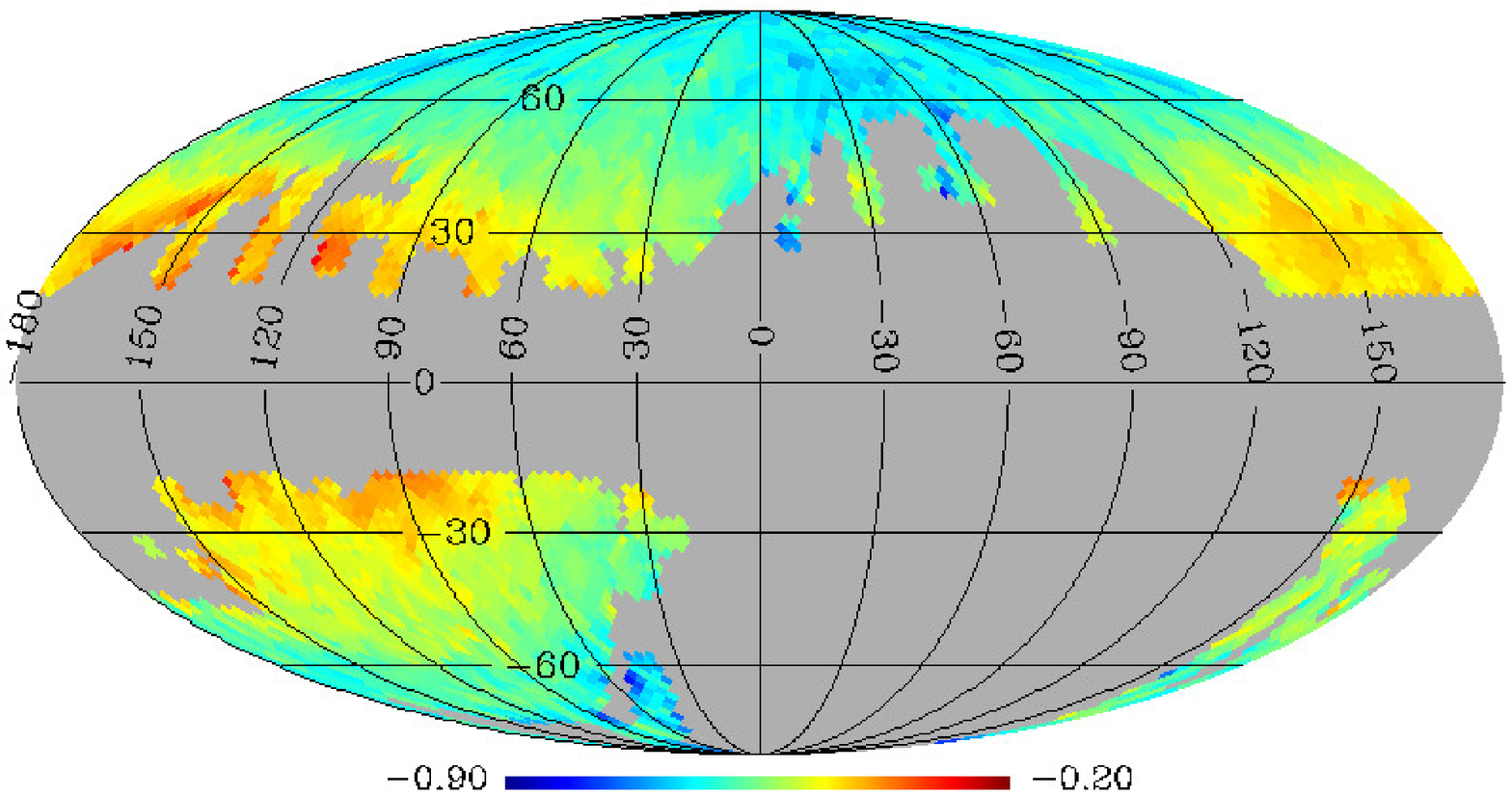}{0.4\textwidth}{\textbf{(a) SDSS $\cap$ Gaia}}
                \fig{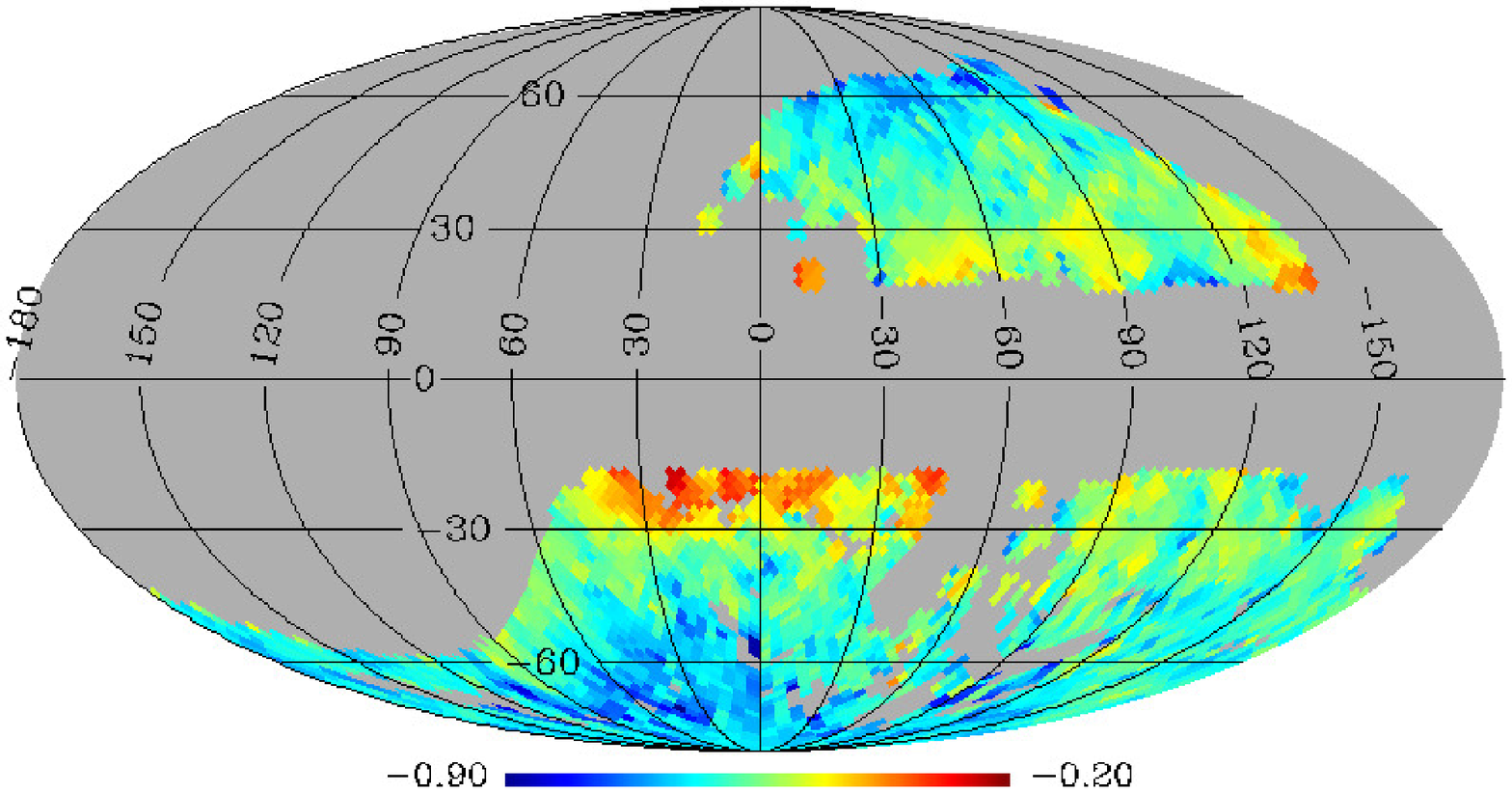}{0.4\textwidth}{\textbf{(b) SMSS $\cap$ Gaia}}}
  \gridline{\fig{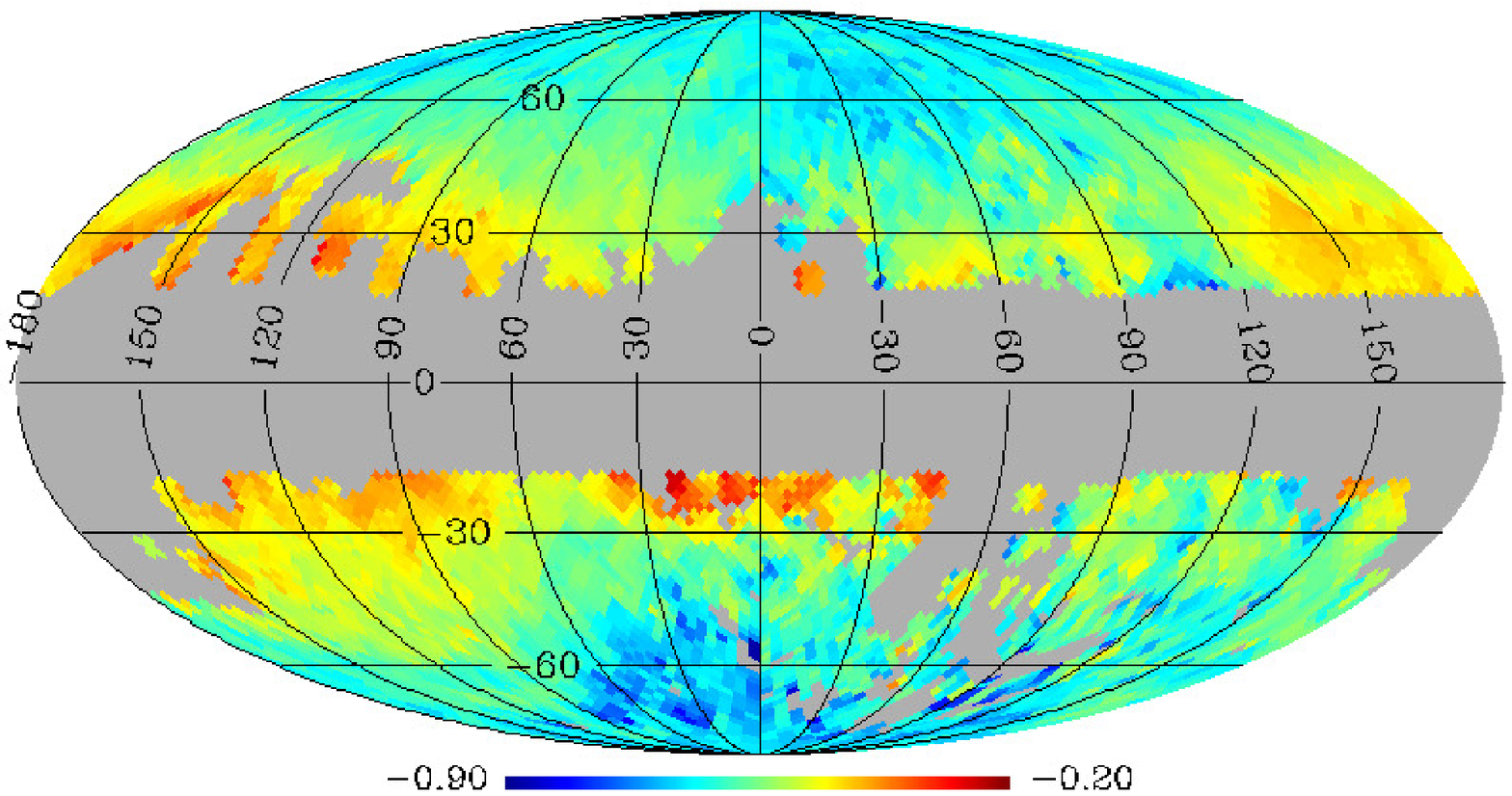}{0.7\textwidth}{\textbf{(c) (SDSS $\cup$ SMSS) $\cap$ Gaia}}}
\caption{A global metallicity map using the Mollweide projection in the Galactic coordinate system. Weighted mean metallicities ([Fe/H]) are shown for stars at $1.2 < d < 3$~kpc from (a) SDSS, (b) SMSS, and (c) SDSS $\cup$ SMSS, where PS1 photometry is used if available. All metallicity estimates are derived using Gaia parallaxes. Only stars with $\sigma_\pi/\pi < 0.3$ and $|b| > 20\arcdeg$ are shown, while high-extinction regions with $\ebv \geq 0.1$ are excluded. Each pixel has an area of $3.3\ {\rm deg}^{2}$ at $N_{\rm side}=32$ in HEALPix.}
\label{fig:map}
\end{figure*}

Figure~\ref{fig:map} shows metallicity maps of stars in the local volume ($1.2 < d < 3$~kpc) in the Galactic coordinate system (Mollweide projection), based on our metallicity estimates for individual stars in each of the three combined catalogs. Panel~(a) displays a map from SDSS $\cap$ Gaia (3.2 million stars), which mostly covers the Northern Galactic Hemisphere, while panel~(b) shows the Southern Hemisphere from SMSS $\cap$ Gaia (0.6 million stars). Panel~(c) displays a full coverage map from SDSS, SMSS, and PS1 (3.7 million stars). All three maps are smoothed using a median filter with a $2\arcdeg$ radius.

There is a limited overlap between SDSS and SMSS, mainly along the celestial equator ($\approx80,000$ stars). In these overlapping areas, photometric [Fe/H] estimates derived from individual catalogs (SDSS $\cap$ Gaia or SMSS $\cap$ Gaia) agree with those based on a combined catalog (SDSS $\cap$ SMSS $\cap$ Gaia) within $0.1$~dex. Metallicity differences for individual stars also do not exhibit a systematic trend with metallicity, which provides a confirmation of the internal consistency in our models. Nonetheless, metallicity distribution functions from these subsets are not identical in Figure~\ref{fig:map}, owing to unequal depths and qualities of these photometric surveys, which result in mild systematic differences in the mean metallicities.

Reassuringly, Figure~\ref{fig:map} reveals that more metal-rich stars are found near the Galactic plane, from which a mean metallicity gradient is evident from the low to high Galactic latitudes, as expected from a simple population gradient. This exercise proves not only that our technique can be used to determine metallicities of stars precisely from multi-band photometry, but also that we can use corrected synthetic spectra to combine data in various filter sets to generate an internally consistent all-sky metallicity map. Our calibration procedure is currently valid for main-sequence stars, and has lower precision for giant stars. Consequently, the above mapping based on main-sequence stars probes a local volume out to $\sim6$~kpc from the Sun. Giant stars are excluded in our sample using color-magnitude relations based on Gaia parallaxes, although a purely photometric approach can also be employed to tag such stars, as demonstrated in Paper~III.

Our photometric technique is a sensitive probe of photometric zero-point errors. It is particularly useful for large photometric surveys because it is non-trivial to have an internally consistent photometric zero point across large areas on the sky. In Appendix~\ref{sec:zp}, we demonstrate the existence of spatially correlated photometric zero-point errors in SDSS $u$ and SMSS $uv$ based on our corrected models. The size of photometric zero-point offsets is a few hundredth of magnitude level, but is as high as $0.1$~mag in some areas. By inverting the problem, zero-point offsets in photometry can be derived to make a uniform mean metallicity of nearby stars on the sky. This backward design on photometric zero-point corrections improves the quality of the metallicity mapping and somewhat narrows the gap in model deviations between the cluster sequences and the SEGUE sample. For this reason, we iterate the calibration procedure (\S~\ref{sec:calib}) using zero-point corrected photometry in SDSS $u$ and SMSS $uv$. In the following analysis, including Figure~\ref{fig:map}, all input photometry is corrected for the spatially correlated zero-point offsets.  These steps closely parallel similar exercises in zero-point corrections for the SMSS $uv$ in \citet{huang:21,huang:22}.

\begin{figure*}
  \centering
  \gridline{\fig{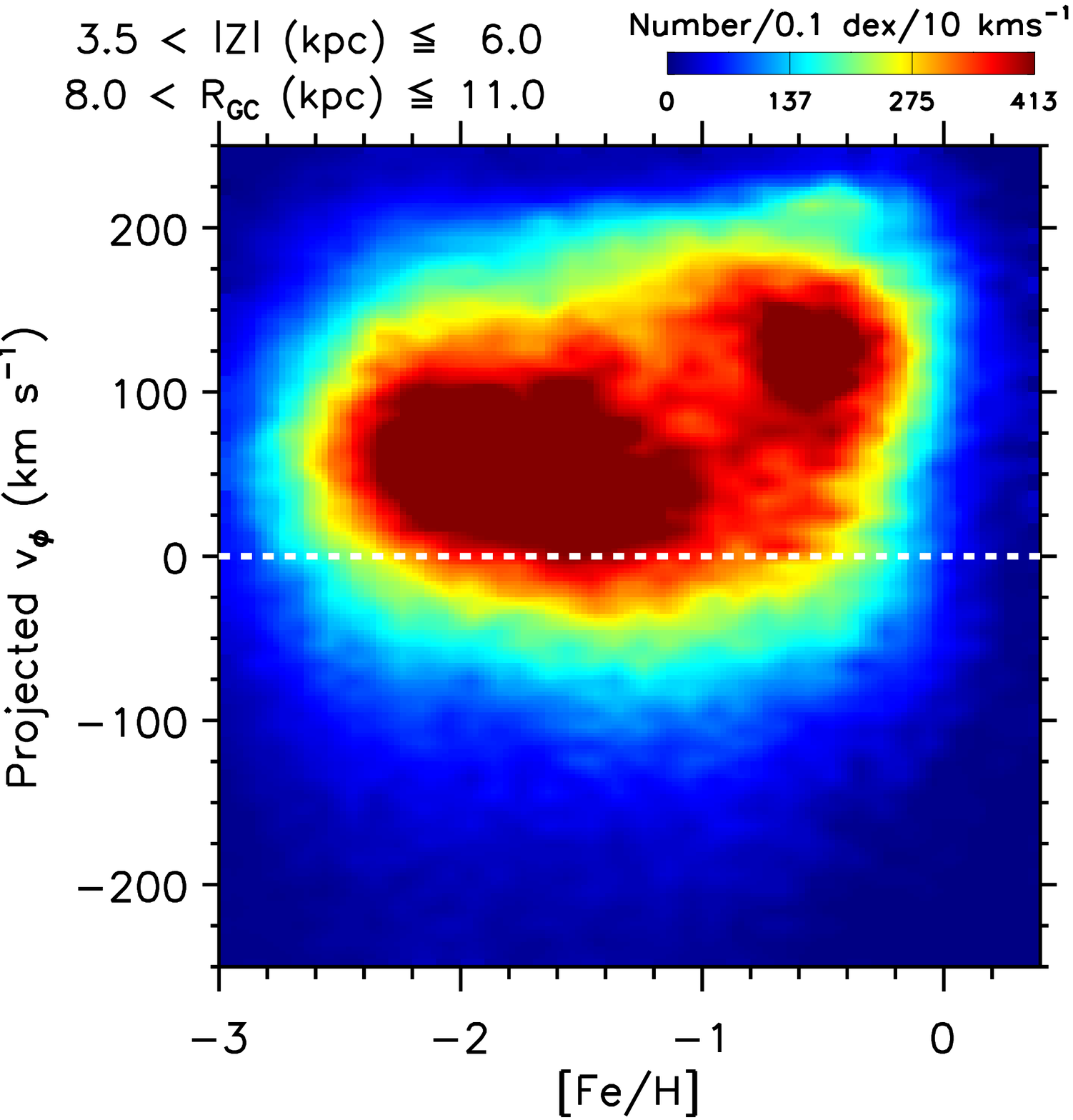}{0.4\textwidth}{\textbf{(a) Full sample at $8 < R_{\rm GC} \leq 11$~kpc}}
                \fig{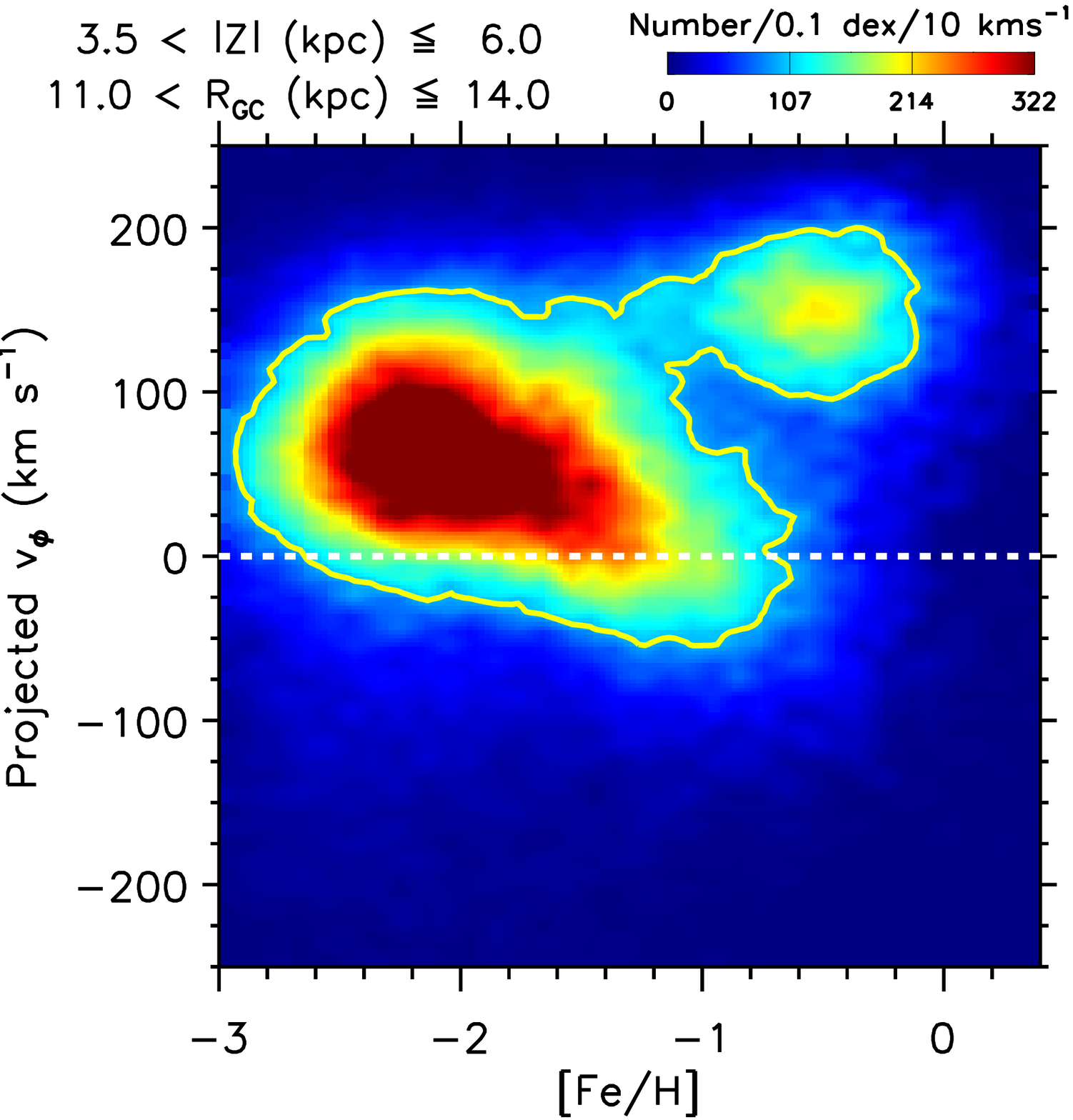}{0.4\textwidth}{\textbf{(b) Full sample at $11 < R_{\rm GC} \leq 14$~kpc}}}
  \gridline{\fig{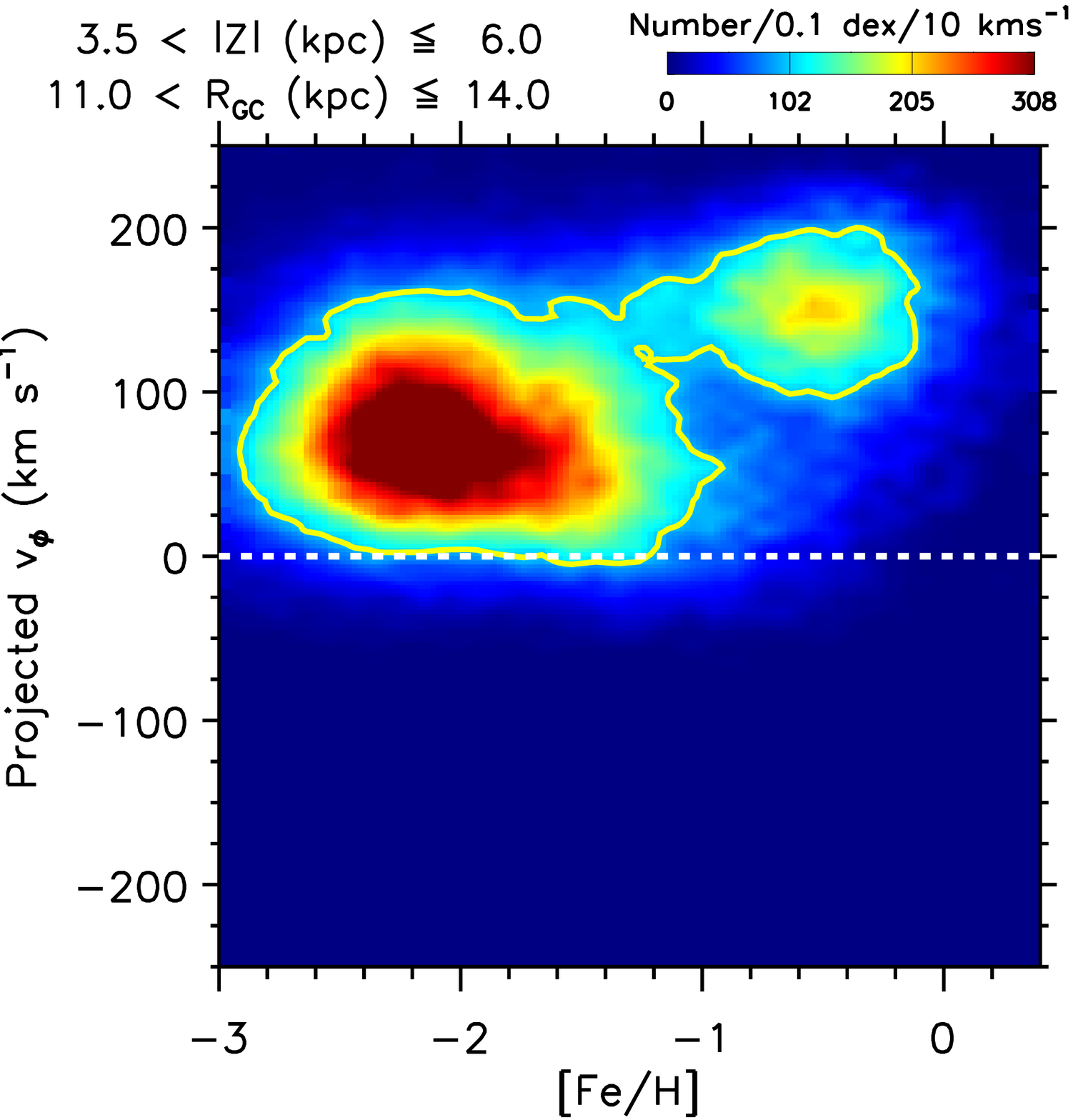}{0.4\textwidth}{\textbf{(c) PM $< 10$~mas yr$^{-1}$ at $11 < R_{\rm GC} \leq 14$~kpc}}
                \fig{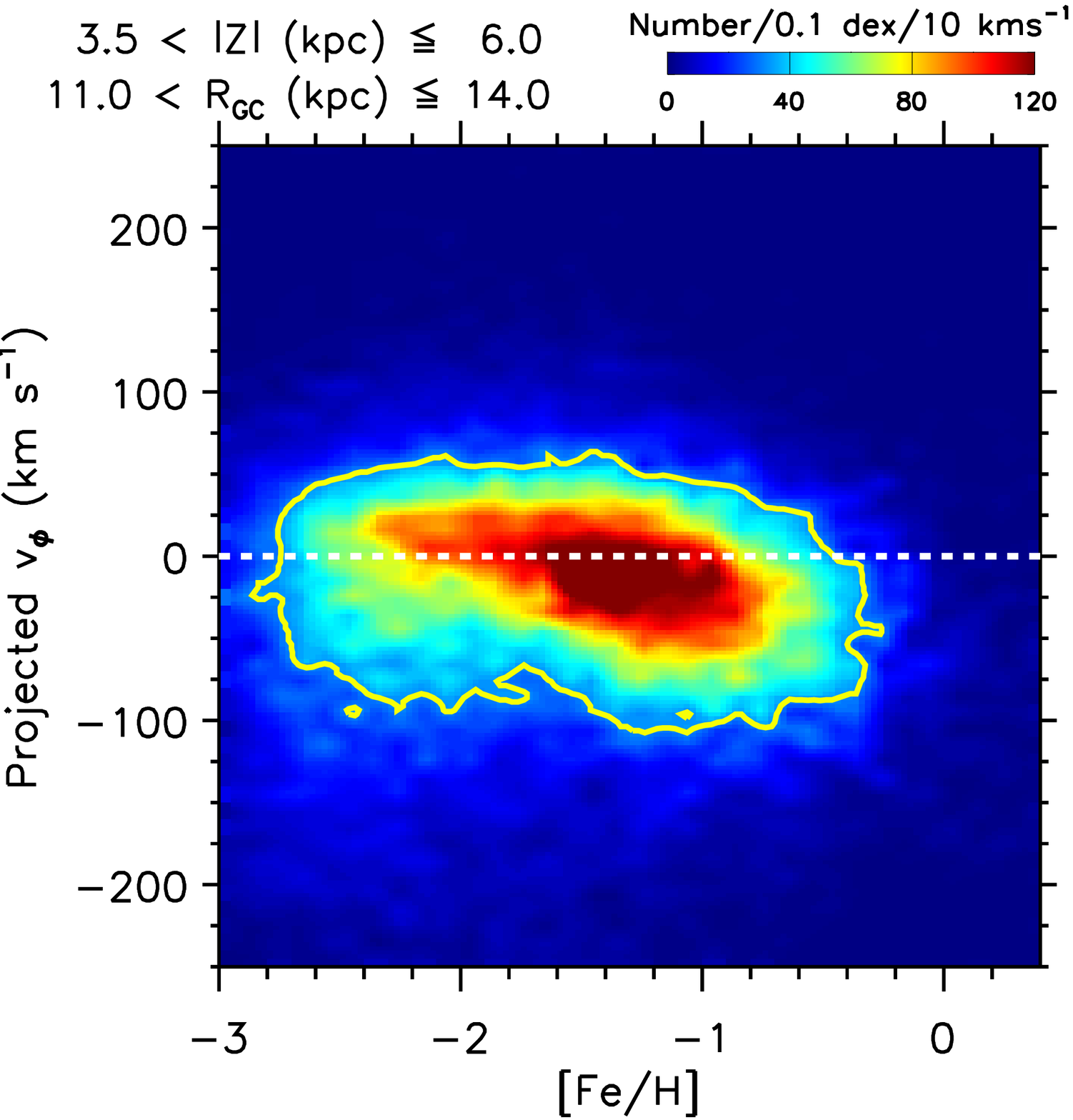}{0.4\textwidth}{\textbf{(d) PM $> 10$~mas yr$^{-1}$ at $11 < R_{\rm GC} \leq 14$~kpc}}}
\caption{Phase-space diagrams based on photometric metallicity and distance estimates from SDSS $\cap$ PS1 photometry. (a) Stars at $3.5 < |Z| \leq 6$~kpc and $8 < R_{\rm GC} \leq 11$~kpc. (b) A full sample of stars at the same $|Z|$ as in panel~(a), but at $11 < R_{\rm GC} \leq 14$~kpc. (c) Same as in panel~(b), but stars having proper motions (PM) less than $10$~mas yr$^{-1}$. (d) Same as in panel~(b), but those having $>10$~mas yr$^{-1}$.}
\label{fig:phase}
\end{figure*}

More quantitative comparisons with previous studies can be made using phase-space diagrams such as shown in Figure~\ref{fig:phase}, in which the number density of stars is displayed as a function of $\vphi$ and [Fe/H]. Positive values of $\vphi$ indicate that stars are moving in the same direction as the Galactic disk. The displayed data are taken at $3.5 < |Z| \leq 6$~kpc, where the contribution of disk stars is minimized (see Paper~III). Since Gaia parallaxes have large uncertainties for our main-sequence star sample at this distance, we use metallicity and distance estimates from fully photometric solutions based on SDSS $\cap$ PS1 photometry, without relying on parallaxes. See Appendix~\ref{sec:phase_error} for the impact of uncertainties in these measurements.

The phase-space diagrams in panels~(a) and (b) of Figure~\ref{fig:phase} are taken from $8 < R_{\rm GC} \leq 11$~kpc and $11 < R_{\rm GC} \leq 14$~kpc, respectively, in Galactocentric spherical coordinates, which reveal complex sub-structures of local stars. The high metallicity, fast-rotating clump at $\langle \vphi \rangle \approx 150\ \kms$ and $\langle {\rm [Fe/H]} \rangle \approx -0.5$ in panel~(b) represents thick-disk stars. Its mean $\vphi$ and [Fe/H] are similar to earlier results in the literature \citep[see][and references therein]{yan:19}. On the other hand, metal-rich ([Fe/H] $>-1$) stars in panel~(a) have a skewed $\vphi$ distribution, owing to an increased contribution of the Splash in the inner Galactic region, as demonstrated in Paper~III.

In panels~(a) and (b) of Figure~\ref{fig:phase}, a group of metal-poor stars is seen at [Fe/H] $\approx-2.2$, which exhibits a slow net prograde rotation ($\langle \vphi \rangle \approx 70\ \kms$). Along with a more metal-rich ([Fe/H] $\approx-1.5$) counterpart in the inner Galactic region, it was seen as one of the main constituents of the Galactic halo in our series of papers. In Paper~I, we considered both of these components as ``inner and outer halos'' in the dual halo paradigm \citep{carollo:07,carollo:10,beers:12}, but we called them ``metal-poor and metal-rich halos'' in Paper~II. On the other hand, we made a presumption in Paper~III that they constitute a main body of GSE, owing to the lack of other analogous structures known in the same phase space.

However, recent evidence suggests that the slowly rotating metal-poor stars are likely a separate entity from GSE. \citet{belokurov:22} used aluminium abundances to separate in situ stars in the halo from accreted stars, and found a large number of metal-poor in situ stars ([Fe/H] $<-1$). Although they span an extreme range in $\vphi$ (from nearly $\sim-150\ \kms$ to $\sim300\ \kms$) and their spectroscopic sample is limited to [Fe/H] $>-1.5$, their approximate mean $\vphi\sim100\ \kms$ and low metallicity ([Fe/H] $<-1$) suggest that these stars (dubbed ``Aurora'') are a part of the structure seen in our previous work. Given its lower metallicity than the Splash, it is likely an old (primordial) in situ halo that formed before the GSE merger at $z =1$--$2$.

In support of this view, we employ a simple proper-motion cut in the sample to separate stars in GSE from the metal-poor halo distribution, as demonstrated in panels~(c) and (d) of Figure~\ref{fig:phase}. They show the same phase-space diagrams of stars as in panel~(b), while having different ranges of proper motion, $< 10$~mas yr$^{-1}$ and $> 10$~mas yr$^{-1}$, respectively. In panel~(c), both the metal-poor in situ halo and thick disk are seen, connected by a narrow band of stars, which we assigned to the Metal-Weak Thick Disk (MWTD) in a Gaussian mixture model in Paper~III. On the other hand, an elongated structure along the $\vphi=0\ \kms$ line stands out from the high proper-motion sample in panel~(d), which encompasses a wide range of metallicity ($-3 < {\rm [Fe/H]} < -0.5$). The chemical and kinematical properties of these high proper-motion stars are analogous to GSE in the original works of \citet{belokurov:18} and \citet{helmi:18}.

Although our $\vphi$ measurement is strongly correlated with proper motions, the above separation can be understood by the highly radial orbits of GSE stars, which contrast with a nearly isotropic velocity distribution of metal-poor in situ halo stars. We also note that the separation does not change appreciably even if the heliocentric distances of the sample are further narrowed down, indicating that smaller (negative) $\vphi$ is not merely caused by systematically shorter distances. In summary, all of these chemical and kinematical properties of individual populations in our phase-space diagrams are consistent with those found from previous (mostly spectroscopic) studies, which essentially validates our photometric [Fe/H] and $\vphi$ estimates.

\subsection{Scale-Height and Scale-Length Distributions}\label{sec:scale}

To examine structural properties of each group of stars, we compute a scale length ($L$) and height ($H$) in bins of $\vphi$ and [Fe/H], with bin sizes set to $25~\kms$ and $0.2$~dex, respectively. The survey data volume has a cone-shaped geometry, so we only use stars within $\pm1$~kpc from the solar radius at $1 < |Z| < 4$~kpc in each Galactic hemisphere to compute the scale height. Similarly, the scale length is determined using stars at $7 < R < 12$~kpc and $1 < |Z| < 3$~kpc, which are limited by the survey volume shape. In each $\vphi$-[Fe/H] bin, we perform a linear least-squares fitting of the logarithmic number density of stars as functions of $Z$ or $R$, where the bin size is fixed at $500$~pc. Although the density profile of bulk halo stars has been known to follow a power-law, we utilize an exponential function as a proxy in the local volume near the Galactic plane to obtain a relative comparison between different populations. From the best-fitting model, we calculate the standard deviation in a logarithmic number density, and assume that all data points have the same uncertainty as this value. We then estimate the uncertainties in $L$ or $H$ by using the best-fitting slope and its estimated error in a successive regression (see Appendix~\ref{sec:error}).

\begin{figure}
\gridline{\fig{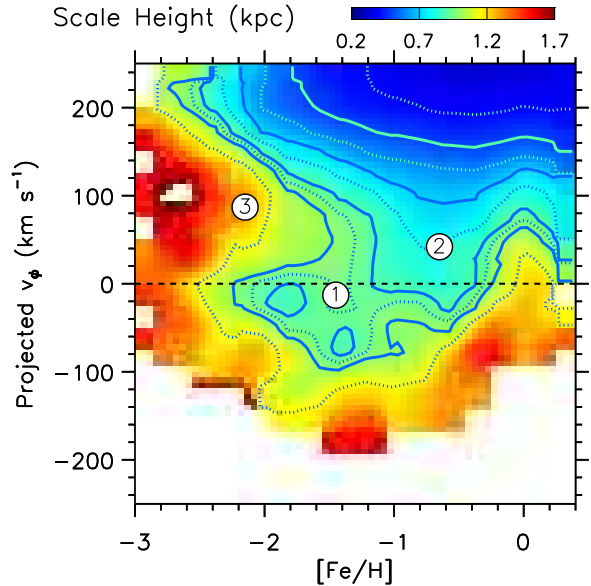}{0.44\textwidth}{\textbf{(a) Northern Galactic Hemisphere}}}
\gridline{\fig{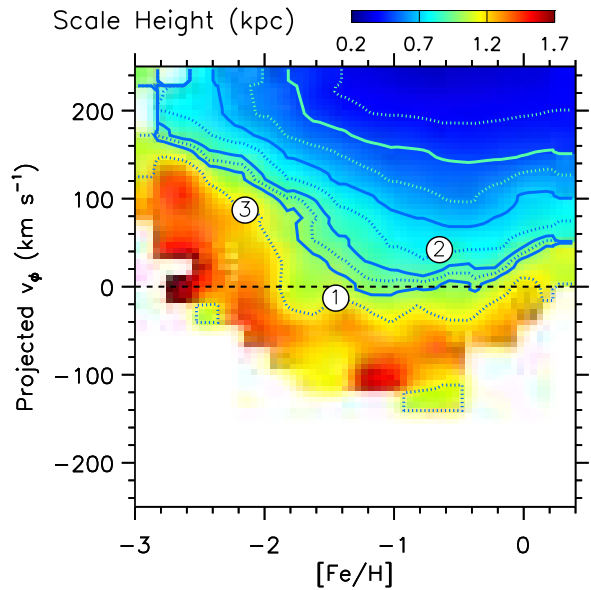}{0.44\textwidth}{\textbf{(b) Southern Galactic Hemisphere}}}
\caption{Distribution of scale height based on the SDSS $\cap$ Gaia sample. Results from the Northern and Southern Hemispheres are shown in the top and bottom panels, respectively. Scale heights are computed within a $2$~kpc-wide zone centered at the Sun ($7.34 \leq R \leq 9.34$~kpc). Solid contour lines represent $H=0.5, 0.7, 0.9, 1.0$~kpc. Notable features seen in Panel~(a) are marked by circled numbers in both panels: \raisebox{.5pt}{\textcircled{\raisebox{-.9pt} {1}}} GSE, \raisebox{.5pt}{\textcircled{\raisebox{-.9pt} {2}}} Splash, and \raisebox{.5pt}{\textcircled{\raisebox{-.9pt} {3}}} metal-poor in situ halo.}
\label{fig:sheight}
\end{figure}

\begin{figure}
\gridline{\fig{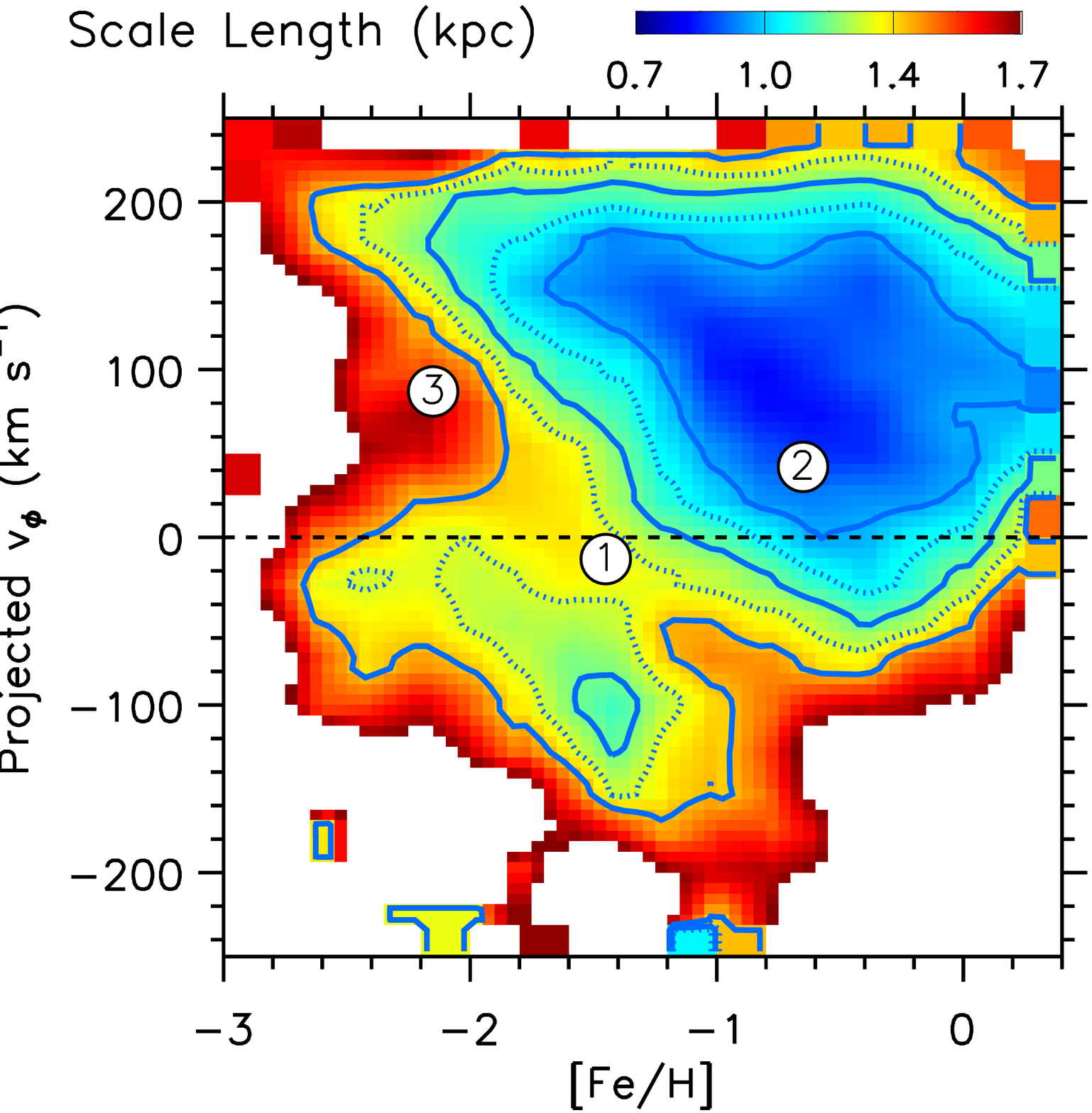}{0.44\textwidth}{\textbf{(a) Northern Galactic Hemisphere}}}
\gridline{\fig{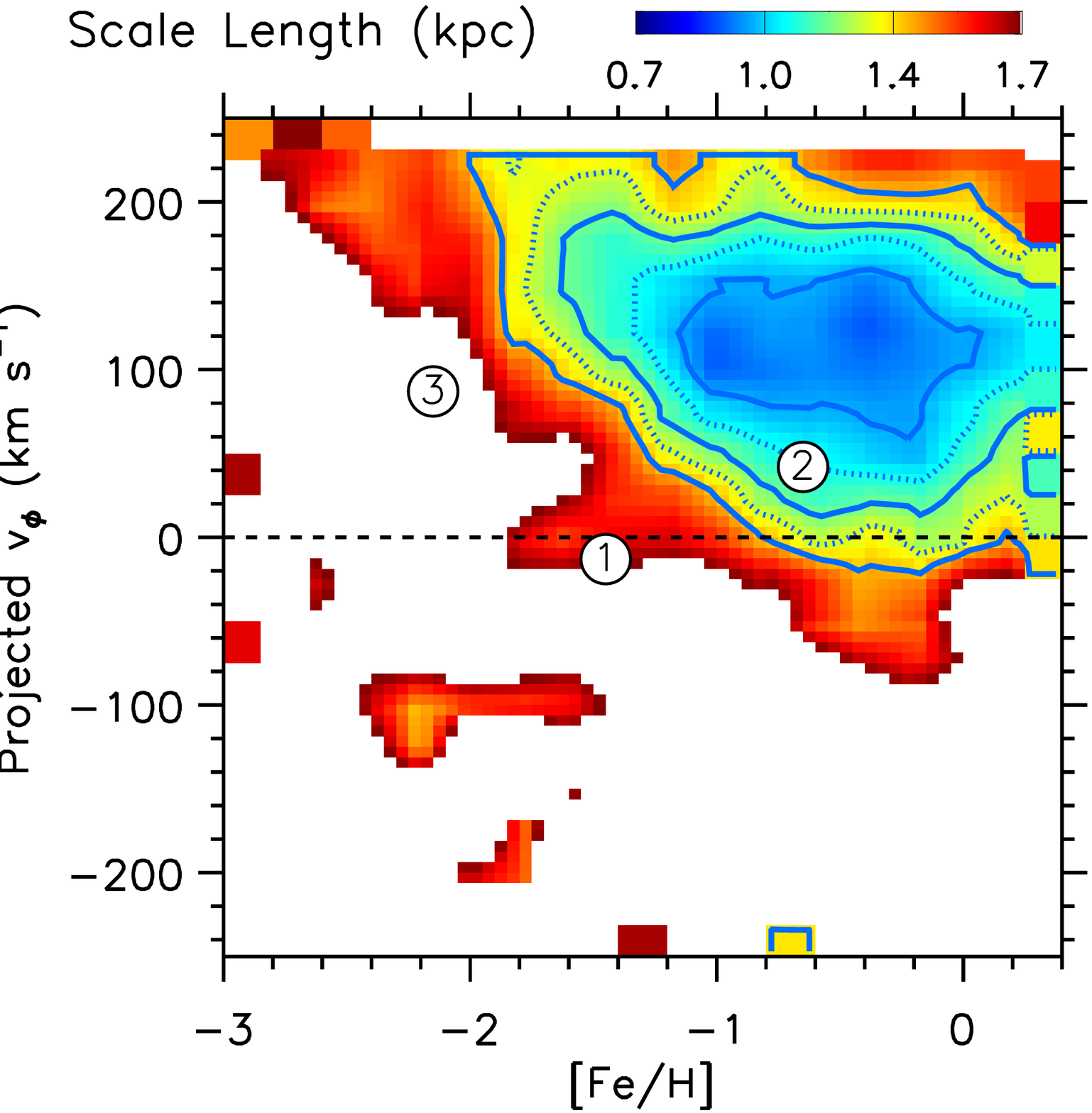}{0.44\textwidth}{\textbf{(b) Southern Galactic Hemisphere}}}
\caption{Distribution of scale length based on the SDSS $\cap$ Gaia sample in the Northern (top) and Southern Hemisphere (bottom). Scale lengths are computed using stars at $7 < R < 12$~kpc and $1 < |Z| < 3$~kpc. Contour lines represent $L=1.0$ to $1.4$~kpc in steps of $0.1$~kpc. The circled numbers are the same as in Figure~\ref{fig:sheight}.}
\label{fig:slength}
\end{figure}

Figure~\ref{fig:sheight} displays a distribution of scale height in each Galactic hemisphere from the SDSS $\cap$ Gaia sample (distances taken from Gaia). Only those pixels with a fractional uncertainty less than $25\%$ are included. A $0.2$~dex $\times\ 25\ \kms$ pixel is subdivided into $4\times4$ sub-pixels, which are smoothed using a $5$-point boxcar average, in order to have a smoothed, global look at changes in the structural properties. Likewise, Figure~\ref{fig:slength} shows a distribution of scale length in both hemispheres, binned and smoothed in the same manner as in Figure~\ref{fig:sheight}. A requirement that stars lie in the stripe along the Galactic prime meridian severely limits the number of stars available in each data set, resulting in $\approx4\times10^6$ and $2\times10^6$ stars in the Northern and Southern Galactic Hemispheres, respectively.

In the top panels of Figure~\ref{fig:sheight} and \ref{fig:slength}, three notable features are seen in the Northern Galactic Hemisphere, which are marked by the circled numbers. The intermediate scale-height and scale-length valley is particularly evident, extending from a region mainly occupied by GSE stars (`\raisebox{.5pt}{\textcircled{\raisebox{-.9pt} {1}}}') to a region populated by the Splash (`\raisebox{.5pt}{\textcircled{\raisebox{-.9pt} {2}}}'). This valley appears even more dramatic, as it contrasts with a scale-height ``highland'' at [Fe/H] $\sim-2.2$ and $\vphi\sim+80\ \kms$ (`\raisebox{.5pt}{\textcircled{\raisebox{-.9pt} {3}}}'). For reference, the circled numbers are also marked in the following figures, including the bottom panels of Figures~\ref{fig:sheight} and \ref{fig:slength}.

The intermediate scale-height valley directly shows that GSE stars are distributed farther from the Galactic plane than disk stars, but not as far as metal-poor in situ halo stars. This observation can be understood by a low inclination, highly radial orbit of the GSE progenitor, which penetrated deep into the primordial Galaxy. In addition, a comparison between the Northern and Southern Hemispheres in Figures~\ref{fig:sheight} and \ref{fig:slength} clearly demonstrates a larger amount of debris in the Northern Hemisphere in the local volume, owing to a pileup of stars at the last apocenter of a highly eccentric orbit of the GSE progenitor \citep{naidu:20}.

\begin{figure}
\gridline{\fig{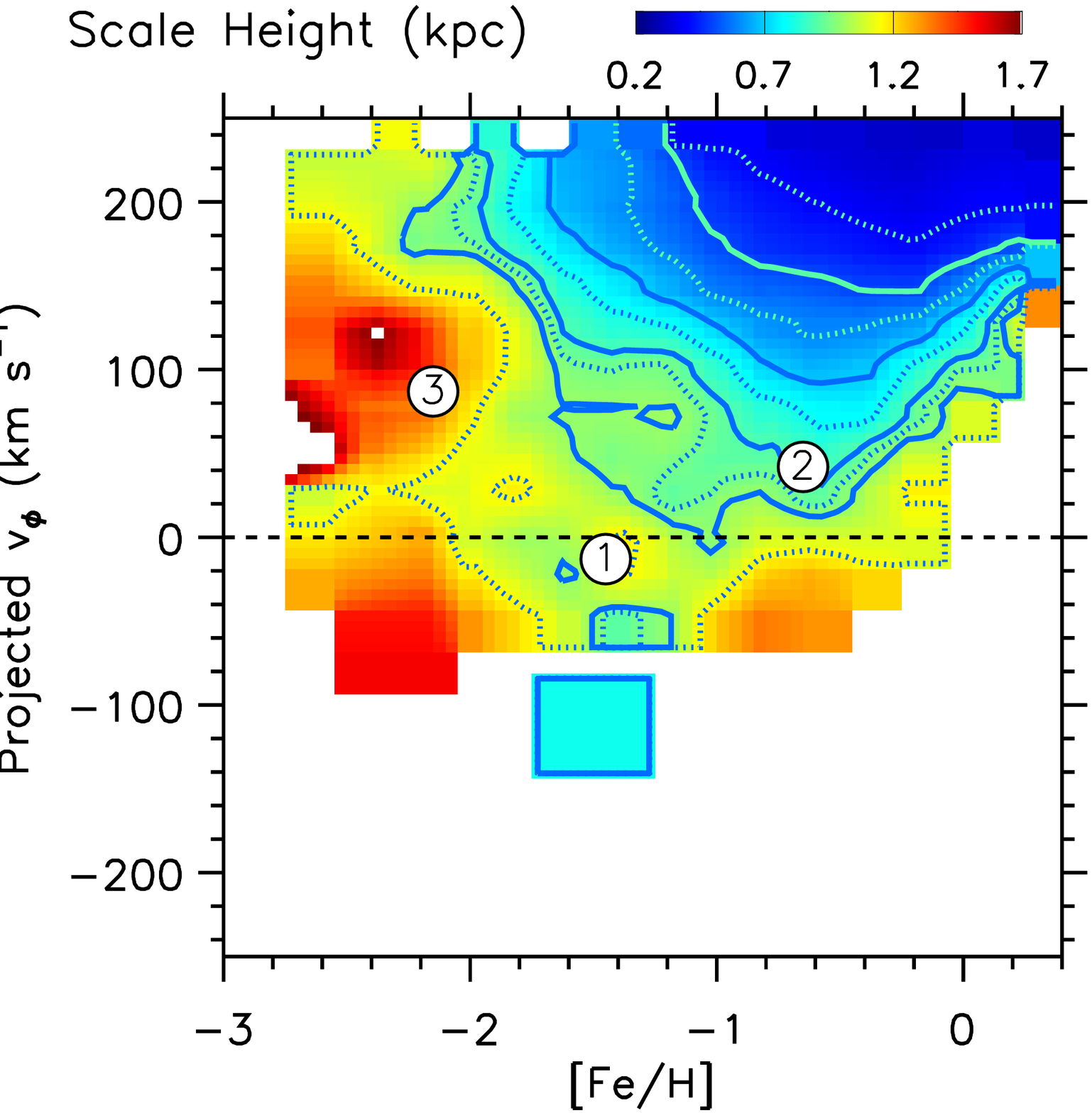}{0.44\textwidth}{\textbf{(a) Scale height in the Southern Galactic Hemisphere}}}
\gridline{\fig{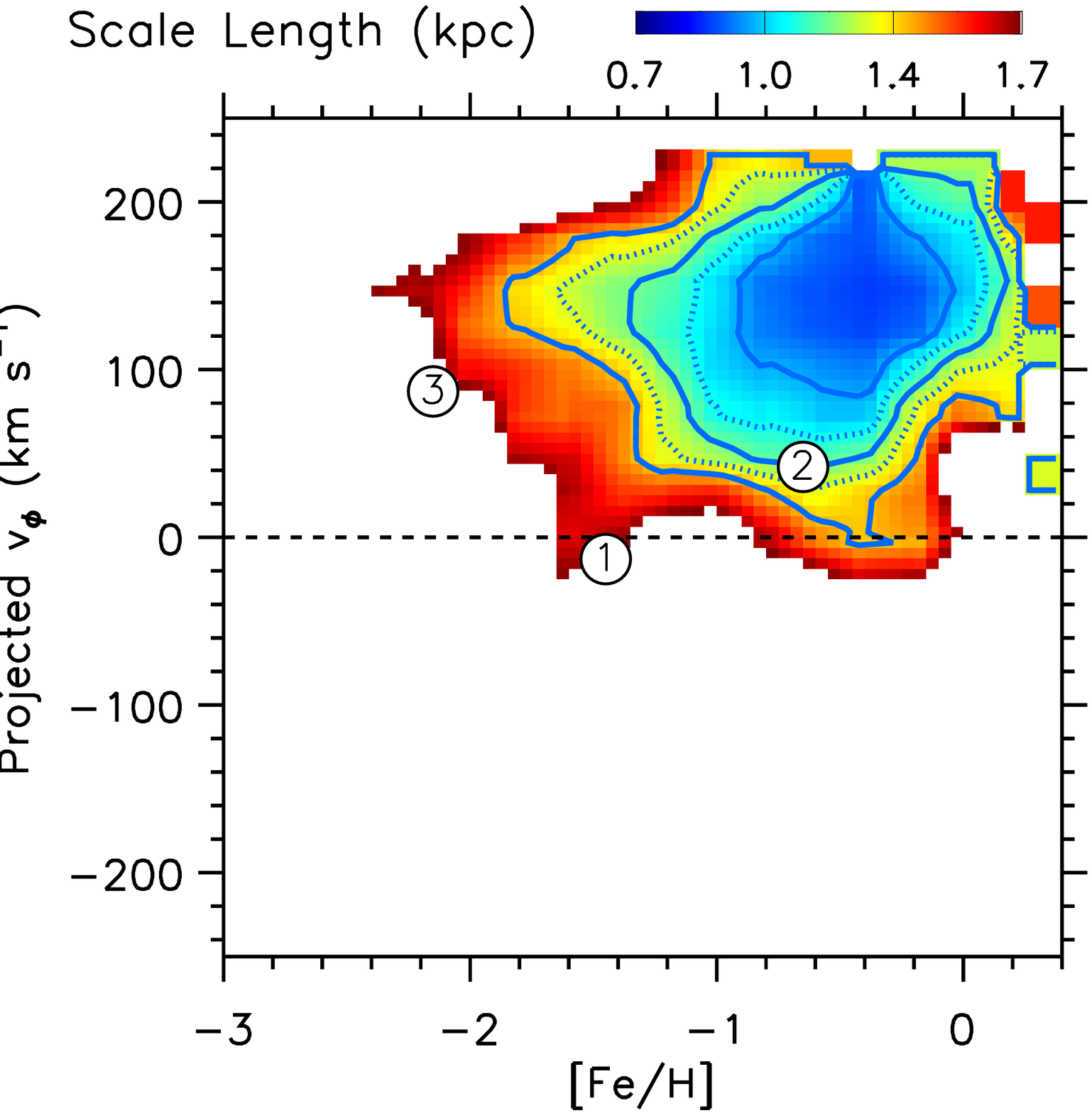}{0.44\textwidth}{\textbf{(b) Scale length in the Southern Galactic Hemisphere}}}
\caption{Distribution of scale height (top) and scale length (bottom) based on the SMSS $\cap$ Gaia sample in the Southern Galactic Hemisphere. The circled numbers are the same as in Figure~\ref{fig:sheight}.}
\label{fig:scale_smss}
\end{figure}

The asymmetric stellar distribution in the halo between the Northern and Southern Galactic Hemispheres can be checked using SMSS $\cap$ Gaia (based on Gaia parallaxes). SDSS mainly covers the Northern Galactic Hemisphere, but has a limited coverage in the South. On the other hand, SMSS covers almost the opposite side of the celestial hemisphere (see Figure~\ref{fig:map}) with a total of $\approx6\times10^5$ stars in the Northern Galactic Hemisphere and $\approx10^6$ stars in the South in our sample. Figure~\ref{fig:scale_smss} displays scale-height and scale-length distributions from SMSS $\cap$ Gaia in the Southern Galactic Hemisphere, generated following the same steps used for Figures~\ref{fig:sheight} and \ref{fig:slength}. The coherent structure in the parameter space covered by GSE is not seen as clearly as in the Northern Galactic Hemisphere from SDSS $\cap$ Gaia, but resembles the distributions in the South, highlighting the highly eccentric orbit of the GSE merger. Nonetheless, we note that, owing to the limited overlap along the Galactic prime meridian, the SDSS $\cap$ Gaia sample is biased toward the Galactic anticenter direction in the Southern Hemisphere, while the sample from SMSS $\cap$ Gaia is more populated toward the Galactic center. Therefore, a trace of the intermediate scale-height valley in the top panel could be real, and may capture the GSE debris in the direction toward the Galactic center. Deeper SMSS photometry in future data releases would be useful to explore this volume in more detail.

In the top panel of Figure~\ref{fig:sheight}, a vertical trough with intermediate scale heights lies at [Fe/H] $\sim-0.8$, extending from $\vphi\sim0\ \kms$ to $\vphi\sim100\ \kms$. Originally, \citet{belokurov:20} defined a region occupied by the Splash as $-0.7 < {\rm [Fe/H]} < -0.2$ and $-150 < \vphi < +100\ \kms$, which overlaps with this trough. The fact that the intermediate scale-height valley from GSE stretches out to a region populated by Splash stars supports a previous claim that the GSE merger has dynamically heated stars in the primordial disk of the Milky Way \citep{bonaca:17,belokurov:20}. In other words, because a nearly in-plane collision with a dwarf galaxy would leave behind heated stars confined to a disk plane, this apparent coincidence supports a view on a causal connection between the Splash and GSE. In addition, since such stars originated from the primordial disk of the Galaxy, which was smaller in size than the current stellar disk, this also naturally explains the short scale length of these stars (Figure~\ref{fig:slength}).

However, our data clearly indicate that dynamical heating took place over a wider range of metallicity than previously considered by \citet{belokurov:20}. If stars were born in the inner region, and then were displaced by mergers, the lowland with small scale lengths in Figure~\ref{fig:slength} shows an approximate extent where the dynamical heating took place. Reassuringly, the lowland in the top panel stretches out to the lower $\vphi$ at the metallicity covered by Splash stars ($-0.7 < {\rm [Fe/H]} < -0.2$), although such feature is not clearly seen in the bottom panel. If the low scale-length region is fenced by a contour line at $L=1$--$1.2$~kpc, the top panel indicates that even lower metallicity stars down to [Fe/H]$\sim-2$ originated from similar excitation mechanisms. This conclusion is further supported by the near coincidence between the horizontal trough along $\vphi \sim +150\ \kms$ at $-2 \la {\rm [Fe/H]} \la -1.2$ and the MWTD, which is considered to be the relic of the primordial disk. Therefore, even if the GSE progenitor was a major source of the dynamical heating of a primordial disk, it seems unlikely that the orbital properties of stars in this region are altered by a single massive merger event. Instead, our result indicates that such a dynamical heating process on the primordial disk was in operation, even before the GSE merger, driven by more numerous minor mergers, as predicted by numerical simulations of the early universe \citep[e.g.,][]{grand:20}.

The other interesting feature is the ``highland'' in the scale-height distribution (`\raisebox{.5pt}{\textcircled{\raisebox{-.9pt} {3}}}' in Figure~\ref{fig:sheight}), which coincides with the metal-poor in situ halo (see Figure~\ref{fig:phase}). Its scale height reaches $H>1.1$~kpc, which is significantly higher than those of heated stars ($\sim 0.8$~kpc); therefore, it is unlikely that it formed through the same dynamical heating process. Moreover, the highland exhibits a gradient in the scale length, as shown in the top panel of Figure~\ref{fig:slength}, in the sense that more metal-rich stars are more strongly concentrated in the inner region of the Galaxy. This implies that more active star formation took place in the deeper potential well, while the Milky Way has grown by chaotic coalescence of numerous small gas-rich dwarf galaxies in the early universe. It is unclear, however, how the Milky Way attained the net angular momentum in the same direction as that of the Galactic disk at this stage. Nonetheless, its small prograde net rotation ($\langle \vphi \rangle \sim +80\ \kms$) suggests that it had maintained a puffy disk-like structure owing to turbulent nature of the gas-rich minor mergers.

\subsection{The Galactic Starburst Sequence (GSS)}\label{sec:starburst}

\begin{figure*}
  \centering
  \gridline{\fig{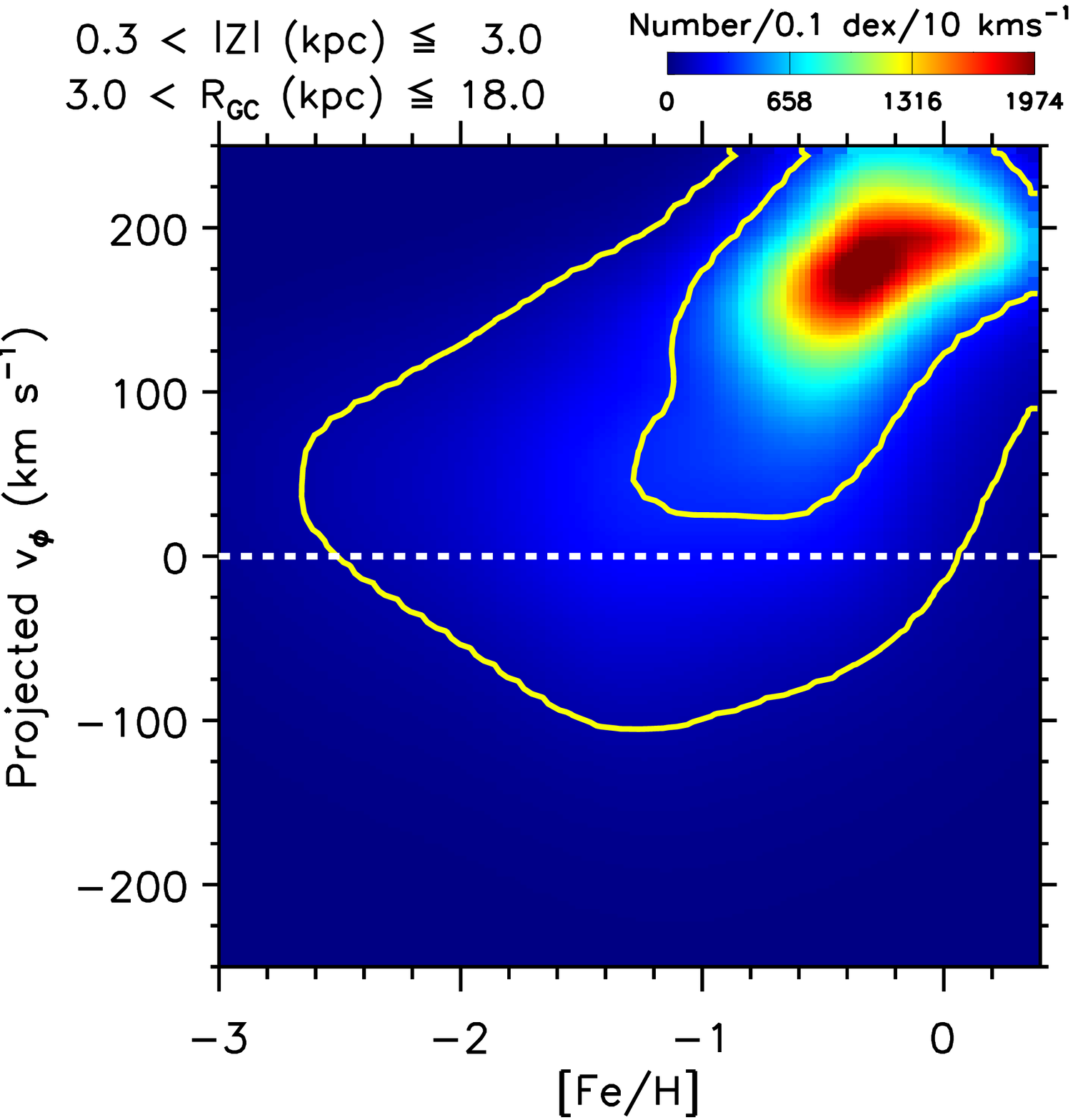}{0.4\textwidth}{\textbf{(a) PM $> 20$~mas yr$^{-1}$}}
                \fig{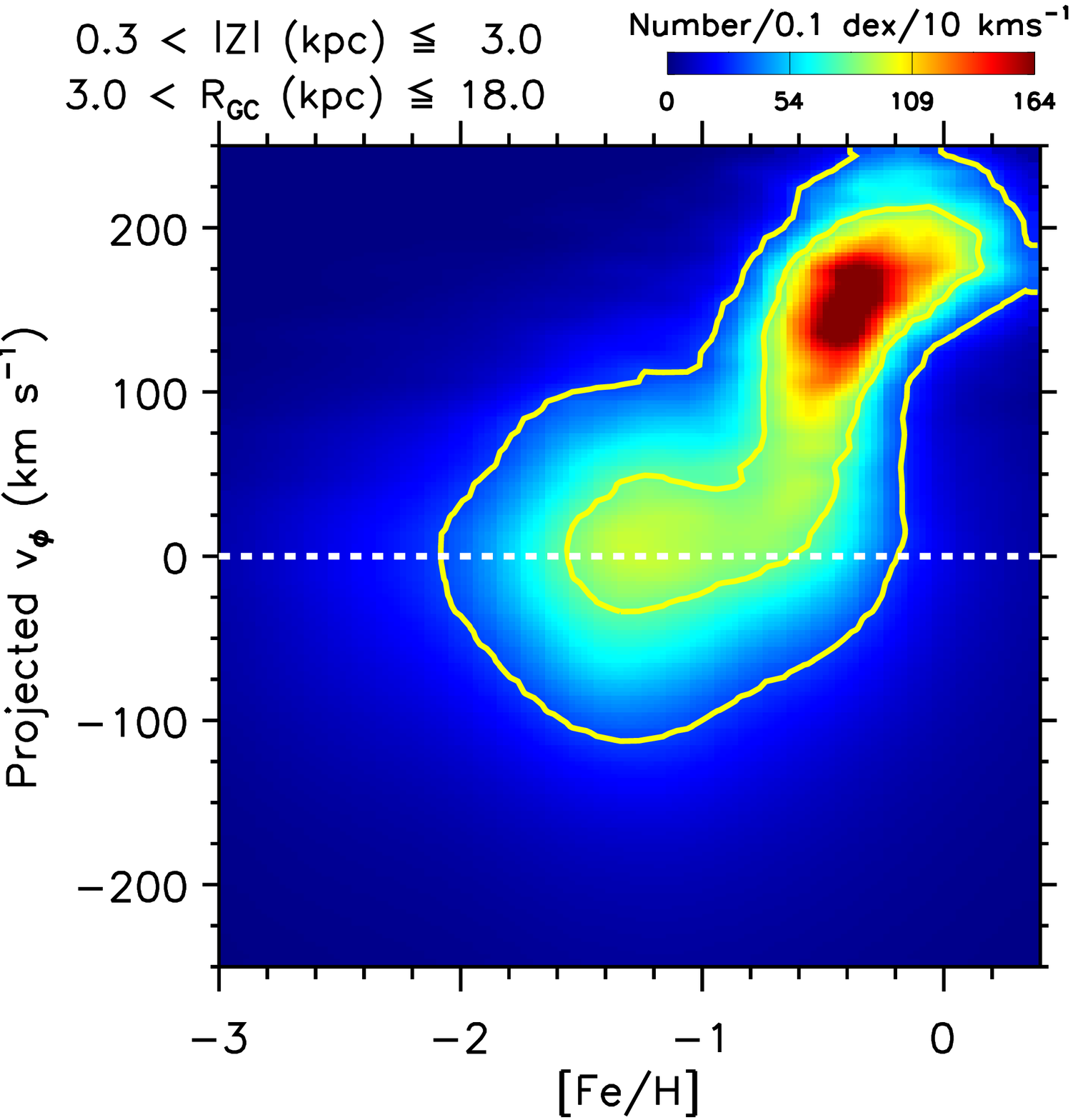}{0.4\textwidth}{\textbf{(b) PM $> 40$~mas yr$^{-1}$}}}
  \gridline{\fig{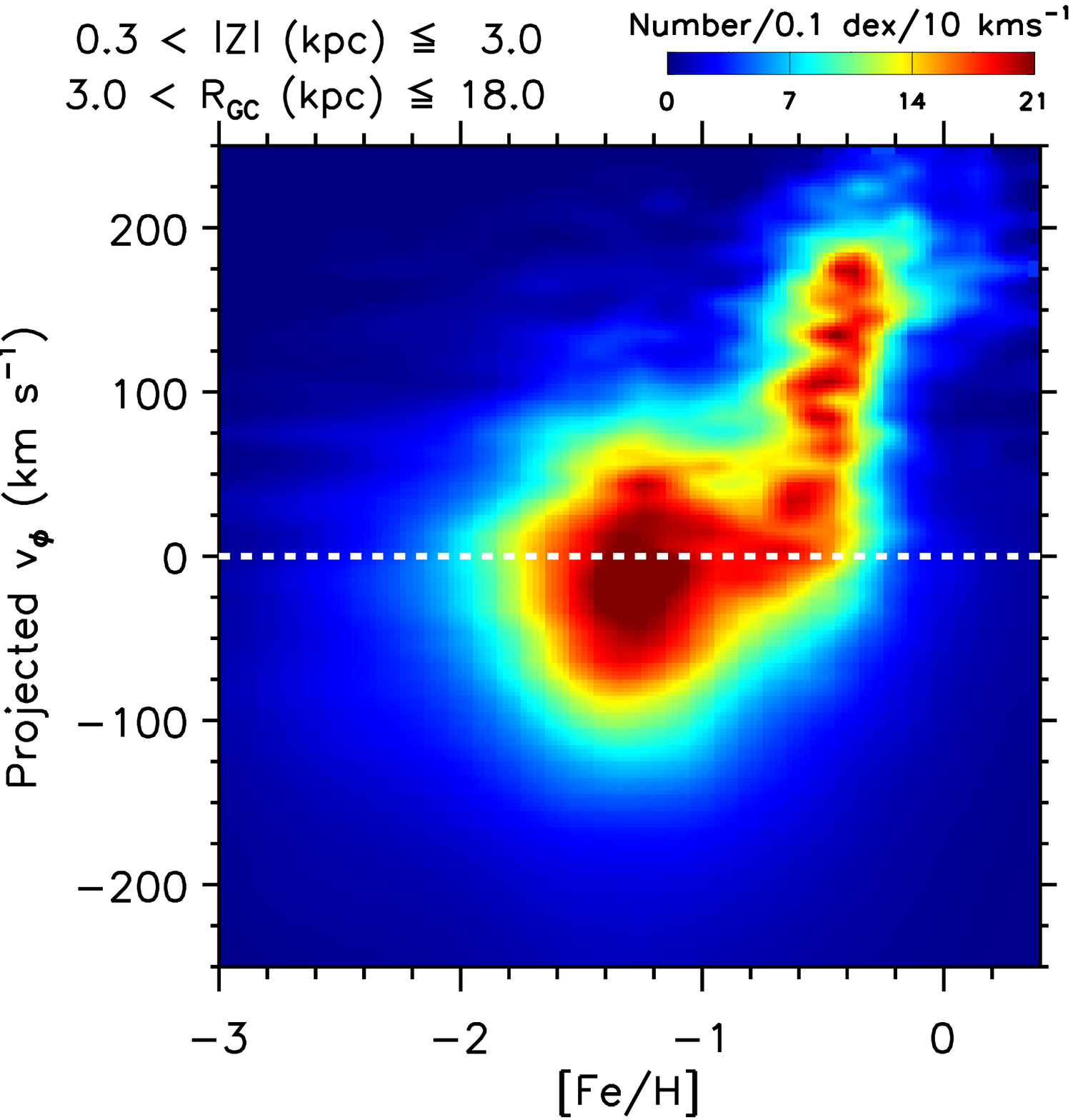}{0.4\textwidth}{\textbf{(c) PM $> 60$~mas yr$^{-1}$}}
                \fig{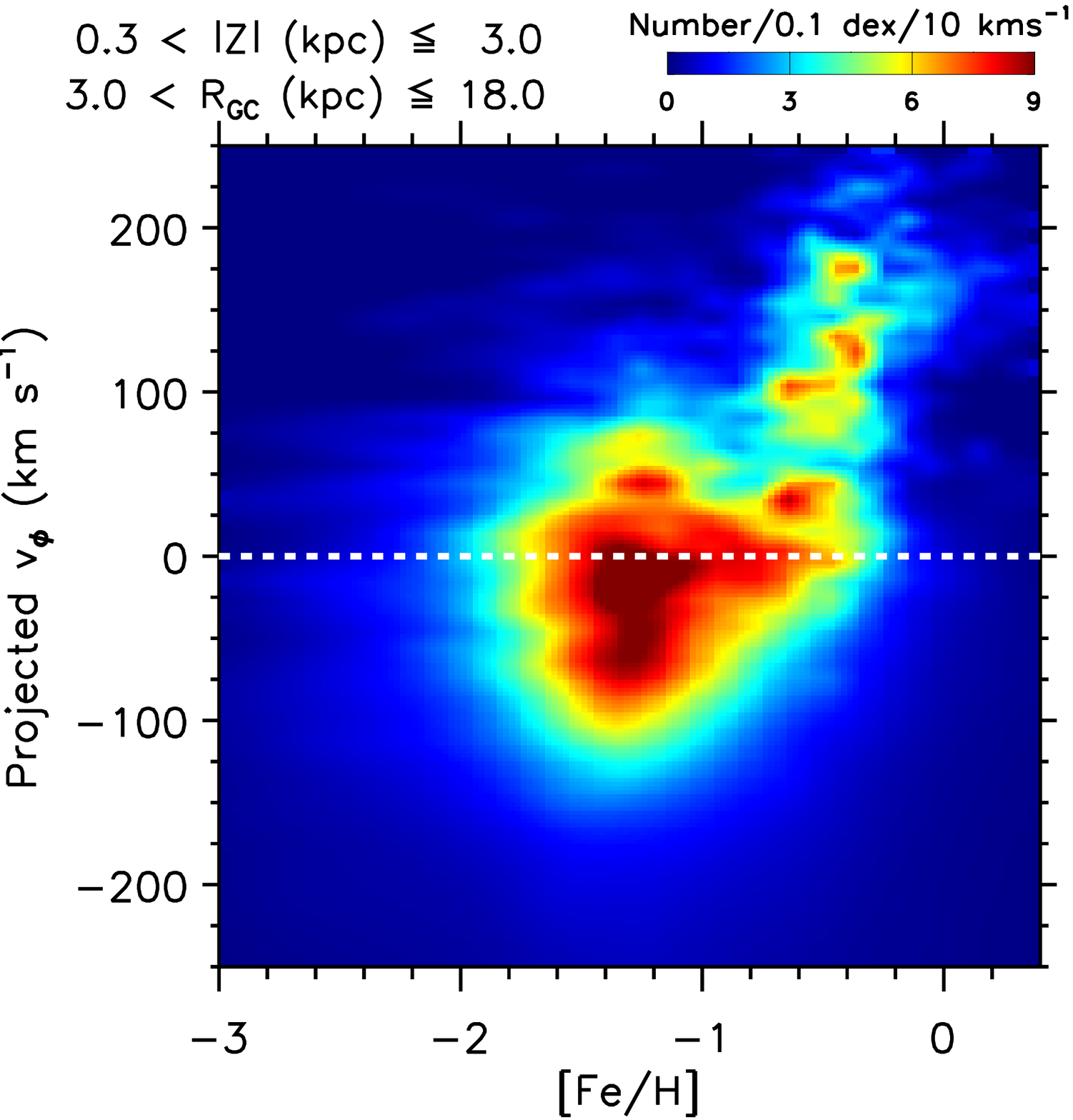}{0.4\textwidth}{\textbf{(d) PM $> 70$~mas yr$^{-1}$}}}
\caption{A phase-space diagram of stars in SDSS $\cap$ Gaia with various proper-motion cuts.}
\label{fig:vphi}
\end{figure*}

Our phase-space diagrams also reveal a long and narrow sequence of stars when a simple cut on proper motion is made. All four panels in Figure~\ref{fig:vphi} are drawn from the same SDSS $\cap$ Gaia sample, but with different cuts on a minimum value of proper motions ($20$, $40$, $60$, and $70$~mas~yr$^{-1}$, respectively). Moreover, we convolve each count of stars using a normalized Gaussian function, with standard deviations set based on the measurement uncertainties in both axes. See Appendix~\ref{sec:phase_error} for the version displaying raw counts and the impact of the upper limits on the uncertainties in distance and metallicity.

The top left panel of Figure~\ref{fig:vphi} is dominated by disk populations (mostly thick-disk stars), but a low $\vphi$, low [Fe/H] tail begins to show up when a mild cut on proper motion ($>25$~mas~yr$^{-1}$) is imposed. At $>40$~mas~yr$^{-1}$, a striking elbow-like feature emerges from these diagrams, where halo and disk stars form a narrow, continuous sequence over a wide range of [Fe/H] and $\vphi$. This feature is characterized by two joint, orthogonal branches; the horizontal arm is nearly parallel to the $\vphi=0\ \kms$ line over a wide range of metallicity ($-2 \la {\rm [Fe/H]} \la -0.6$), while the vertical arm has a narrow metallicity range ($-0.6 \la {\rm [Fe/H]} \la -0.4$). The sequence passes through GSE and the Splash, and is eventually connected to the disk. Although our $\vphi$ estimates depend on distance, the elbow-like feature becomes stronger with higher proper-motion cuts, but changes little with distance from the Sun.

\begin{figure*}
  \centering
  \gridline{\fig{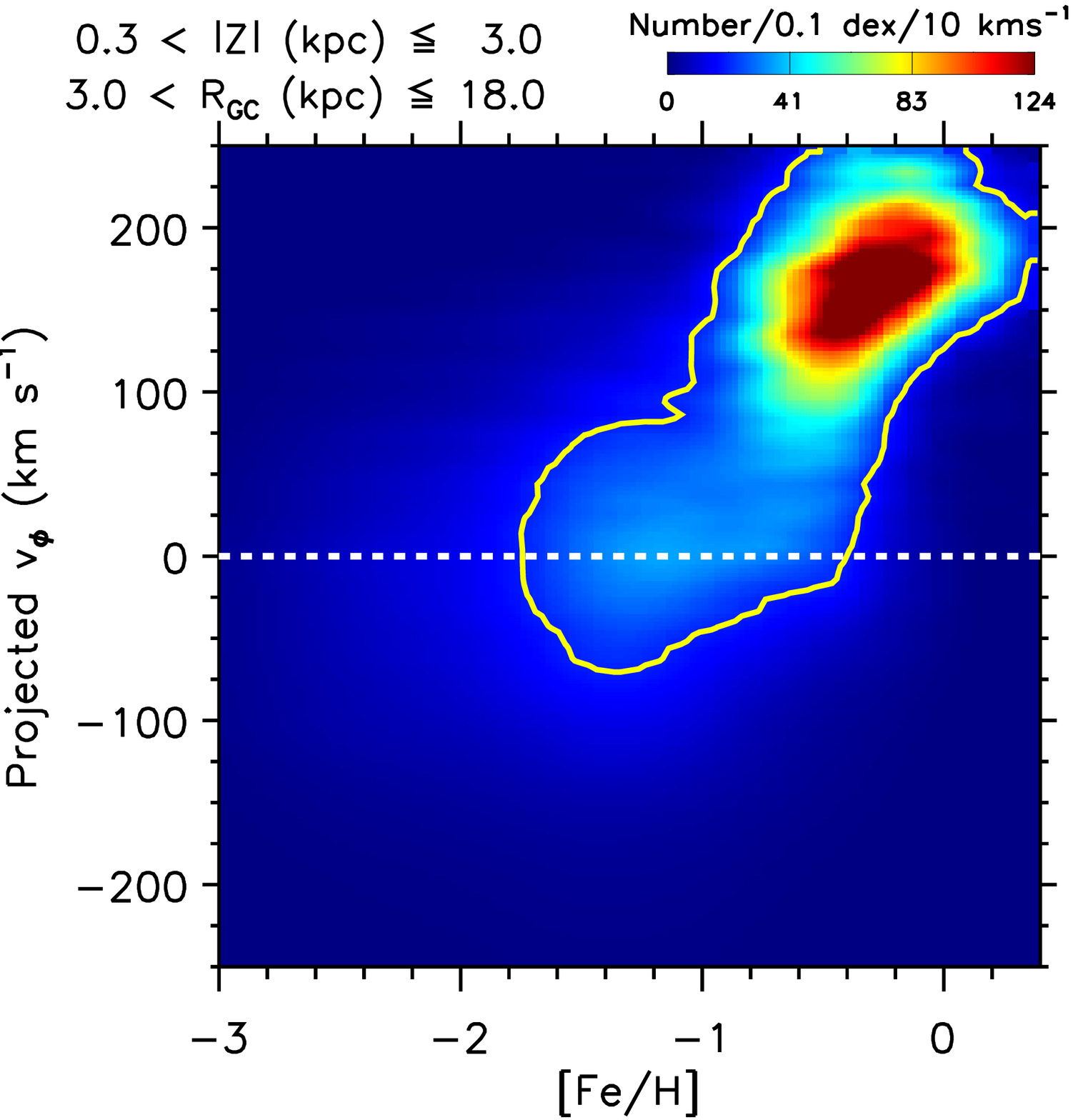}{0.4\textwidth}{\textbf{(a) PM $> 40$~mas yr$^{-1}$}}
                \fig{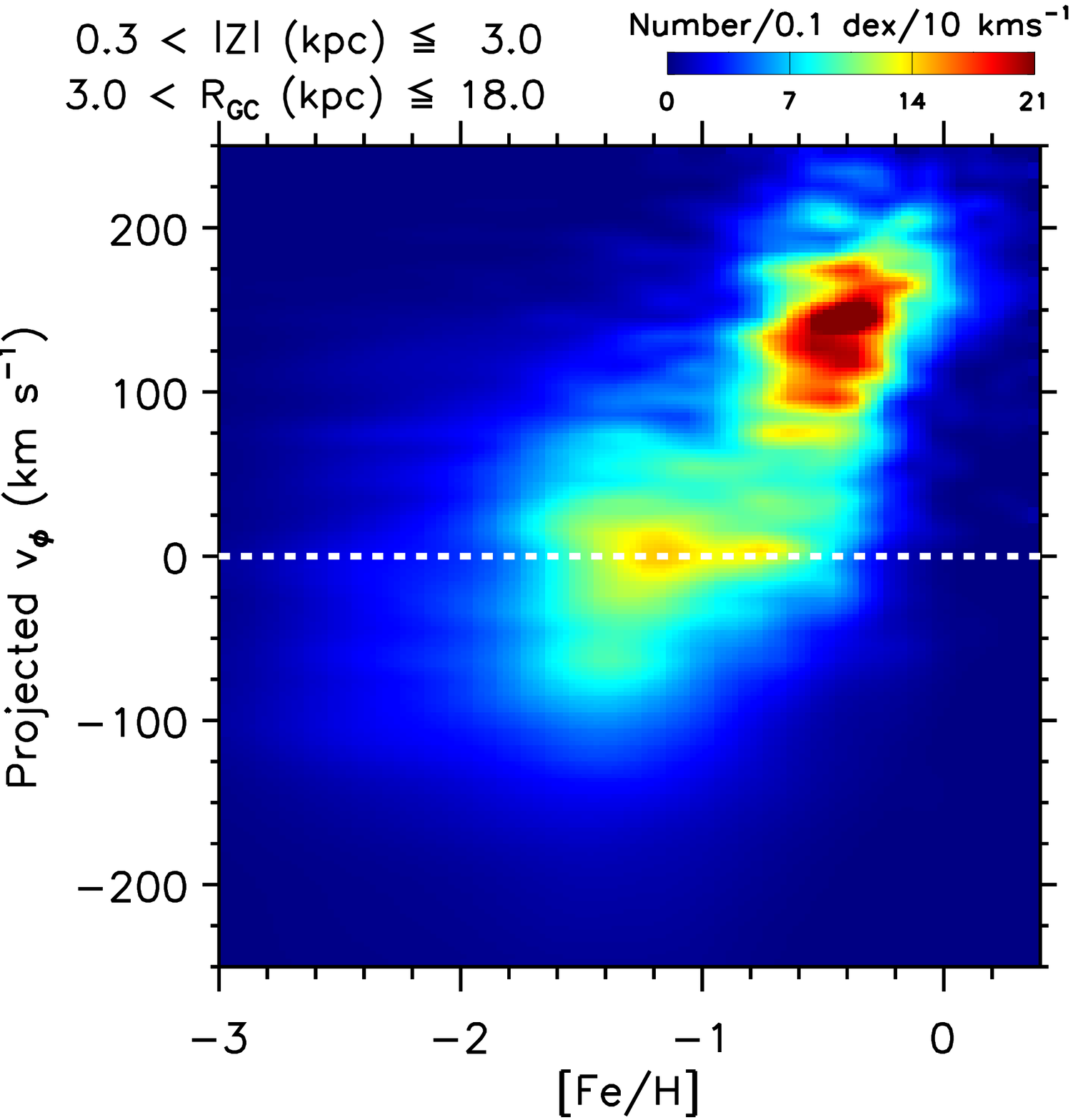}{0.4\textwidth}{\textbf{(b) PM $> 60$~mas yr$^{-1}$}}}
\caption{Same as in Figure~\ref{fig:vphi}, but based on SMSS $\cap$ Gaia.}
\label{fig:vphi_smss}
\end{figure*}

\begin{figure*}
  \centering
  \gridline{\fig{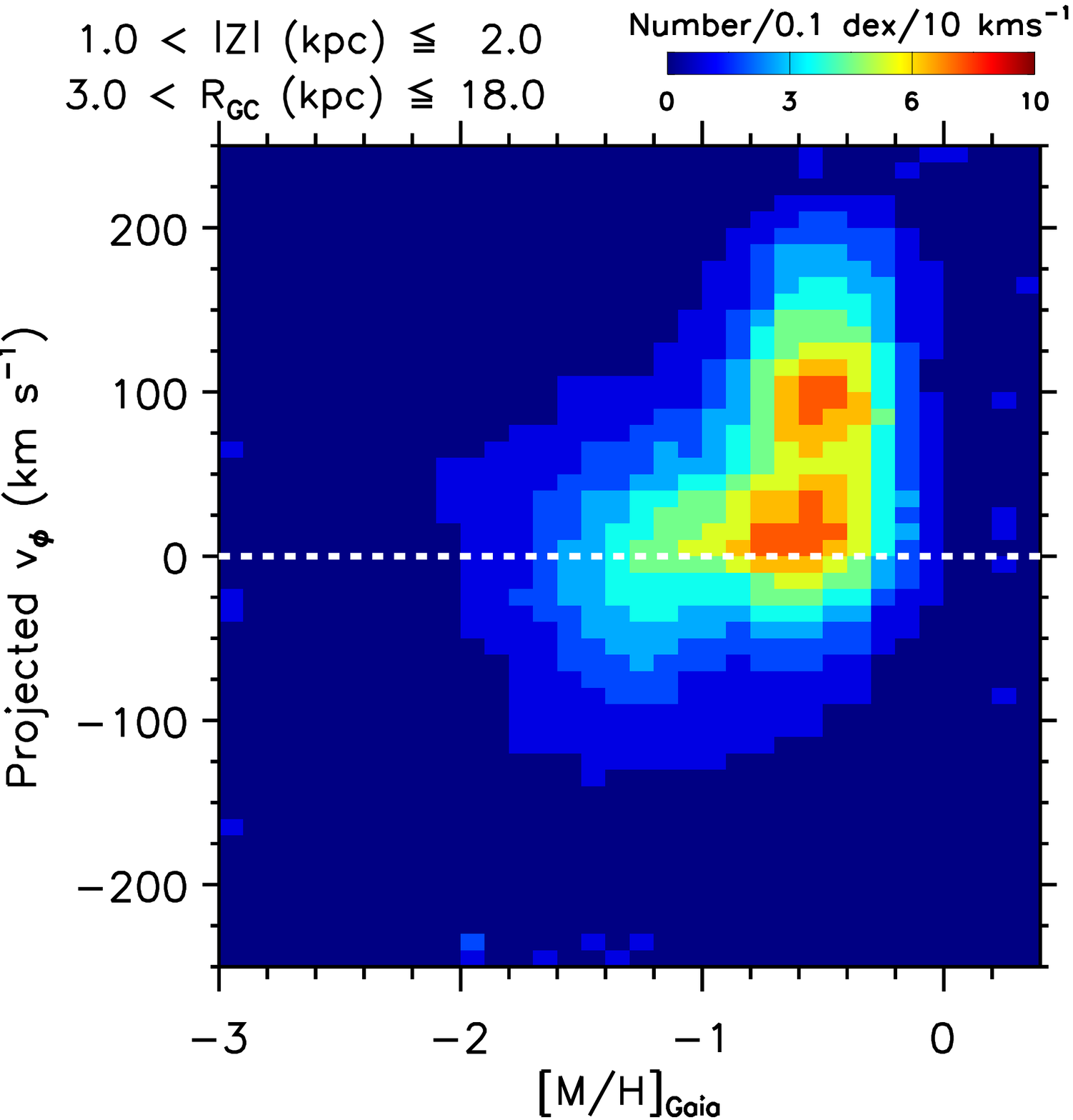}{0.4\textwidth}{\textbf{(a) Projected velocity}}
                \fig{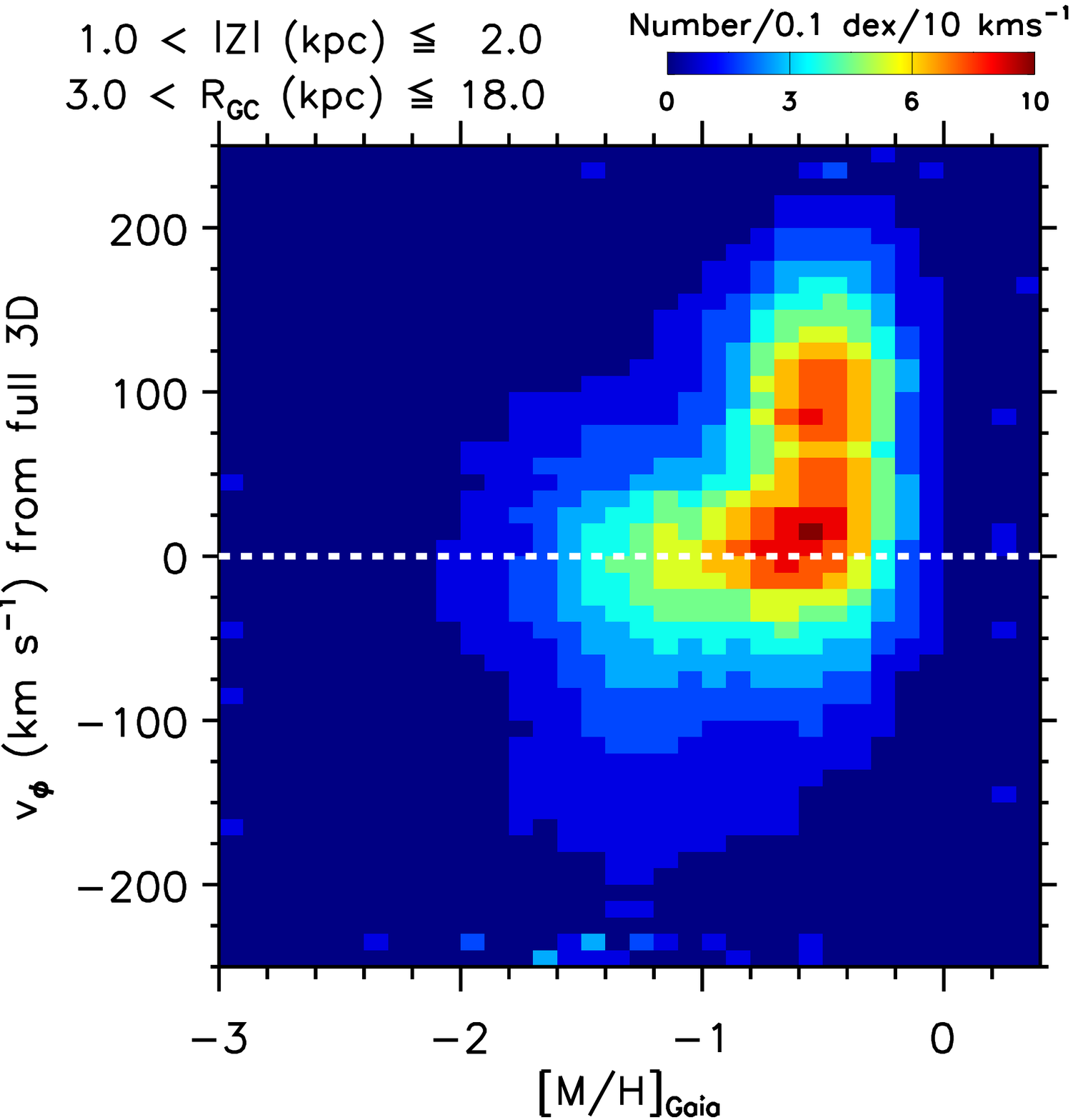}{0.4\textwidth}{\textbf{(b) Full 3-D kinematics}}}
\caption{(a) Same as in Figure~\ref{fig:vphi}, but based on spectroscopic metallicity estimates in Gaia DR3. A vertical distance range ($1 < |Z| < 2$~kpc) and a proper-motion cut ($> 25$~mas yr$^{-1}$) are adjusted to emphasize the observed sequence. (b) Same as in panel~(a), but using $\vphi$ based on three dimensional motions in Gaia.}
\label{fig:vphi_gaia}
\end{figure*}

As shown in Figure~\ref{fig:vphi_smss}, the sequence persists even when SMSS $\cap$ Gaia is used, indicating that it is present in both hemispheres. Aside from our photometric metallicities, the same sequence can also be seen in Figure~\ref{fig:vphi_gaia}, based on spectroscopic metallicities in Gaia DR3 using the General Stellar Parameteriser-Spectroscopy (GSP-Spec) module \citep{recioblanco:22} from the Radial Velocity Spectrometer (RVS) spectra ($\lambda/\Delta\lambda=11,500$). Because these spectroscopic observations are available for bright stars ($G < 14$~mag), the sample is limited to relatively nearby stars ($1 < |Z| < 2$~kpc); the lower $|Z|$ cut is made to exclude numerous disk stars. The proper-motion cut is lowered to $25$~mas~yr$^{-1}$ to retain as many stars as possible along the sequence. On the other hand, the constraint on Galactic latitudes is lifted, as the spectroscopic metallicities are only weakly dependent on foreground reddening. While the left panel in Figure~\ref{fig:vphi_gaia} is based on the projected $\vphi$ as for our photometric samples along the prime meridian, the right panel is based on full three-dimensional space motions from both radial velocities and proper motions in Gaia. The sequence appears almost identical in both panels, which not only supports the existence of this coherent structure, but also validates our approach for obtaining $\vphi$ from proper motions only.

Our phase-space diagram presents a continuous chain of stars, suggesting that they formed through successive metal enrichment along this pathway. As seen in panels (b)--(d) of Figure~\ref{fig:vphi}, the sequence runs along ([Fe/H], $v_\phi$) $\approx$ \{(-2, 0), (-0.6, 0), (-0.4, $180\ \kms$)\}. Although the complete sequence covers nearly $2$ dex in [Fe/H], the vertical arm of the sequence has a remarkably narrow range of metallicities. This implies that these stars formed quickly, with insufficient time for successive metal enrichment (except $\alpha$-element enhancement from core-collapse supernovae), while star-forming clouds were collapsing or reorienting their angular momentum vectors rapidly from nearly zero to $\sim180\ \kms$ in $v_\phi$. In other words, our result indicates that the young Milky Way went through a phase of starburst activity. For this reason, we refer to this structure as the Galactic Starburst Sequence (GSS). The GSS traverses through known stellar populations and structures, including the GSE at the lower metallicity range and disk stars at the metal-rich end, implying a chronological order of formation of these Galactic components that can be traced to a common origin.

\begin{figure}
\epsscale{1.1}
\plotone{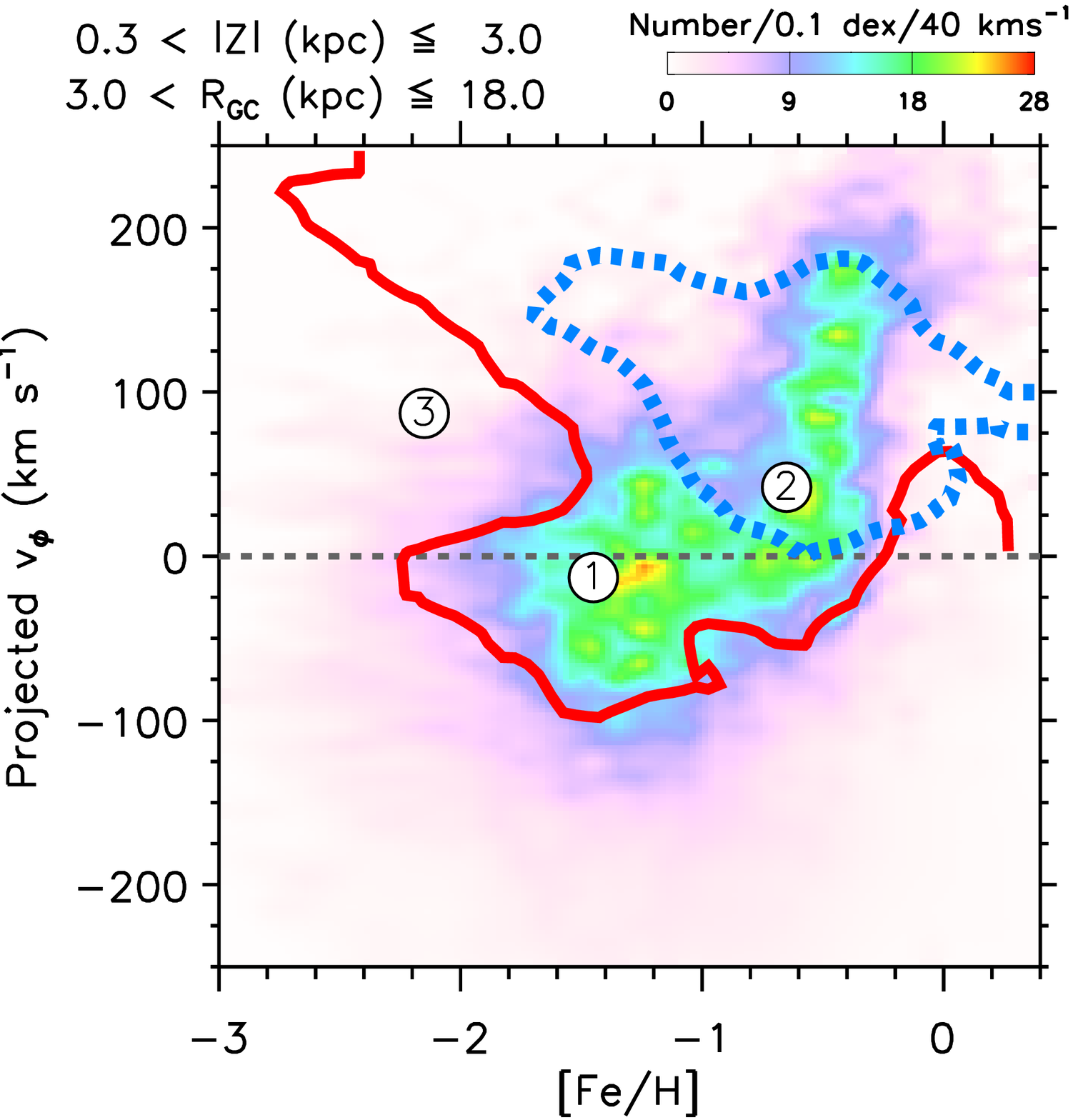}
\caption{Contours of equal scale height (red solid line; $H=1$~kpc) and length (blue dotted line; $L=1$~kpc), plotted on top of a phase-space diagram of stars with large proper motions ($> 60$~mas yr$^{-1}$; bottom left panel in Figure~\ref{fig:vphi}).}
\label{fig:comb}
\end{figure}

To better understand the properties of the GSS, we put the above results together in Figure~\ref{fig:comb}. The red solid contour shows a region with a scale height $H=1$~kpc from the top panel of Figure~\ref{fig:sheight}, while the blue dashed contour delineates a scale length $L=1$~kpc from the top panel of Figure~\ref{fig:slength}. They are overlaid on top of contours of stars with high proper motions ($> 60$~mas yr$^{-1}$) in Figure~\ref{fig:vphi}. First, the width of the vertical arm of the GSS is narrower than the metallicity range of the heated population, the extent of which can be delineated by the blue dotted enclosure with a small scale length. This indicates that the Splash, which can be defined as a group of metal-rich stars with halo-like kinematics, is in fact composed of two distinct groups of stars. The first is a group of dynamically heated stars, and the other is the starburst population. Heated stars formed in the inner region of the primordial Galaxy and then were displaced to the current location in the halo, while starburst stars formed in a top-down fashion, as indicated by a weak positive correlation between [Fe/H] and $\vphi$.

Secondly, both GSE and the GSS lie along the intermediate scale-height valley, as shown by the red solid line in Figure~\ref{fig:comb}. The simplest explanation is that the GSE merger has triggered a starburst in the mixture of gas from the primordial Milky Way and gas donated by the GSE merger, since mergers were likely gas rich at high redshifts. According to this scenario, the radially biased orbit of GSE should be responsible for the relatively high transverse motions of stars in the GSS, which contrast with a nearly isotropic velocity distribution of metal-poor in situ halo stars, because these stars formed out of metal-enriched gas having similar orbital properties with the GSE progenitor. Therefore, they are expected to show up in the high proper-motion sample, even though their orbits have evolved significantly over time from halo-like to disk-like orbits.

Notably, the full GSS is not seen in panel~(d) of Figure~\ref{fig:phase}, although it is also made using high proper-motion stars. This difference is manifested by a high central concentration of Splash(-like) stars in the Galaxy, as shown in Paper~III, and more clearly demonstrated by the short scale lengths in Figure~\ref{fig:slength}. The GSS is not a dominant structure at large Galactocentric distances.

\section{Summary and Discussion}\label{sec:summary}

\subsection{Summary}

Thanks to Gaia and large-area photometric and spectroscopic databases, it is now possible to perform accurate comparisons of theoretical models to extensive, high-quality data. In this work, we use models generated using YREC and MARCS, and quantify model deviations as a function of $\teff$, metallicity, and wavelength. The current approach relies on a comparison with multi-wavelength ultra-violet, optical, and near-infrared color-$T_{\rm eff}$ relations derived from Galactic cluster sequences, Gaia's double sequence, and a sample of spectroscopic data in SEGUE and GALAH, using photometric data in broad-band filters from SDSS, SMSS, and APASS and standard-star photometry in the literature. Mean flux deviations are derived as a function of wavelength, which amount to up to $\sim20\%$ at $< 4000$~\AA, but significantly less in longer wavelengths. We find that no single factor can remove the observed offsets, but it is more likely a problem arising from a combination of various sources of errors in the models and/or observational data. Subsequently, we define the model offset as an empirical correction function for our specific choice of models.

By combining our technique with proper-motion measurements in Gaia, we construct phase-space ($\vphi$ versus [Fe/H]) diagrams of stars to provide a global perspective on the stellar populations in the Milky Way. In this way, we identify a long and narrow sequence of stars in a phase-space diagram, which we call the GSS. The GSS is not a representation of a single stellar population, but rather consists of several previously known Galactic stellar populations or components, arranged like pearls on a string. In particular, it overlaps with GSE in a valley with intermediate scale heights, suggesting that GSE has likely triggered successive formation of these stars. It also passes through the Splash, showing rapid evolution of $\vphi$ of star-forming clouds within a narrow metallicity range, which testifies to a starburst event in the young Milky Way. The wide metallicity range of dynamically heated stars as traced by small scale-length regions in a phase-space diagram indicates that the Splash is likely composed of two stellar populations with distinct origins -- dynamically heated and starburst populations.

The red sequence of stars with high transverse motions ($>200\ \kms$) in Gaia \citep{gaiahrd} is possibly another manifestation of the GSS. As shown in Paper~II, the blue and red sequences are separated by a few tenths of a magnitude in $griz$ colors, and more strongly when the $u$-band is included, indicating a metallicity offset by $\sim1$~dex. Based on their corresponding metallicity ranges, stars on the blue sequence belong to GSE, while stars in the red sequence constitute the Splash \citep[e.g.,][]{gallart:19}. According to our work, a fair fraction of stars in the red sequence should constitute the vertical arm of the GSS.

\subsection{GSE Merger-driven Starburst}

Our new perspective into the phase-space diagram of stars in the local halo reveals two consecutive modes of chemo-kinematical evolution: a rapid chemical enrichment along the orbit of GSE, followed by starburst in rapidly evolving orbits of gas clouds. Our finding is in line with recent numerical simulations in the literature, which explicitly predict the existence of a merger-driven starburst in the Milky Way-size galaxies \citep{cooper:15,bignone:19,grand:20,renaud:21}. For instance, in \citet{grand:20}, a merger with a gas-rich GSE-like progenitor triggers starburst activity owing to the increased compression of gas clouds in both galaxies. The star-formation rate suddenly increases by a factor of two, which lasts less than $1$~Gyr. Strikingly, their simulations reproduce some of the observed key features in our data, in that both starburst and heated populations produced by the merger event are present in a narrow metallicity bin, but span a wide range of $v_\phi$. Furthermore, the heated population shows a correlation between [Fe/H] and $\vphi$ in their simulation, while there is essentially no such dependence for the starburst population (considering an excessively large scatter in $\vphi$ at a given metallicity). This may well be explained by a narrow range in [Fe/H] of the vertical arm of the GSS. Another intriguing aspect is that the starburst population in the simulation exhibits a radially concentrated, rotationally supported disk. This prediction is also in agreement with the fact that the GSS lies along the small scale-length, intermediate scale-height valley.

If the vertical and horizontal arms of the GSS are attributed to stars formed during a merger-driven star formation and those accreted from the GSE progenitor, respectively, we also find a good agreement in the fraction of such stars with numerical simulations. We count each group of stars in our phase-space diagram in Figures~\ref{fig:vphi} and \ref{fig:vphi_smss} using a simple box criterion: $-50 < \vphi \leq 200\ \kms$ and $-0.9 < {\rm [Fe/H]} \leq 0$ for the starburst, and $-60 < \vphi \leq +60\ \kms$ and $-2 < {\rm [Fe/H]} \leq -0.9$ for the accreted population, respectively. The observed ratios between these two populations are $1.1$--$3.9$ from SDSS $\cap$ Gaia (for proper-motion cuts at $60$ and $20$~mas~yr$^{-1}$, respectively) and $1.7$--$3.1$ from SMSS $\cap$ Gaia (for $60$ and $40$~mas~yr$^{-1}$ cuts, respectively), in that the total mass of stars formed during a merger-driven star formation in the host galaxy exceeds the stellar mass of accreted stars. They are quantitatively in agreement with a stellar-mass ratio of $\sim2$--$5$ found in Milky Way-like simulations with radially anisotropic stellar halos in \citet[][except the case of a merger with the lowest stellar mass]{grand:20}; see also \citet{orkney:22}.

On observational grounds, \citet{myeong:22} argued for a starburst event triggered by the GSE merger by analyzing spectroscopic databases from APOGEE and GALAH (including the $\alpha$-elements, Al, and Ce), and dynamical information from Gaia (orbital energy). Based on unsupervised Gaussian mixture models, they showed that the local halo populations could be described by four Gaussian components, of which three were previously known -- the GSE, the Splash, and the in situ halo (``Aurora'') -- while the former was further divided into metal-poor and metal-rich parts. The remaining component was found to reside between the GSE and the low-$\alpha$ (thin) disk in the chemical space. The authors argued that the stars belonging to this component named ``Eos'' were formed from a starburst. The [Fe/H] range of $\sim170$ possible members is bounded by a $2\sigma$ uncertainty to approximately $-1.0$ and $-0.3$, with a mean of $-0.7$, where we found a rapid change of dynamical properties of GSS stars. These results provide supporting evidence that Eos could represent a sub-component of the GSS, assuming it formed during the same starburst event. However, there is currently a lack of available dynamical information on Eos members, and further investigations are needed to confirm the relationship between Eos and the GSS.

Using the framework of the GSS, other recent observational studies have uncovered additional pieces of the puzzle that can be put together in a coherent manner. For instance, \citet{lee:23} delved into a chemical space originally occupied by the Splash in the spectroscopic database from the SEGUE and the Large Sky Area Multi-Object Fiber Spectroscopic Telescope \citep[LAMOST;][]{cui:12}. They found that the sample in this narrow metallicity bin can be split into the low- and high-[$\alpha$/Fe] groups, given the large dispersion in radial velocity ($v_R$) of the former in the Galactocentric coordinate system. They used systematic changes in kinematics (orbital inclinations and eccentricities) and $v_R$-[Fe/H] relations to argue that about half of the low-[$\alpha$/Fe] population and the majority of the high-[$\alpha$/Fe] population are from the GSE progenitor and the dynamically heated-disk population, respectively, while the rest are of different origin, most likely from starburst activity.

In addition, \citet{ciuca:22} utilized precise stellar-age estimates based on the asteroseismic measurements in APOKASC-2 \citep{pinsonneault:18}, and discovered a rapid chemical enrichment of stars from [Fe/H] $\approx-0.5$ to $\approx+0.2$ at a look-back time of $10$--$12$~Gyr, which they dubbed the ``Blob''. It is accompanied by an increase of [Mg/Fe], after a small decrease in [Fe/H] (``Dip''), which qualitatively agrees with mixing of fresh gas by a gas-rich merger in numerical simulations. This ``Blob'' feature is most clearly seen in the inner Galactic region, which is in line with our finding that the GSS has a small scale length.

In other studies, age-metallicity relations also reveal a chain of metal-rich stars, establishing a link between the halo and the disk \citep{haywood:13,nissen:20,xiang:22}. Most recently, \citet{xiang:22} used $\alpha$-element-rich stars with low orbital angular momenta, and demonstrated the existence of a narrow and continuous age-metallicity relation in the Milky Way. According to their analysis, the Milky Way achieved a high metallicity floor ([Fe/H] $\sim-1$) about $13$~Gyr ago, with successive metal enrichment over the following $\sim5$~Gyr. The majority of stars are found in a narrow metallicity range ($-0.7 < {\rm [Fe/H]} < -0.2$), which leads to its possible connection to the vertical arm of the GSS. Importantly, its narrow and continuous channel of stars in \citet{xiang:22} and \citet{ciuca:22} reinforces the physical nature of the GSS.

At large $\vphi$, our data show that the GSS is connected to the disk, providing a direct evidence that disk stars were formed in part from gas clouds left behind after the starburst episode. Interestingly, according to the dichotomy of disk stars into high- and low-[$\alpha$/Fe] sequences \citep[e.g.,][and references therein]{hayden:15}, the tip of the GSS has a metallicity ([Fe/H] $\approx-0.4$) similar to that of the low-metallicity end of the low-$\alpha$ disk. This implies that gas clouds in the protogalactic disk were diluted by fresh, low-metallicity materials accreted by the GSE merger, as has often been invoked in numerical simulations \citep{brook:07,buck:20}; see also \citet{chiappini:97}. Inflow from the circumgalactic medium \citep{grand:20} or a gaseous outer disk \citep{renaud:21} after the merger may also be plausible \citep[also see discussions in][]{myeong:22}. Intermediate [$\alpha$/Fe] is then a consequence of core-collapse supernovae during the starburst. The inflow of metal-enriched gas onto the thin disk can naturally explain the G-dwarf problem as well --- the apparent excess of metal-rich stars in the local disk compared to the prediction from a simple closed-box chemical model \citep[e.g.,][]{greener:21}.

In this context, we conjecture that the ``Nyx'' stream identified by \citet{necib:20} is closely related to the vertical arm of the GSS. In their study, a group of $\sim100$ stars in the solar vicinity shows coherent radial ($134\ \kms$) and azimuthal ($130\ \kms$) motions. Their metallicities peak at [Fe/H] $=-0.55$, with a dispersion of $0.13$~dex, and are mostly confined to a plane with a maximum vertical distance of $Z = 1.7$~kpc. The latter value is consistent with a small scale height of stars having similar [Fe/H] and $\vphi$ as Nyx stars ($H\approx0.6$~kpc; Figure~\ref{fig:sheight}). Nonetheless, most of the Nyx stars have $100 \la \vphi < 250\ \kms$, with a few extreme cases, and therefore trace only the upper half of the vertical arm of the GSS. \citet{necib:20} postulated that the Nyx stream is a remnant of a disrupted dwarf galaxy. However, a more recent analysis in \citet{zucker:21} argues against their extragalactic origin, based on the fact that a subset of these stars are indistinguishable from thick-disk stars in the elemental-abundance space, having larger $\alpha$-element abundances than those of accreted stars at a given metallicity.

\subsection{Future Prospects}

The extensibility of calibrated synthetic spectra presented in this work will enable accurate prediction of stellar magnitudes for filter passbands in various photometric surveys, such as the Javalambre/Southern Photometric Local Universe Survey \citep[J/S-PLUS;][]{cenarro:19,splus} and the Legacy Survey of Space and Time \citep[LSST;][]{ivezic:19}. Our metallicity-mapping technique based on the empirically calibrated isochrones will also serve as a useful resource for studying the demographics of stellar populations that are yet to be discovered in the local universe from the upcoming surveys.

\acknowledgements\

We thank Donald M.\ Terndrup and Christopher A.\ Onken for carefully checking our manuscript and providing a number of helpful comments. We express our gratitude to Vasily Belokurov for a valuable conversation regarding Eos and its possible connection to the GSS. D.A.\ thanks Anirudh Chiti, Robert Grand, and Ioana Ciuc{\u{a}} for useful discussions. T.C.B.\ and Y.S.L.\ thank Andreia Carrillo for an enlightening discussion on the Galactic starburst. D.A.\ acknowledges support provided by the National Research Foundation (NRF) of Korea grant funded by the Ministry of Science and ICT (No.\ 2021R1A2C1004117). T.C.B. and Y.S.L. acknowledge partial support for this work from grant PHY 14-30152; Physics Frontier Center/JINA Center for the Evolution of the Elements (JINA-CEE), and OISE-1927130: The International Research Network for Nuclear Astrophysics (IReNA), awarded by the US National Science Foundation. Y.S.L.\ acknowledges support from the NRF of Korea grant funded by the Ministry of Science and ICT (NRF-2021R1A2C1008679). T.M.\ acknowledges financial support from the Spanish Ministry of Science and Innovation (MICINN) through the Spanish State Research Agency, under the Severo Ochoa Program 2020-2023 (CEX2019-000920-S) as well as support from the ACIISI, Consejer\'{i}a de Econom\'{i}a, Conocimiento y Empleo del Gobierno de Canarias and the European Regional Development Fund (ERDF) under grant with reference  PROID2021010128.

Funding for the Sloan Digital Sky Survey IV has been provided by the Alfred P.\ Sloan Foundation, the U.S.\ Department of Energy Office of Science, and the Participating Institutions. SDSS-IV acknowledges support and resources from the Center for High Performance Computing  at the University of Utah. The SDSS website is www.sdss.org.

SDSS-IV is managed by the Astrophysical Research Consortium for the Participating Institutions of the SDSS Collaboration including the Brazilian Participation Group, the Carnegie Institution for Science, Carnegie Mellon University, Center for Astrophysics | Harvard \& Smithsonian, the Chilean Participation Group, the French Participation Group, Instituto de Astrof\'isica de Canarias, The Johns Hopkins University, Kavli Institute for the Physics and Mathematics of the Universe (IPMU) / University of Tokyo, the Korean Participation Group, Lawrence Berkeley National Laboratory, Leibniz Institut f\"ur Astrophysik Potsdam (AIP),  Max-Planck-Institut f\"ur Astronomie (MPIA Heidelberg), Max-Planck-Institut f\"ur Astrophysik (MPA Garching), Max-Planck-Institut f\"ur Extraterrestrische Physik (MPE), National Astronomical Observatories of China, New Mexico State University, New York University, University of Notre Dame, Observat\'ario Nacional / MCTI, The Ohio State University, Pennsylvania State University, Shanghai Astronomical Observatory, United Kingdom Participation Group, Universidad Nacional Aut\'onoma de M\'exico, University of Arizona, University of Colorado Boulder, University of Oxford, University of Portsmouth, University of Utah, University of Virginia, University of Washington, University of Wisconsin, Vanderbilt University, and Yale University.

This work presents results from the European Space Agency (ESA) space mission Gaia. Gaia data are being processed by the Gaia Data Processing and Analysis Consortium (DPAC). Funding for the DPAC is provided by national institutions, in particular the institutions participating in the Gaia MultiLateral Agreement (MLA). The Gaia mission website is https://www.cosmos.esa.int/gaia. The Gaia archive website is https://archives.esac.esa.int/gaia.

This paper makes use of data from the AAVSO Photometric All Sky Survey, whose funding has been provided by the Robert Martin Ayers Sciences Fund and from the NSF (AST-1412587).

\facilities{AAVSO, PS1, SkyMapper, Sloan}

\software{GNU parallel (Tange 2021)}

\appendix

\section{Construction of the Gaia Double Sequence and Model Comparisons}\label{sec:cteff}

\begin{figure*}
  \centering
  \gridline{\fig{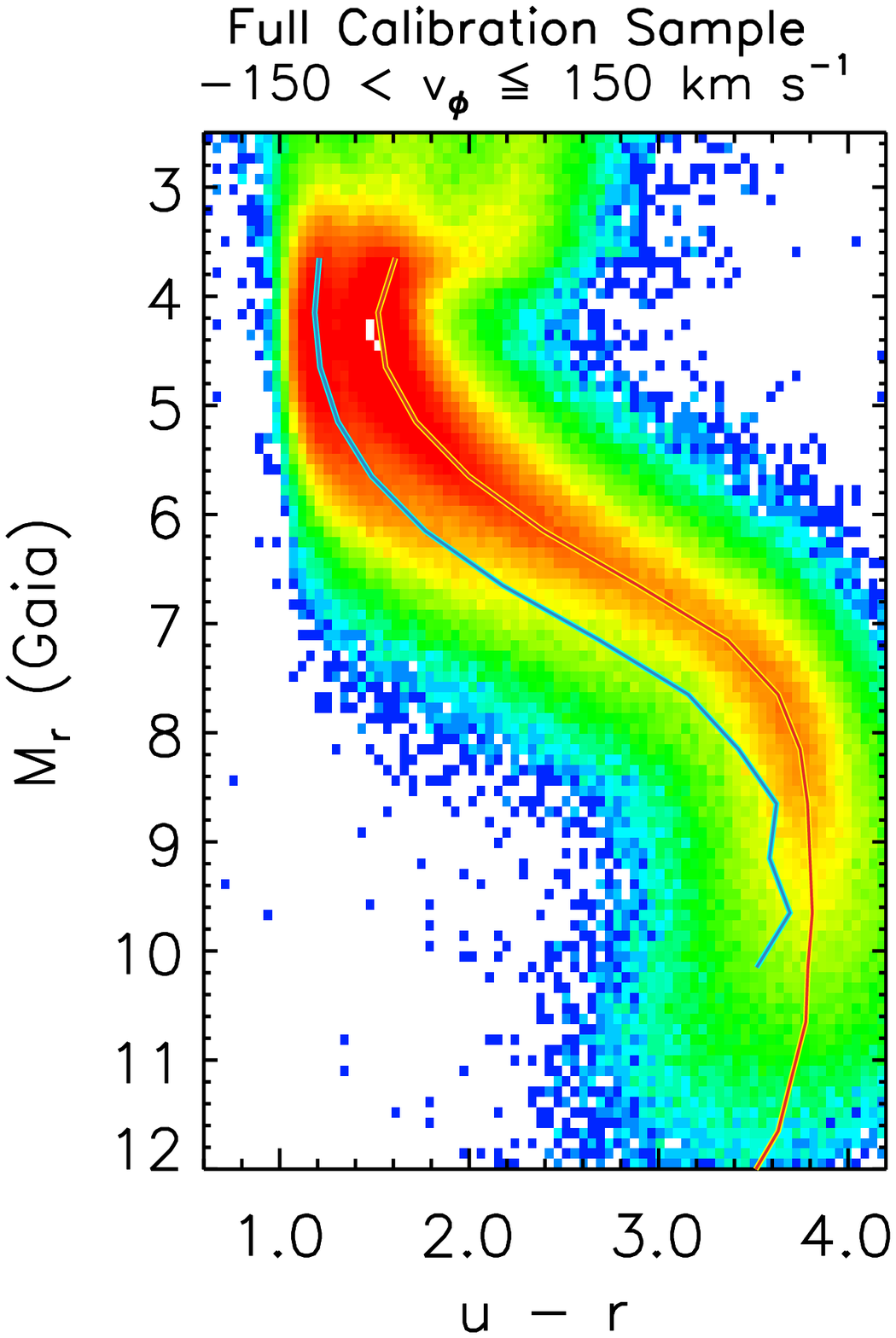}{0.34\textwidth}{\textbf{(a) SDSS $ur$}}
                \fig{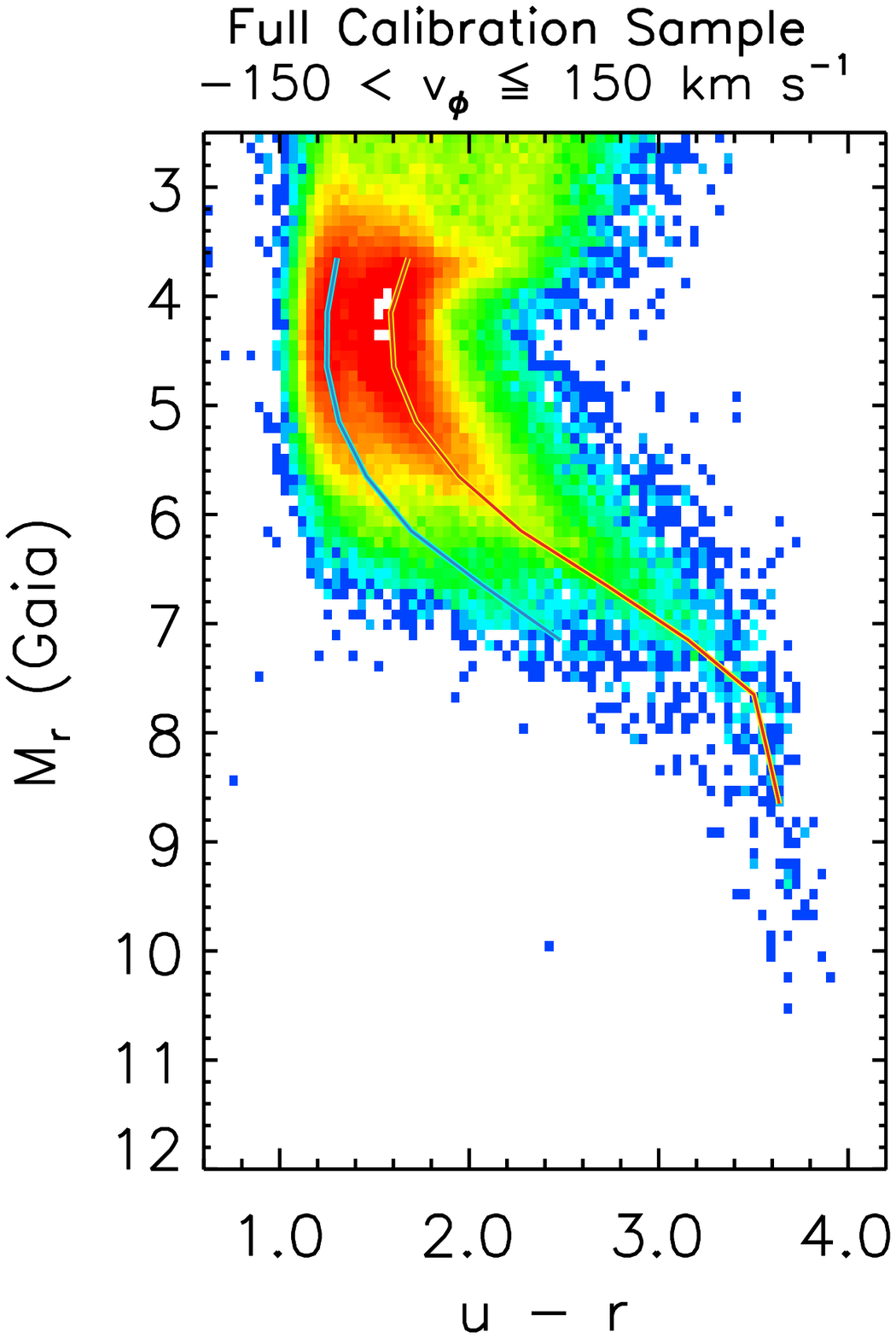}{0.34\textwidth}{\textbf{(b) SMSS $ur$}}}
  \gridline{\fig{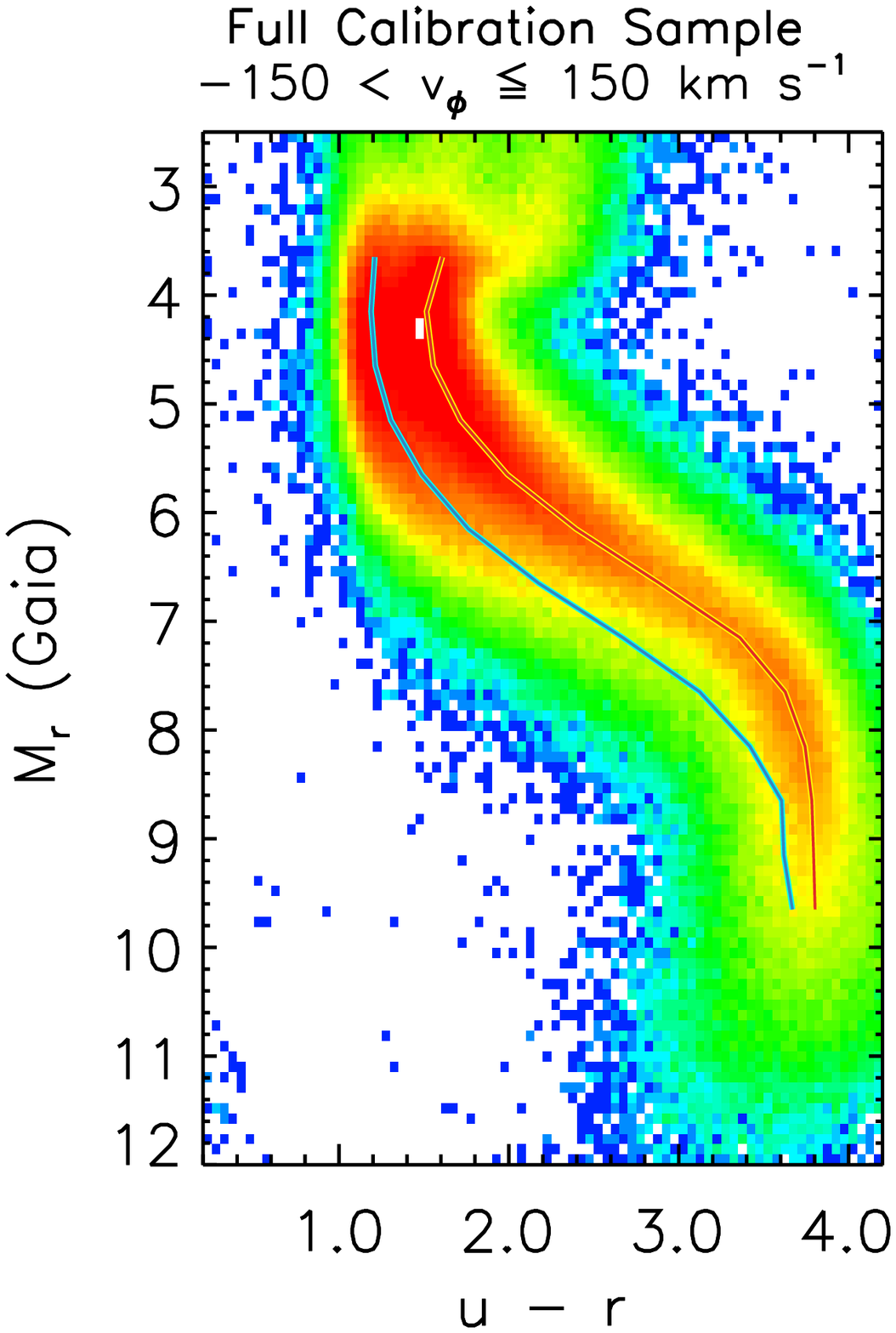}{0.34\textwidth}{\textbf{(c) SDSS $u$ and PS1 $r$}}
                \fig{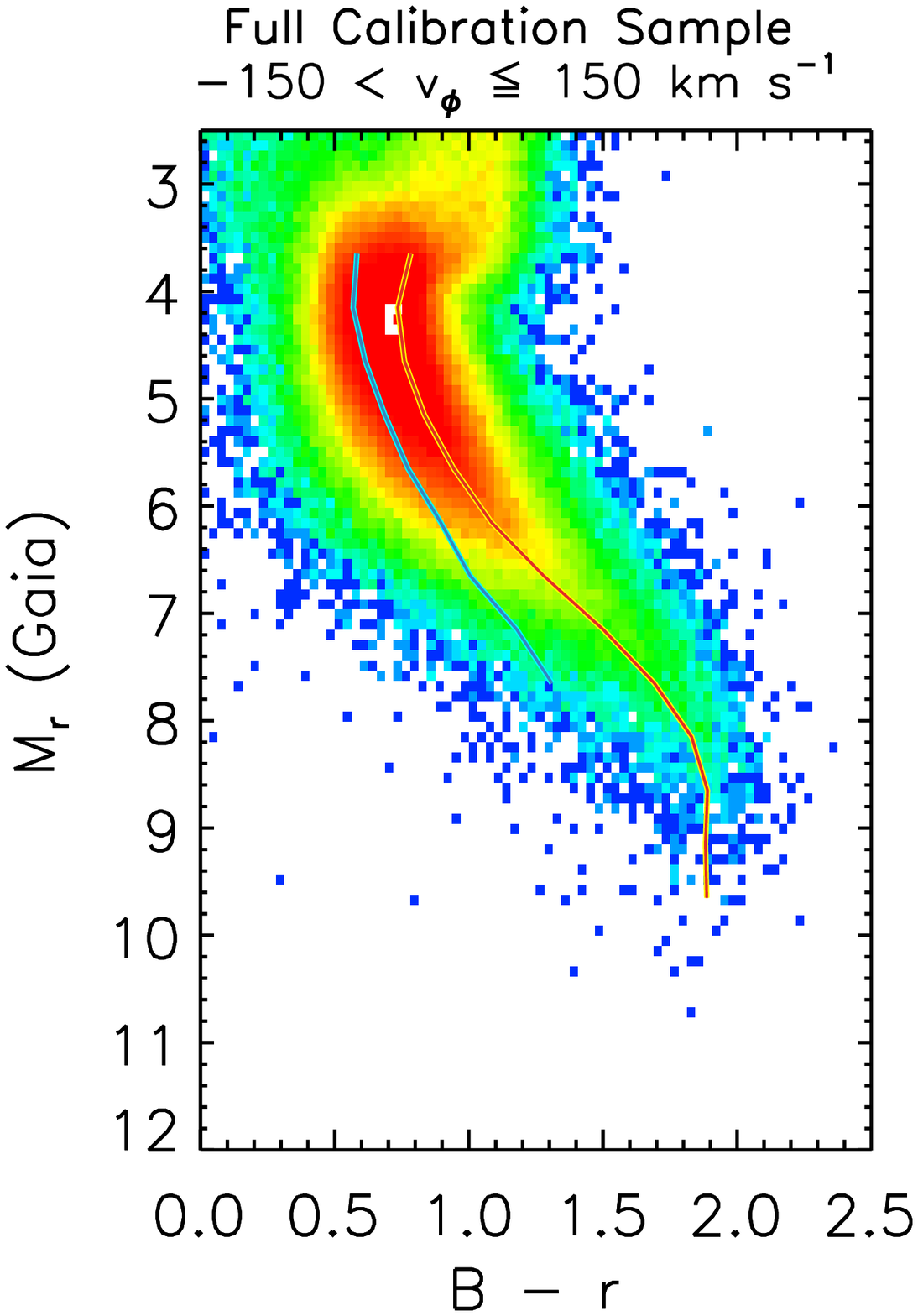}{0.34\textwidth}{\textbf{(d) APASS $B$ and SDSS $r$}}}
  \caption{Gaia's double sequence in selected passbands. The colored histogram shows a number density of stars with $-150 < \vphi \leq +150\ \kms$. The blue and red lines indicate fiducial sequences derived in this study.}
  \label{fig:gaiadouble}
\end{figure*}

\begin{figure*}
  \centering
  \gridline{\fig{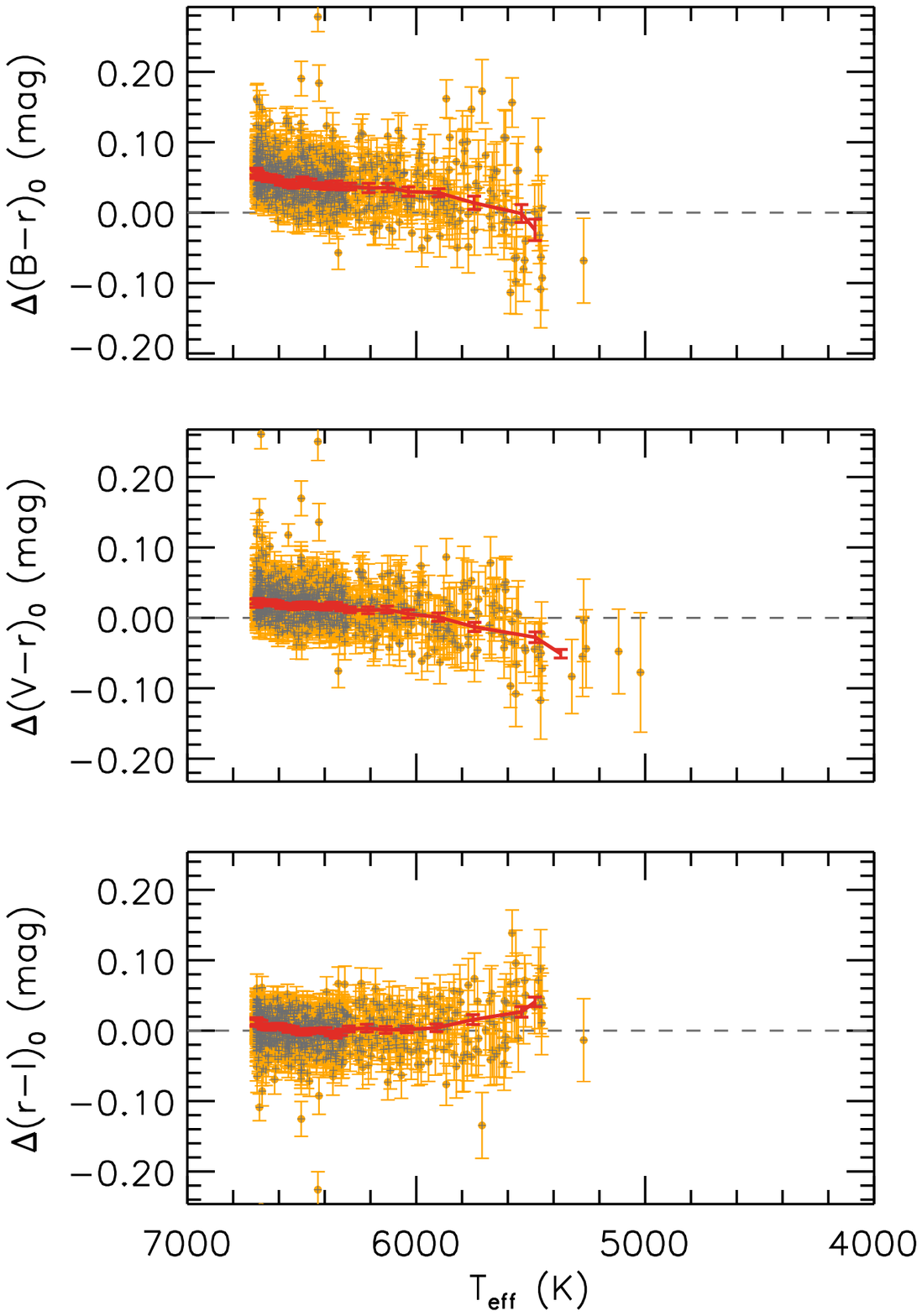}{0.4\textwidth}{\textbf{(a) M92}}
                \fig{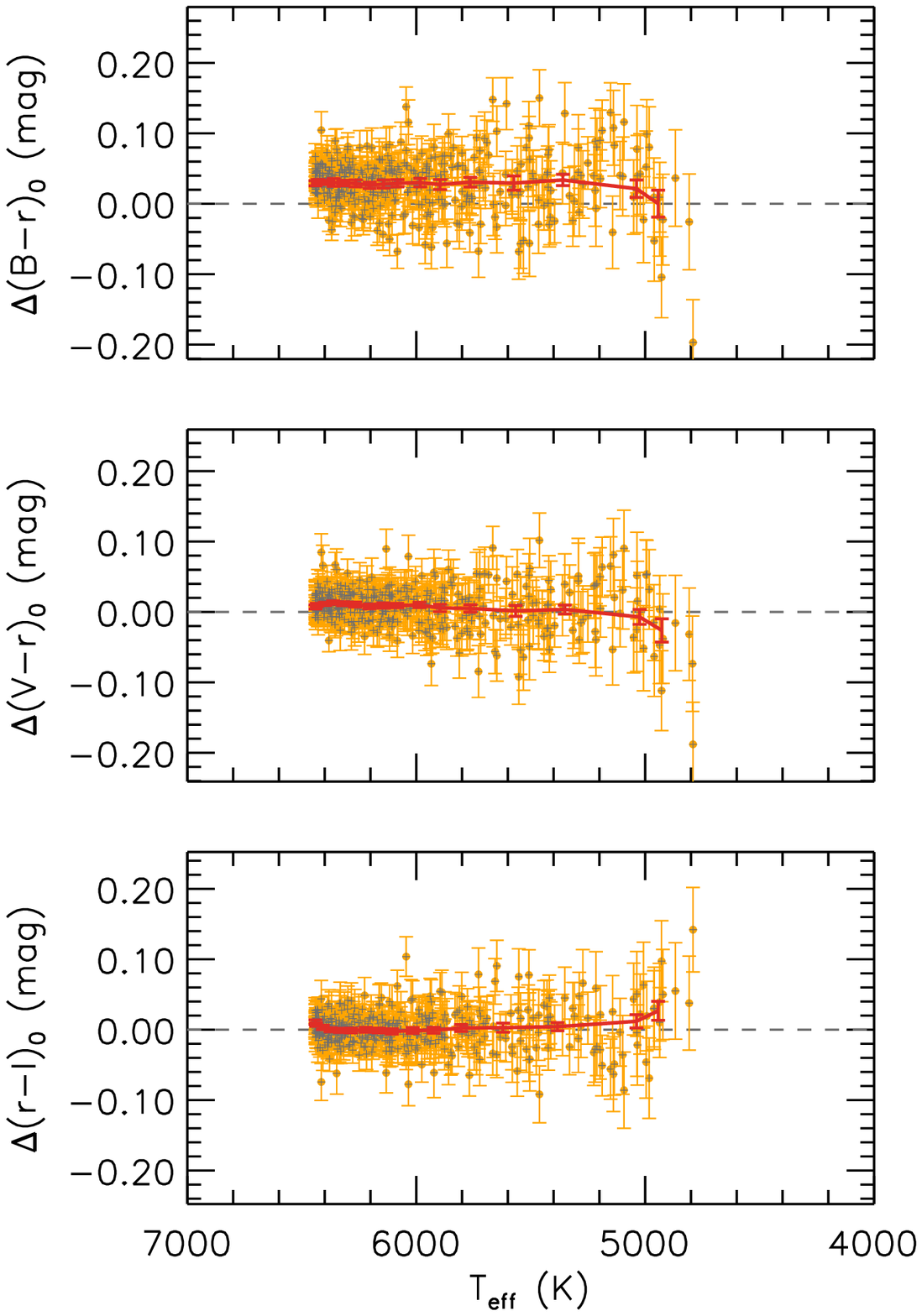}{0.4\textwidth}{\textbf{(b) M13}}}
  \gridline{\fig{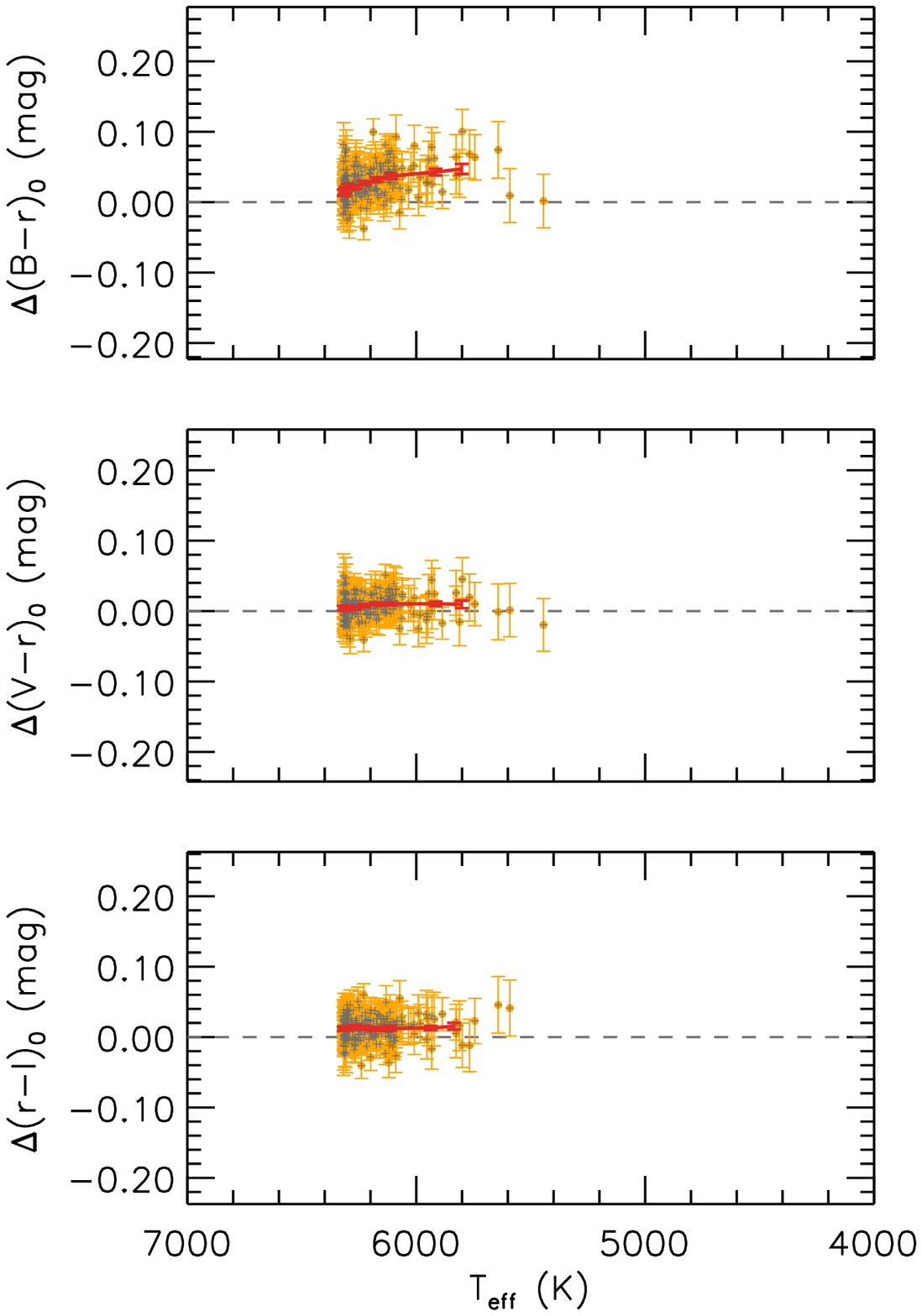}{0.4\textwidth}{\textbf{(c) M5}}
                \fig{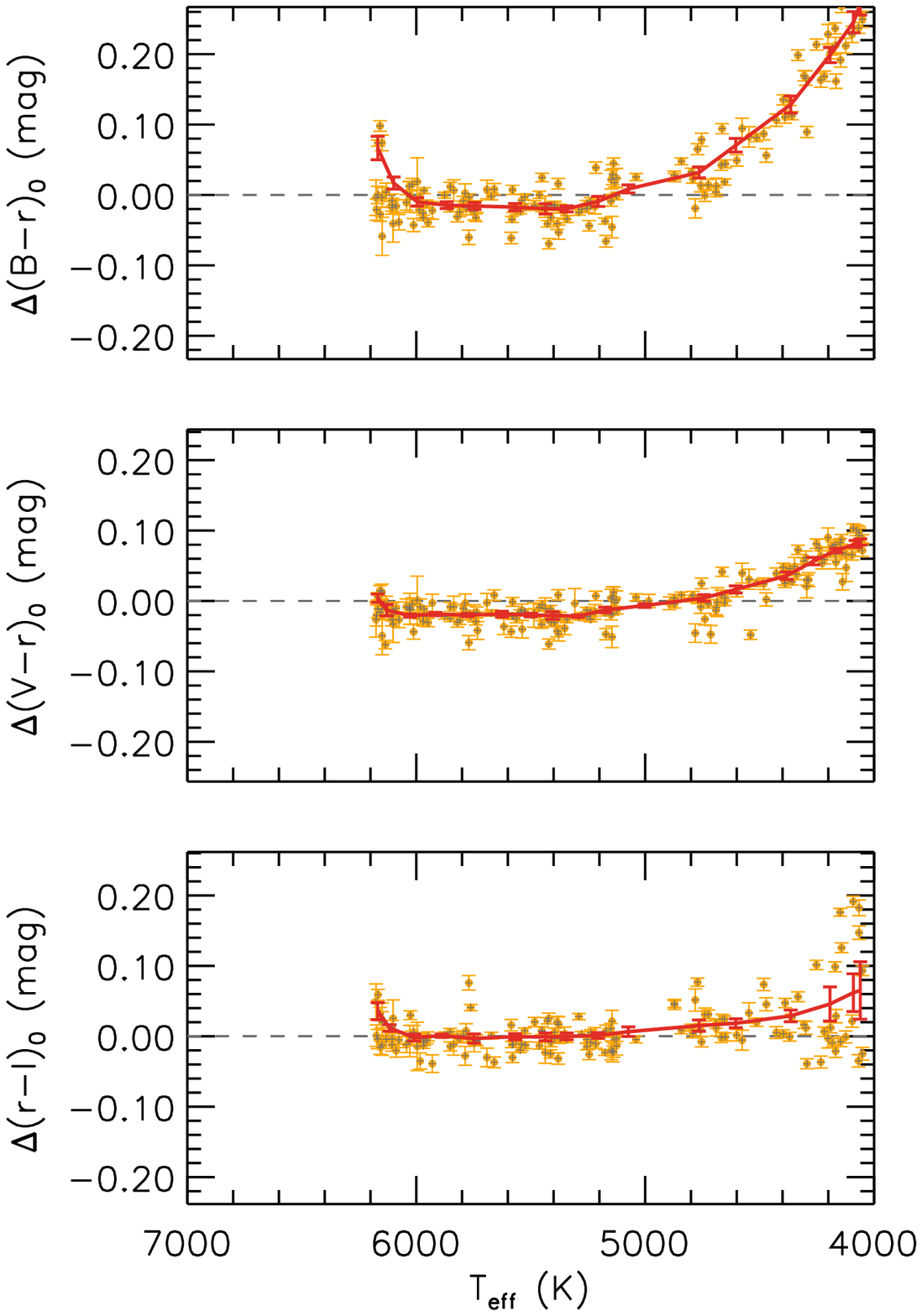}{0.4\textwidth}{\textbf{(d) M67}}}
  \caption{Comparisons of Stetson's photometry with theoretical models for selected clusters.}
  \label{fig:stetson}
\end{figure*}

\begin{figure*}
\epsscale{0.9}
\plotone{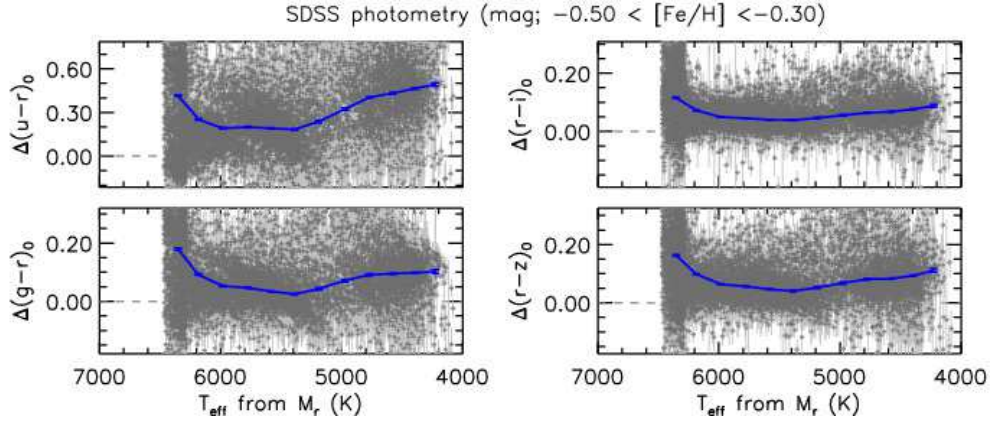}
\caption{Comparisons between the observed and original model colors for the SEGUE sample in the SDSS passbands. The above example shows differences of extinction-corrected colors using stars at $-0.5 < {\rm [Fe/H]} < -0.3$ as a function of photometric $\teff$. The solid line shows average differences in moving boxes (with error bars indicating uncertainties of the mean values).}
\label{fig:cteff_spec1}
\end{figure*}

\begin{figure*}
\epsscale{0.9}
\plotone{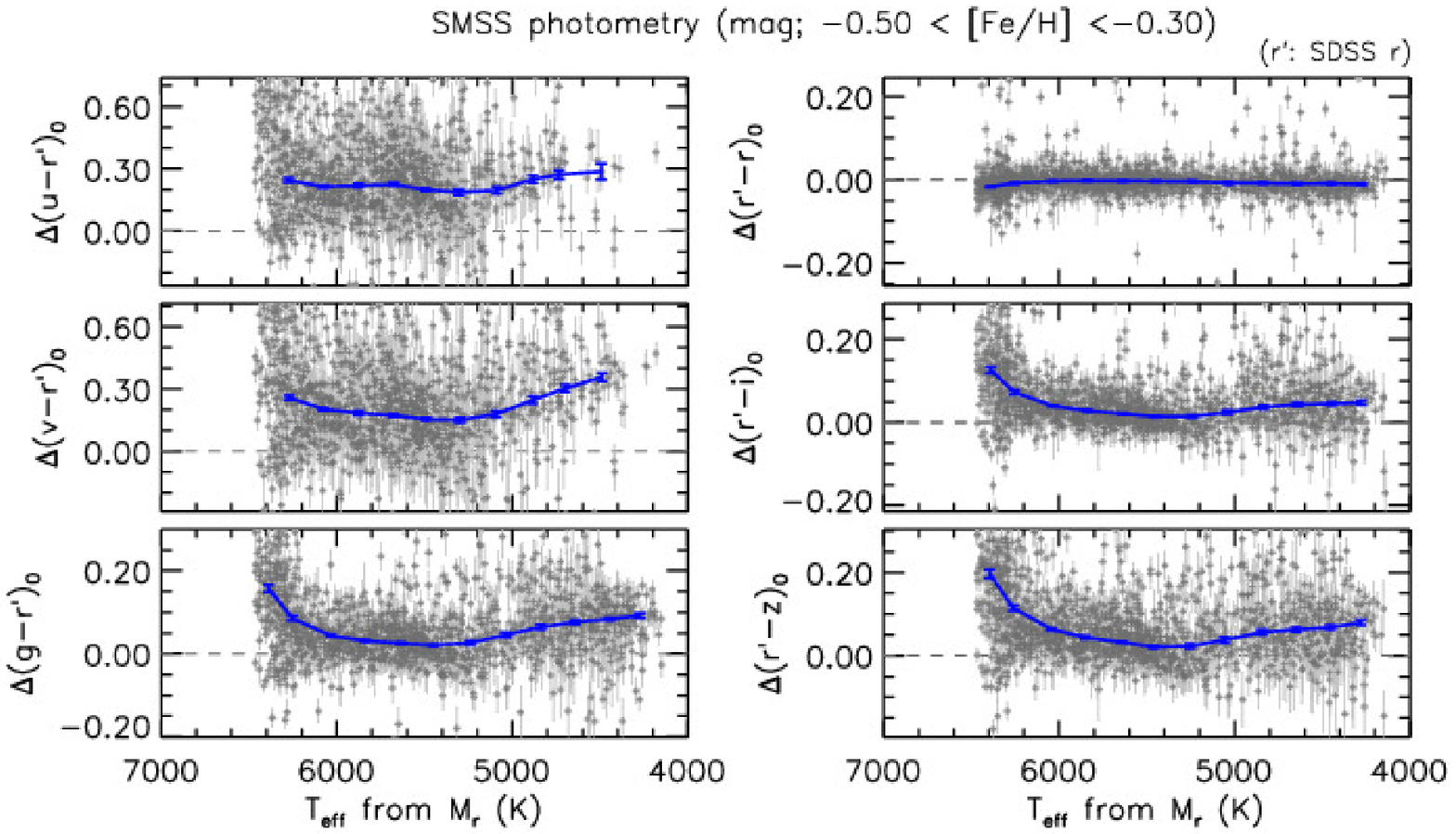}
\caption{Same as in Figure~\ref{fig:cteff_spec1}, but in the SMSS filters. SDSS $r$ is shown as $r'$.}
\label{fig:cteff_spec2}
\end{figure*}

\begin{figure*}
\epsscale{0.9}
\plotone{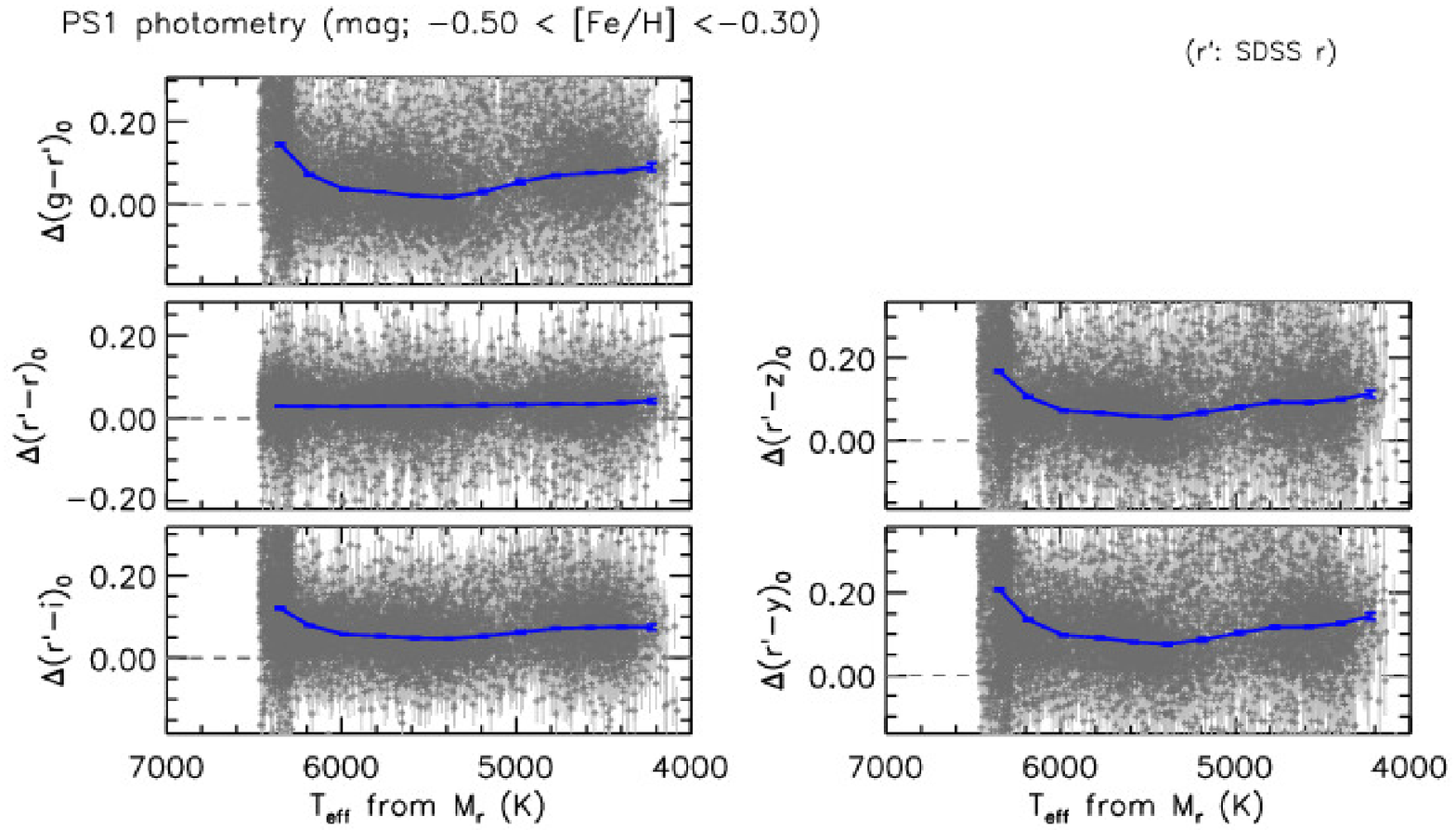}
\caption{Same as in Figure~\ref{fig:cteff_spec1}, but in the PS1 filters. SDSS $r$ is shown as $r'$.}
\label{fig:cteff_spec3}
\end{figure*}

\begin{figure*}
\epsscale{0.5}
\plotone{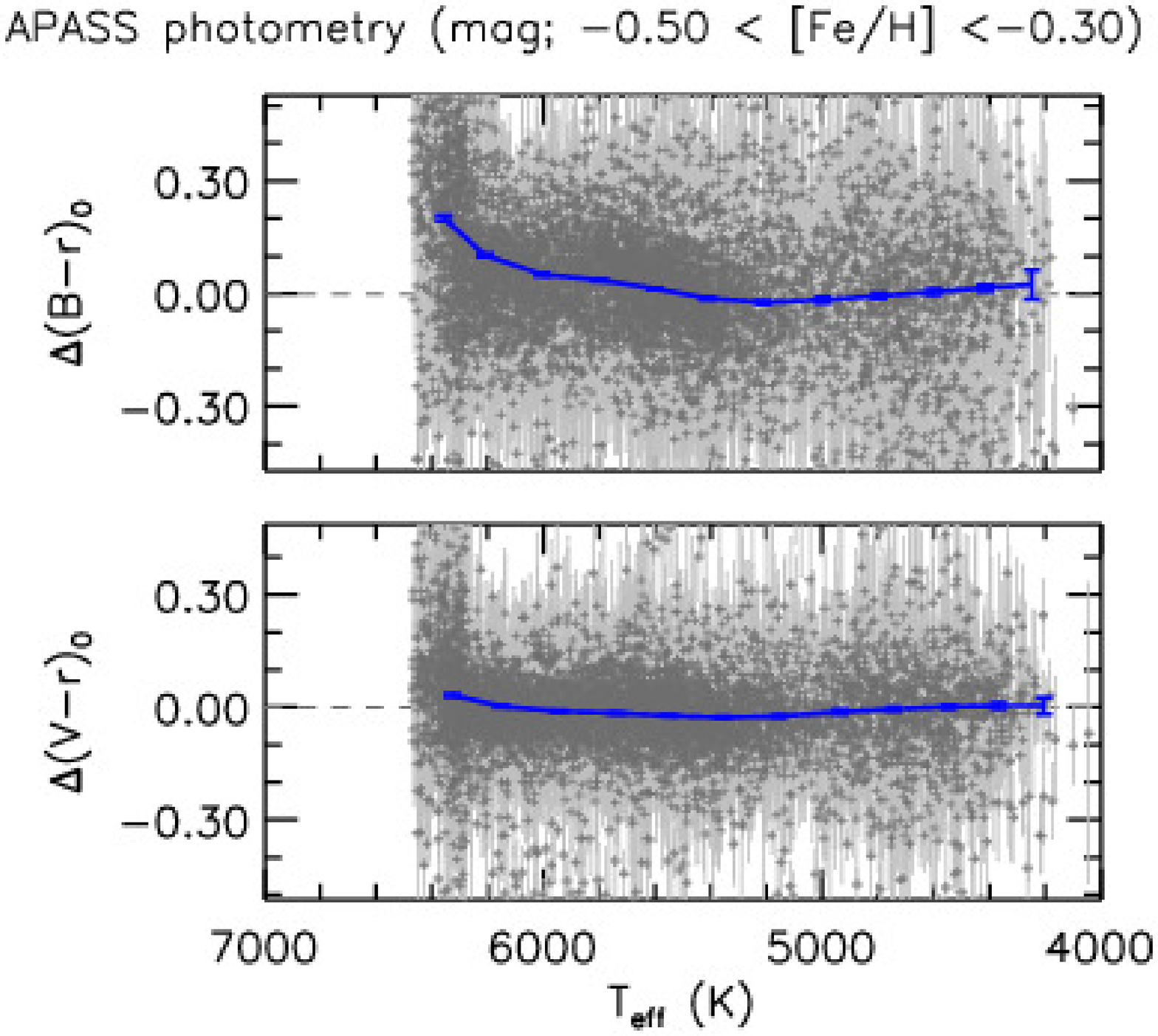}
\caption{Same as in Figure~\ref{fig:cteff_spec1}, but in the APASS filters (with SDSS $r$).}
\label{fig:cteff_spec4}
\end{figure*}

In this Appendix, a number of complementary plots are provided for some selected data sets, which are used in our model calibration (\S~\ref{sec:calib}). Figure~\ref{fig:gaiadouble} displays Gaia's double sequences in SDSS, SMSS, PS1, and APASS using color indices containing $u$ or $B$ passbands, which best separate the blue and red sequences of stars having $-150 < \vphi \leq 150\ \kms$. Figure~\ref{fig:stetson} displays comparisons of the original models (without empirical corrections) with Stetson's cluster photometry. See Paper~II for more information on the construction of the Gaia's double sequence and model comparisons with the cluster and the Gaia double sequence in the SDSS filter set. Model comparisons with the SEGUE sample ($-0.5 < {\rm [Fe/H]} < -0.3$) are shown in Figures~\ref{fig:cteff_spec1}--\ref{fig:cteff_spec4} in SDSS, SMSS, PS1, and APASS photometry, respectively.

\section{Photometric Properties of SDSS, SMSS, and PS1}\label{sec:zp}

\subsection{Random and Zero-Point Uncertainties}

\begin{figure}
\epsscale{1.1}
\plottwo{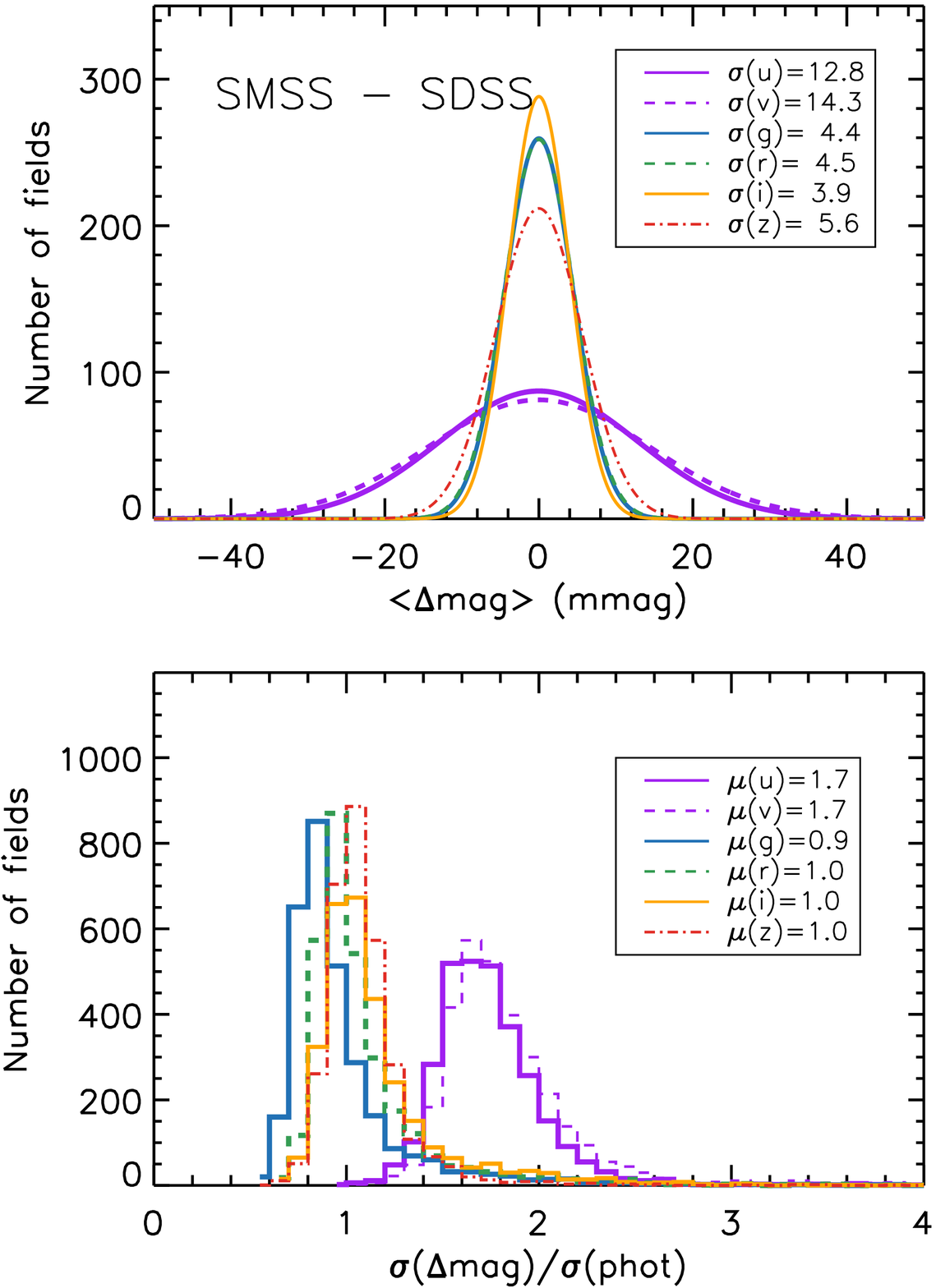}{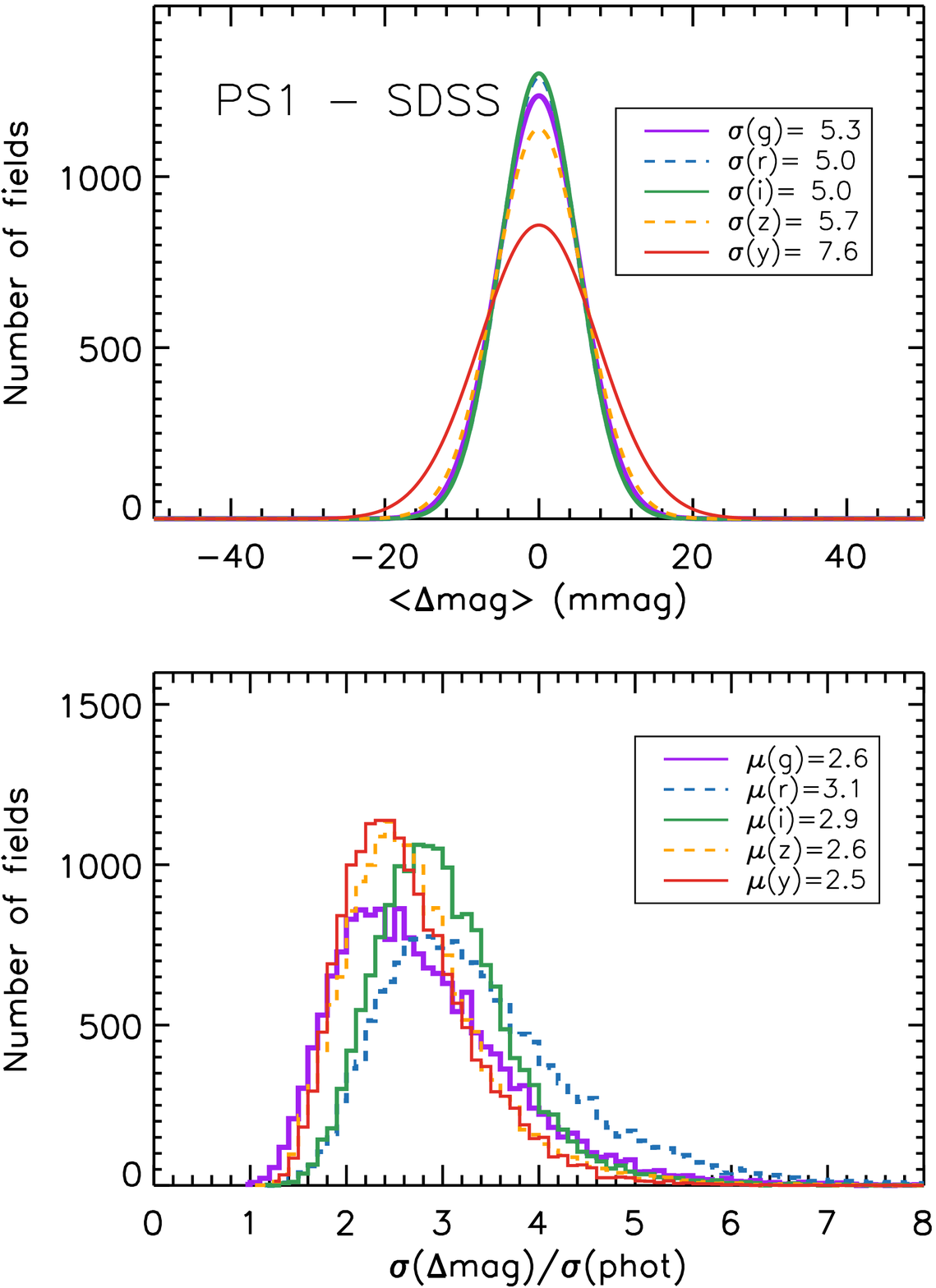}
\caption{Statistical properties of SMSS (left) and PS1 photometry (right). Top: A best-fitting Gaussian distribution of mean magnitude offsets from SDSS photometry ($\langle\Delta {\rm mag}\rangle$) based on comparisons in $(\Delta \alpha, \Delta \delta) = (1\arcdeg, 1\arcdeg$) tiles. The mean difference is computed after subtracting a color-dependent difference between different passbands, where both SMSS $u$- and $v$ bands are matched to SDSS $u$, and PS1 $y$ to SDSS $z$. A standard deviation of each distribution is shown in the inset. Bottom: A distribution of the ratio between a standard deviation of a magnitude difference [$\sigma (\Delta {\rm mag})$] and a propagated uncertainty [$\sigma ({\rm phot})$] in each $(1\arcdeg, 1\arcdeg$) tile. The mean ratio in each filter passband is shown in the inset.}
\label{fig:phot}
\end{figure}

Because photometric data in various filter passbands are combined to derive a set of stellar parameters in this study, it is necessary to adopt accurate photometric uncertainties in the $\chi^2$ statistics. Here, we compare SMSS and PS1 photometry with SDSS, and compute a mean magnitude difference and a dispersion to infer the size of the true uncertainties. The comparison with SMSS is limited to narrow regions, since the imaging stripes of SDSS overlaps only a little with SMSS footprints along the celestial equator. The overlap with PS1 is more extensive, as both SDSS and PS1 cover the Northern Hemisphere.

Figure~\ref{fig:phot} shows statistical properties of the comparison of SDSS photometry with SMSS (left) and PS1 (right), respectively. The top panels display a distribution of a magnitude difference of relatively bright ($14.5 < r < 18$~mag) stars in each passband. To take into account non-negligible color terms between different filter passbands (i.e., color transformations), the magnitude differences in each $1\arcdeg$-wide strip in R.A. are fit using a third-order polynomial as a function of $g\, -\, r$ in $0.3 < g\, -\, r < 1.1$, and its mean trend is removed. The SMSS $v$ is compared to SDSS $u$, and PS1 $y$ to SDSS $z$.

A weighted median difference is computed in each $(\Delta \alpha, \Delta \delta) = (1\arcdeg, 1\arcdeg$) region, and an ensemble of these differences are fit using a Gaussian function, as shown by solid lines in the top panels of Figure~\ref{fig:phot}. Standard deviations of the best-fitting Gaussian functions provide a measure of the spatial variation of the photometric zero points across the sky. They are
$\sigma_{\rm zp}^{\rm SMSS} (u,v,g,r,i,z) = \{13, 14, 4, 4, 4, 6\}$~mmag and $\sigma_{\rm zp}^{\rm PS1} (g,r,i,z,y) = \{5, 5, 5, 6, 8\}$~mmag, respectively, which are comparable to the quoted systematic uncertainties in these surveys.

The bottom panels of Figure~\ref{fig:phot} show ratios between the standard deviation of a magnitude difference and a propagated uncertainty for each filter passband in each $(\Delta \alpha, \Delta \delta) = (1\arcdeg, 1\arcdeg$) patch. Again, SMSS $v$ is compared to SDSS $u$, and PS1 $y$ to SDSS $z$. Some of the brightest objects are rejected in this comparison, owing to unrealistically small uncertainties in the SDSS PSF $i$ magnitude ($<0.005$~mag). As shown in the bottom left panel, the ratios for the SMSS and SDSS $grz$ are near unity, indicating that photometric uncertainties are comparable to the observed scatter, while photometric uncertainties in $uv$ are likely under-estimated. As shown in the bottom right panel, the differences between propagated uncertainties and observed dispersions are even larger for PS1 passbands.

Based on the above comparisons, an uncertainty `floor' ($\sigma_f$) is computed in each passband in order to make a median of a standard deviation equal to a median of propagated uncertainties in all $1\arcdeg \times 1\arcdeg$ patches. Assuming that quoted uncertainties in SDSS are correct, we find $\sigma_f^{\rm SMSS} (u,v,g,r,i,z) = \{42, 42, 10, 9, 9, 10\}$~mmag and $\sigma_f^{\rm PS1} (g,r,i,z,y) = \{43, 44, 42, 44, 43\}$~mmag for SMSS and PS1, respectively. In all cases, zero-point uncertainties ($\sigma_{\rm zp}$) are overwhelmed by the uncertainty floors ($\sigma_f$). We add both uncertainties in quadrature to the original photometric uncertainties in SMSS and PS1, and use them throughout this work.

\subsection{Re-calibration of SMSS $uv$-Band Photometry}

By design, the zero point of SMSS DR2 photometry was set based on synthetic $griz$ photometry in the PS1 system, calculated from all-sky Gaia photometry \citep{onken:19}. The wavelengths of SMSS $griz$ passbands overlap with Gaia and PS1 passbands, so in principle one can tie them together without losing information on the properties of stellar spectra. On the contrary, photometric zero points in $u$-band and $v$-band remain largely unconstrained due to the lack of short-wavelength passbands in Gaia and PS1.

\begin{figure*}
  \centering
  \gridline{\fig{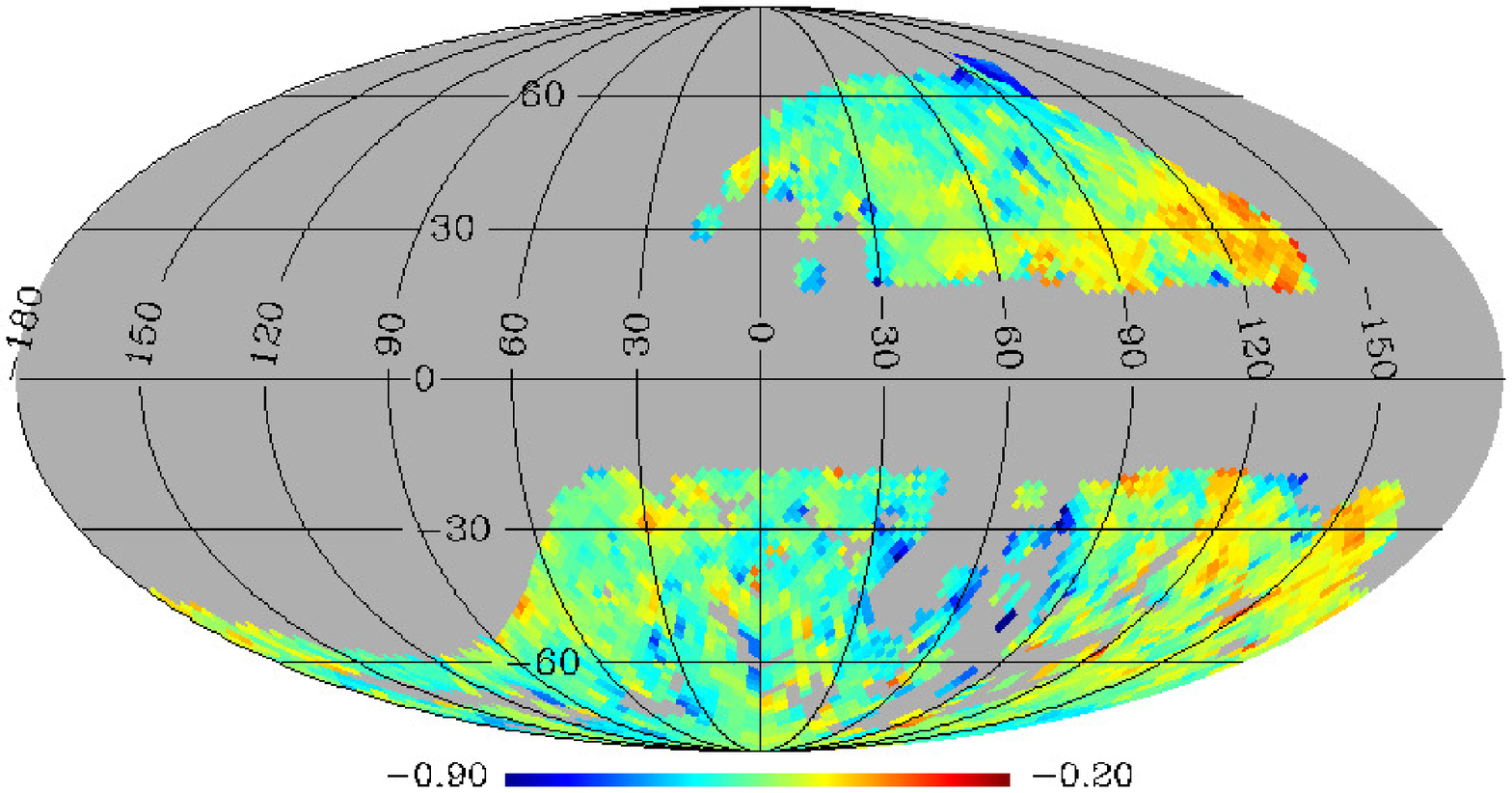}{0.4\textwidth}{\textbf{(a) Original SMSS photometry}}
                \fig{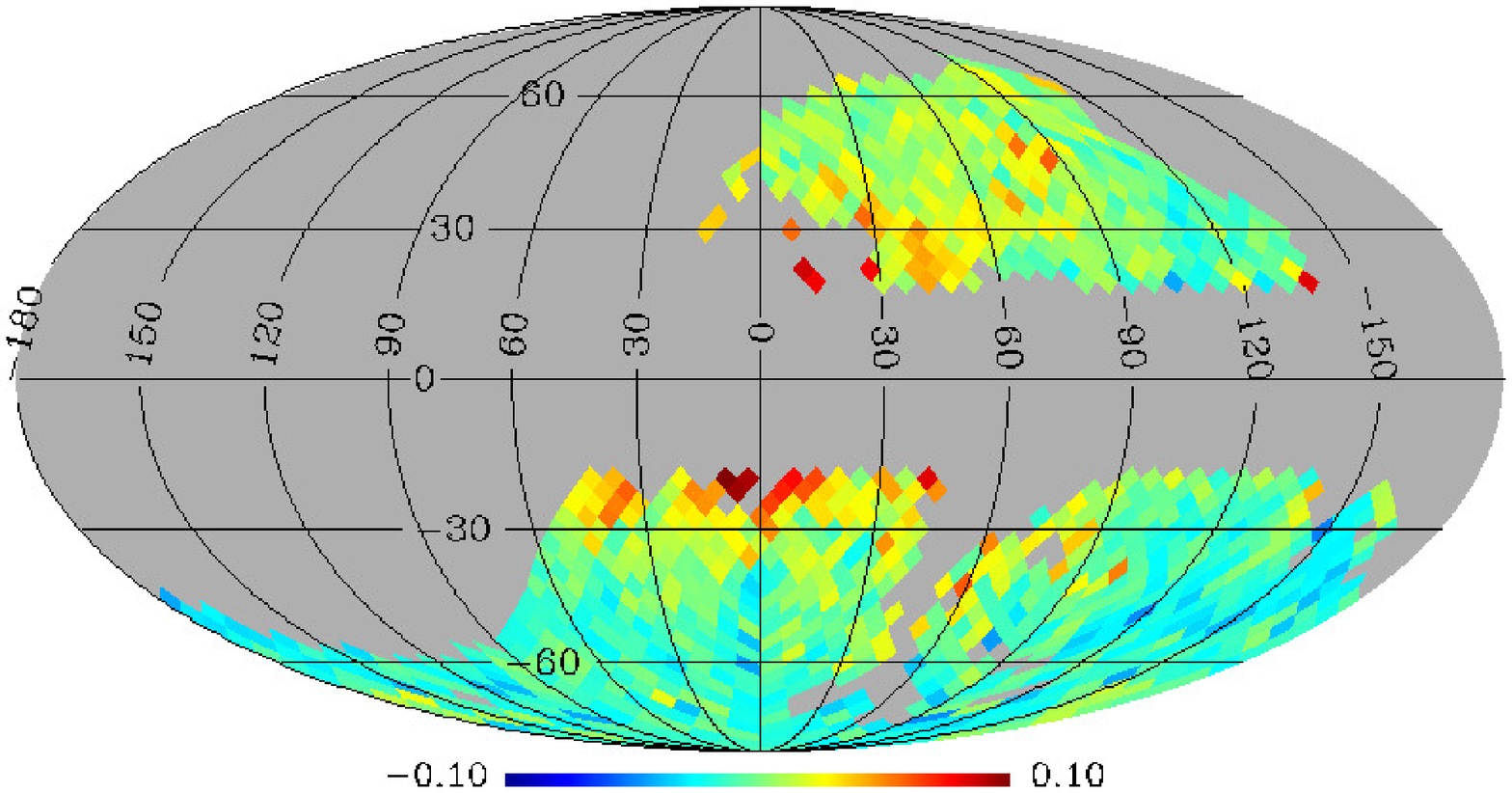}{0.4\textwidth}{\textbf{(b) Zero-point correction in $uv$.}}}
  \gridline{\fig{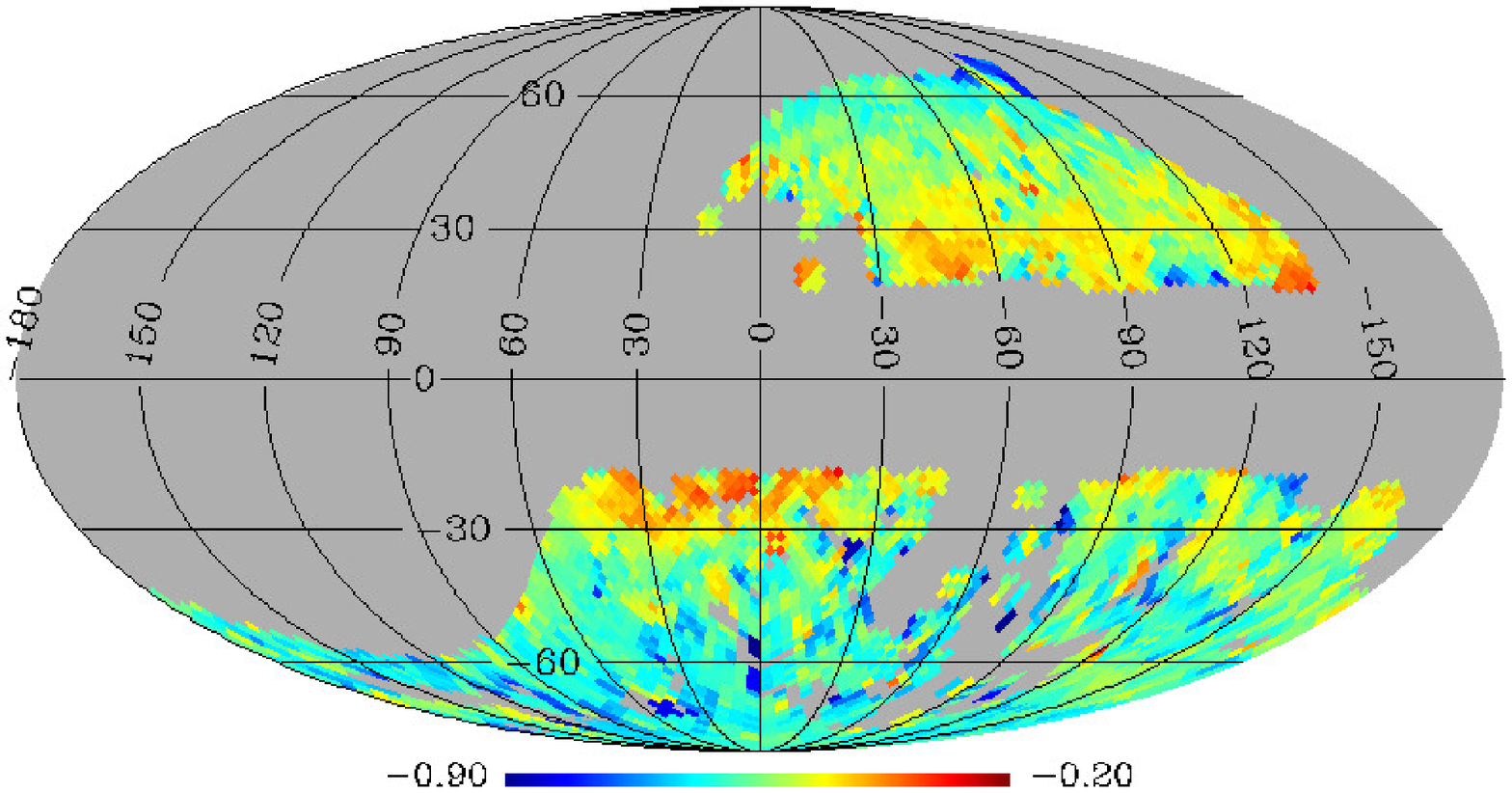}{0.4\textwidth}{\textbf{(c) Re-calibrated SMSS photometry}}
                \fig{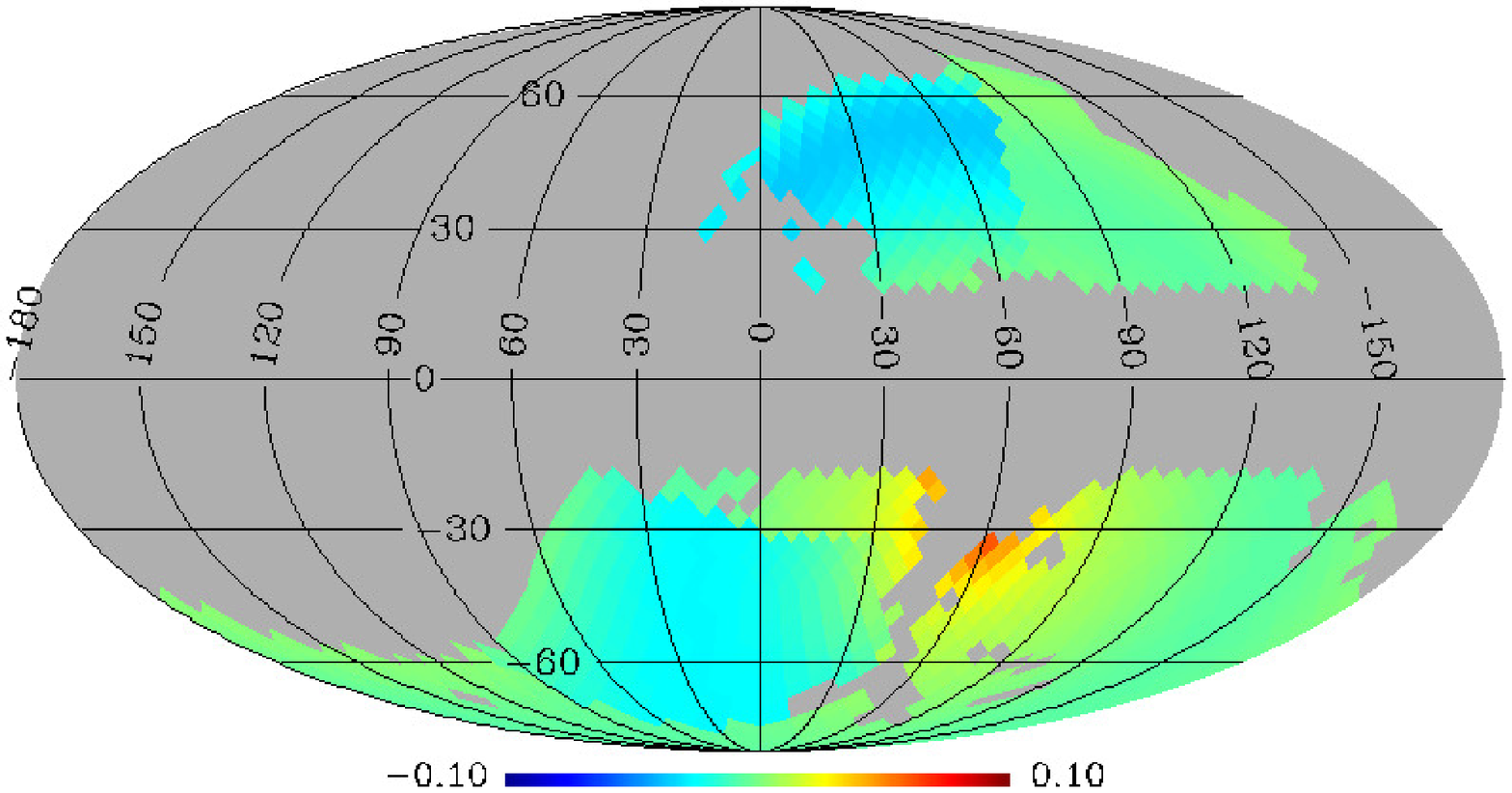}{0.4\textwidth}{\textbf{(d) Zero-point correction map in $u$ from \citet{huang:22}}}}
  \caption{Panel~(a): A projection map of mean metallicities in the Galactic coordinate system based on the original photometry. The color scheme shows a mean metallicity of stars at $1.2 < d < 3$~kpc from a generalized histogram. Panel~(b): Zero-point corrections on $u$-band and $v$-band photometry, in order to match the peak metallicities of nearby stars ($0.5 < d < 1$~kpc) to a reference value (see text). Panel~(c): Same as in panel~(a), but based on the re-calibrated $uv$ photometry. Panel~(d): Zero-point corrections in the $u$-band in \citet{huang:22}. In all panels, the Galactic center is at the center, and the North Galactic Pole is to the top. Areas at low Galactic latitudes ($|b| < 20\arcdeg$) and large cumulative extinction [$\ebv \geq 0.1$] are excluded. Note that the above metallicity maps [panels~(a) or (c)] are not exactly the same as in Figure~\ref{fig:map}, owing to different ways of estimating metallicities without and with Gaia parallaxes, respectively.}
  \label{fig:smsszp}
\end{figure*}

Panel~(a) of Figure~\ref{fig:smsszp} shows the mean metallicity distribution of stars ($1.2 < d < 3$~kpc) from the SMSS in the Galactic coordinate system, which demonstrates the necessity for a second-order photometric zero-point correction. Here, we use metallicities from a fully photometric solution (0.6 million stars), as it is more prone to photometric errors than the case based on Gaia parallaxes. The mean metallicity in each pixel is derived from a generalized histogram of photometric metallicities, which accounts for an uncertainty in metallicity by taking it as a standard deviation of a normal probability distribution. We implement HEALPix \citep{gorski:05} in the Galactic coordinate system, for which we set a resolution parameter $N_{\rm side}$ to $32$, corresponding to a constant pixel size of $3.3\ {\rm deg}^{2}$. Although the bright survey limit restricts the sample to relatively nearby stars, nearly uniform metallicities of stars are contrary to what is expected in the local volume.

To obtain zero-point corrections on SMSS $uv$ photometry, we assume that nearby stars ($0.5 < d < 1$~kpc) have the same metallicity in every direction, as they are mostly thin-disk stars. Any deviation in the mean metallicity is attributed entirely to a zero-point error in the $u$- and $v$-band, because of their larger zero-point uncertainties (Figure~\ref{fig:phot}) and stronger sensitivities on metallicity than other passbands. We also assume the same amount of offset in both passbands. The sensitivity of our metallicity estimate on zero point is estimated using high-latitude stars ($|b| > 60\arcdeg$) by comparing to a case assuming an arbitrary $0.06$~mag offset in the $u$- and $v$-band, from which we find $\Delta {\rm [Fe/H]}/\Delta {\rm mag}=-2.3$. Using this, the metallicity map is forced to match a reference metallicity ([Fe/H] $=-0.28$), which is taken from the average metallicity in the above volume. As shown in panel~(b) of Figure~\ref{fig:smsszp}, we adopt $N_{\rm side}=16$ (a pixel area of $13.4\ {\rm deg}^{2}$) for the zero-point correction map, because a higher spatial resolution results in the loss of some pixels with small numbers of stars, while information on spatial dependence is lost on a lower resolution map. In a new metallicity map based on zero-point corrections [panel~(c)], a global change of the mean metallicity from low- to high-latitude regions is evident, as expected from a simple population gradient from the disk to the halo.

For comparison, panel~(d) in Figure~\ref{fig:smsszp} shows the zero-point corrections in \citet{huang:21,huang:22}, who used spectroscopic estimates from the GALAH survey and Gaia parallaxes to derive photometric offsets in each of the SMSS filters. It shows $u$-band corrections, but similar patterns and amplitudes are seen in the $v$-band. The median difference from our map is negligible ($0.004$~mag and $0.001$~mag in $u$- and $v$-band, respectively, in the sense of our study minus their values), and a standard deviation amounts to $0.025$~mag in both bands. However, high-latitude regions ($|b| > -50\arcdeg$) are only sparsely populated by GALAH targets, and therefore their correction functions are weakly constrained in the Galactic pole region. In this respect, our correction map provides more complete information for our chemo-kinematic sample along the Galactic prime meridian.

\subsection{Re-calibration of SDSS $u$-Band Photometry}

\begin{figure*}
  \centering
  \gridline{\fig{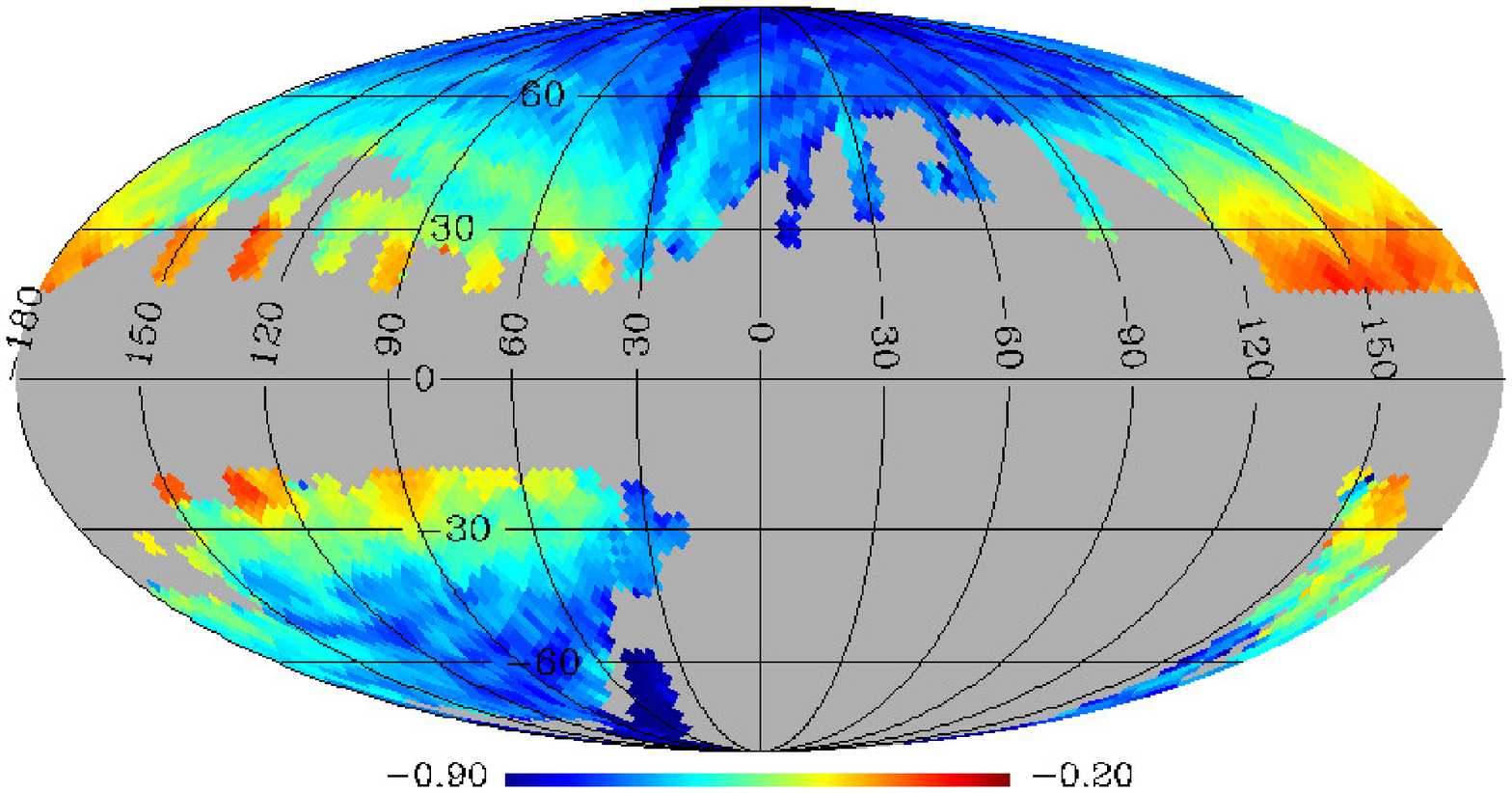}{0.4\textwidth}{\textbf{(a) Original SDSS photometry.}}
                \fig{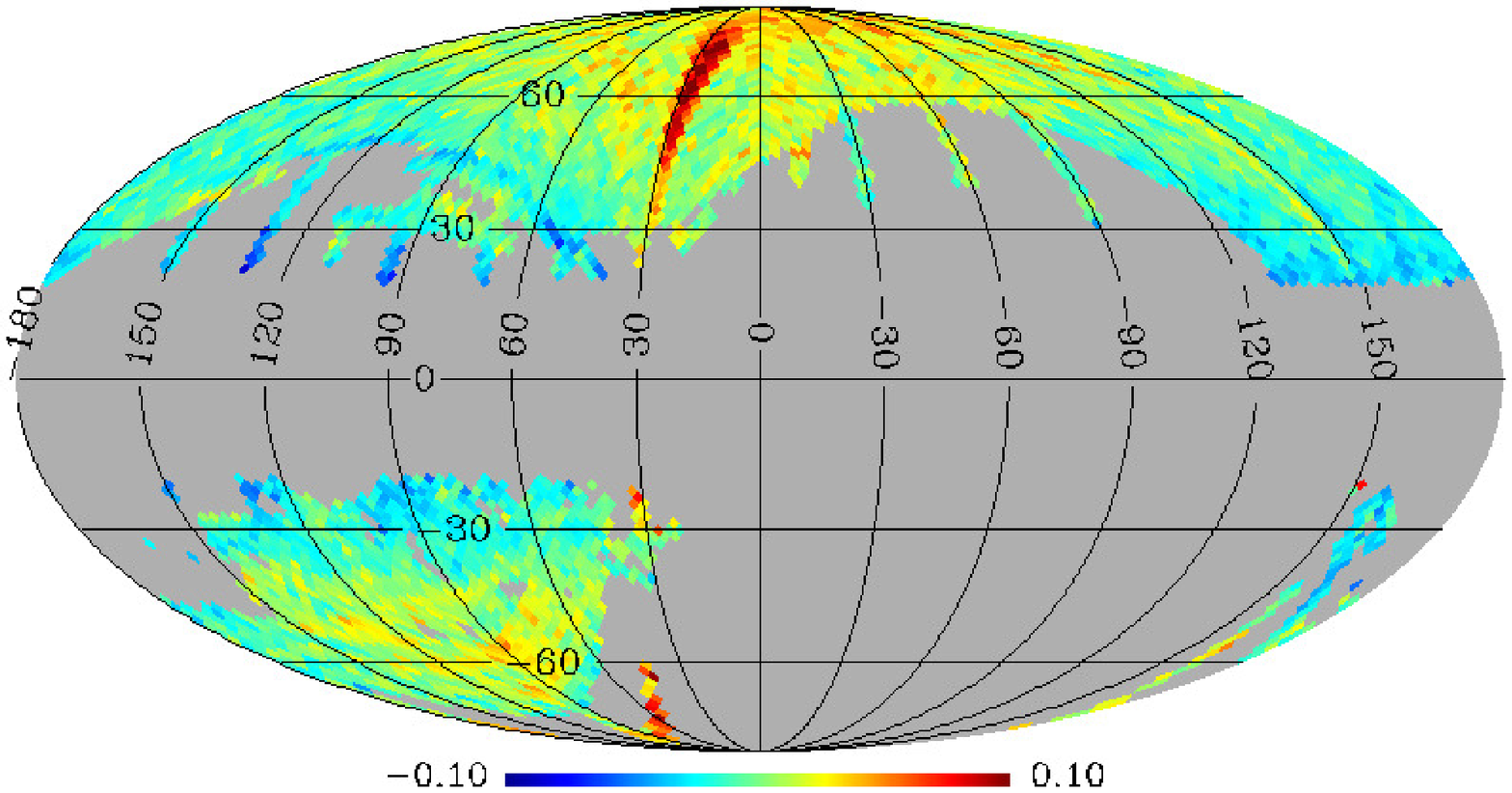}{0.4\textwidth}{\textbf{(b) Zero-point corrections in the $u$-band.}}}
  \gridline{\fig{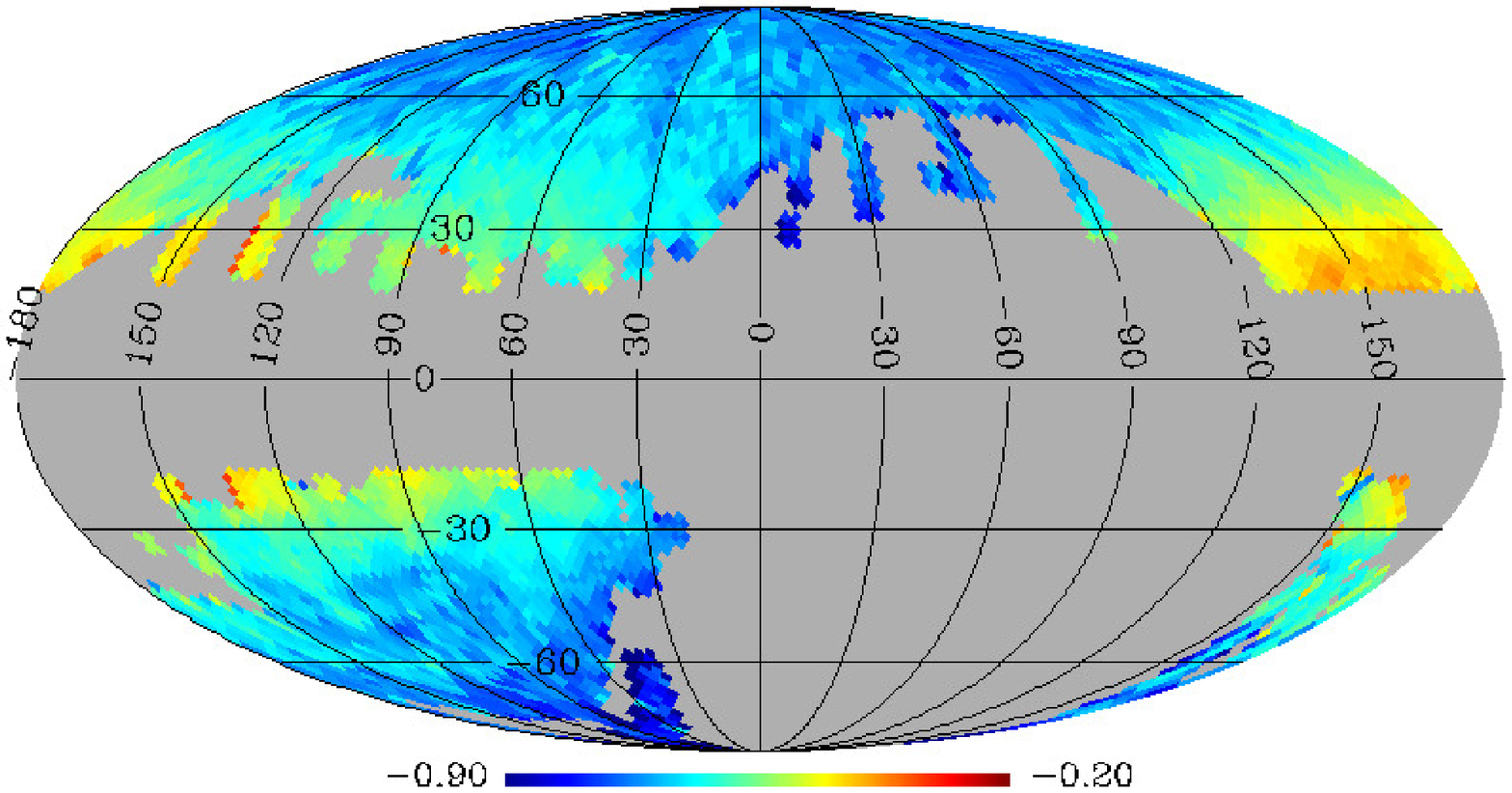}{0.4\textwidth}{\textbf{(c) Re-calibrated SDSS photometry.}}}
  \caption{Same as in panels~(a)--(c) in Figure~\ref{fig:smsszp}, but for photometric metallicity estimates from SDSS photometry (without relying on Gaia parallaxes).}
  \label{fig:sdsszp}
\end{figure*}

In the same manner as for the SMSS $u$ and $v$ photometry, SDSS $u$-band photometry is re-calibrated for a small, but significant offset in our metallicity map. Panel~(a) in Figure~\ref{fig:sdsszp} shows a mean metallicity distribution from stars at $1.2 < d < 3$~kpc (5.1 million stars). There are strips with distinctly lower or higher metallicities than surrounding areas (e.g., the stripe along $l=30\arcdeg$; see also Figure~1 in Paper~I). These strips are parallel to the scanning direction of the SDSS imaging footprints (each $2.5\arcdeg$ wide and $\sim120\arcdeg$ long), suggesting that the spatially correlated offsets in metallicity are induced by photometric zero-point errors in the metallicity-sensitive $u$-band.

The zero-point correction map in panel~(b) is derived in the same way as for the SMSS $u$ and $v$ passbands. As there are more stars available for the construction of a metallicity distribution function in SDSS, we adopt a finer pixel size for HEALPix, $N_{\rm side}=32$. The metallicity sensitivity $\Delta {\rm [Fe/H]}/\Delta {\rm mag}=-2.6$ is obtained from a case with a $0.06$~mag offset in $u$, and is used to make a uniform metallicity distribution of nearby stars ($0.5 < d < 1$~kpc) at the ensemble average ([Fe/H] $=-0.25$). In this way, only the relative zero-point offsets are rectified, while the global mean remains intact. The revised metallicity map shown in panel~(c) is significantly more smooth than in panel~(a), and no longer shows artificial structures.

\section{Uncertainties in Parameter Estimates}\label{sec:error}

\begin{figure*}
  \centering
  \gridline{\fig{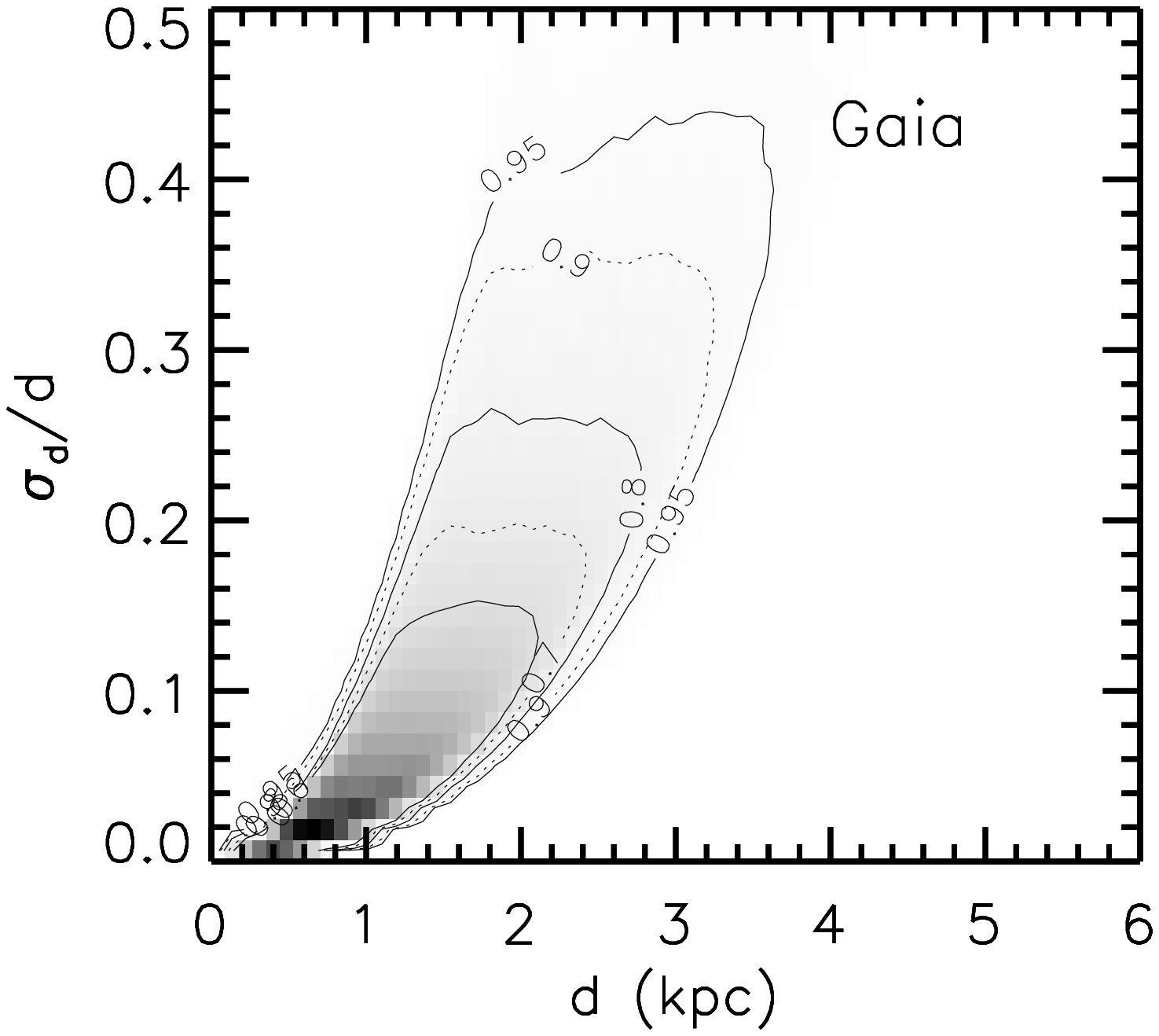}{0.36\textwidth}{\textbf{(a) Gaia parallax uncertainty.}}
                \fig{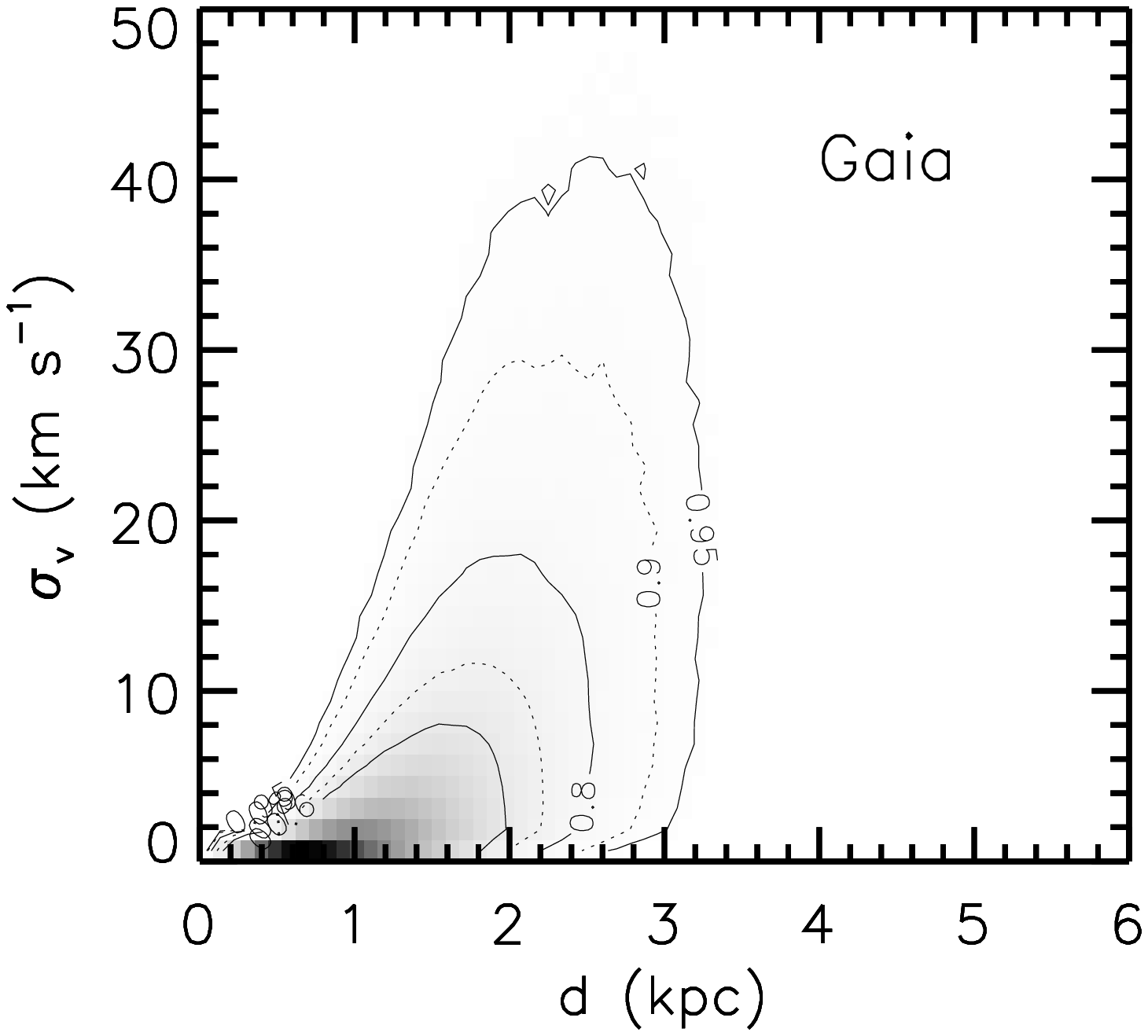}{0.36\textwidth}{\textbf{(b) Projected $\vphi$ uncertainty from Gaia.}}}
  \gridline{\fig{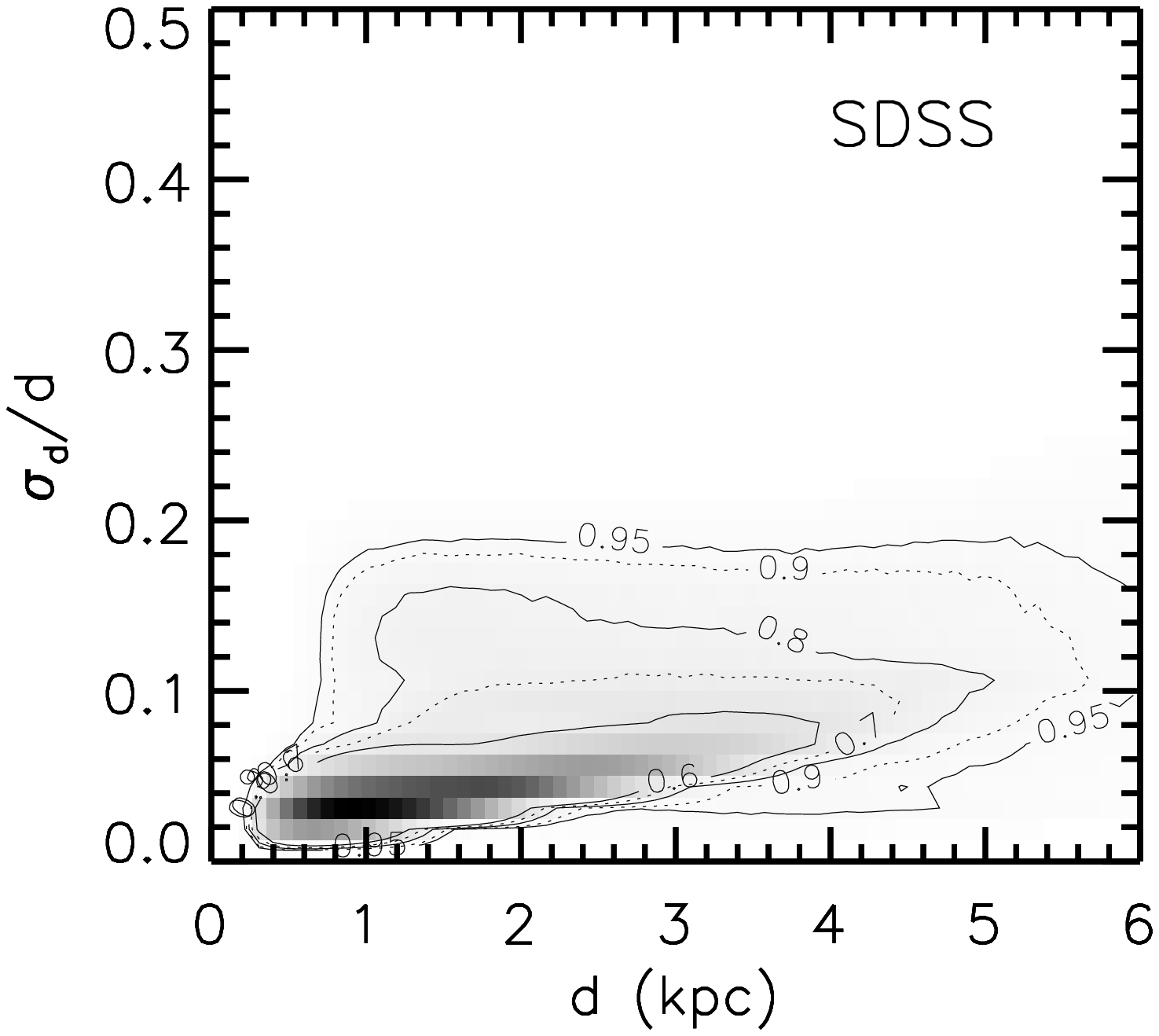}{0.36\textwidth}{\textbf{(c) Photometric distance uncertainty from SDSS.}}
                \fig{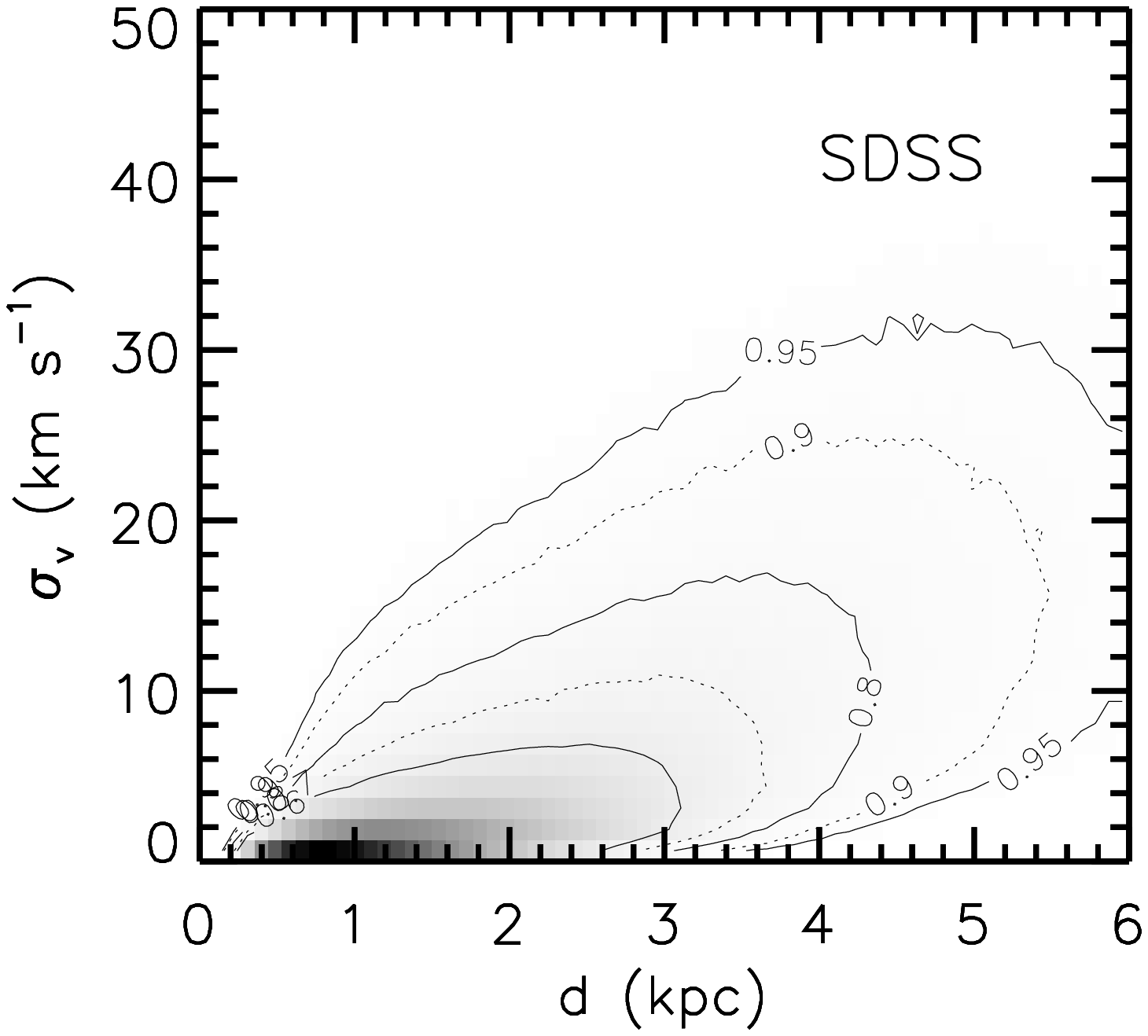}{0.36\textwidth}{\textbf{(d) Projected $\vphi$ uncertainty using SDSS distance.}}}
  \gridline{\fig{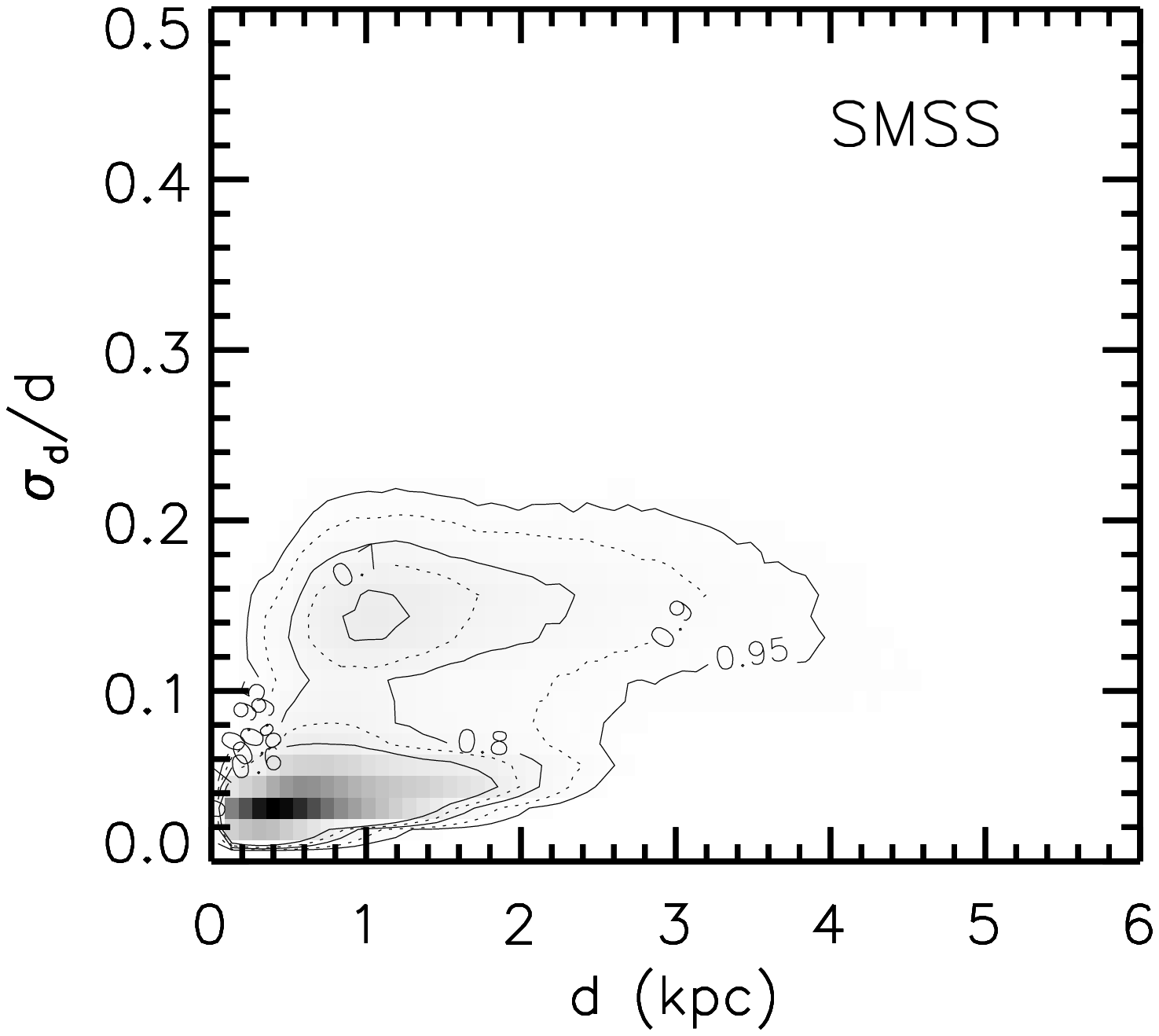}{0.36\textwidth}{\textbf{(e) Photometric distance uncertainty from SMSS.}}
                \fig{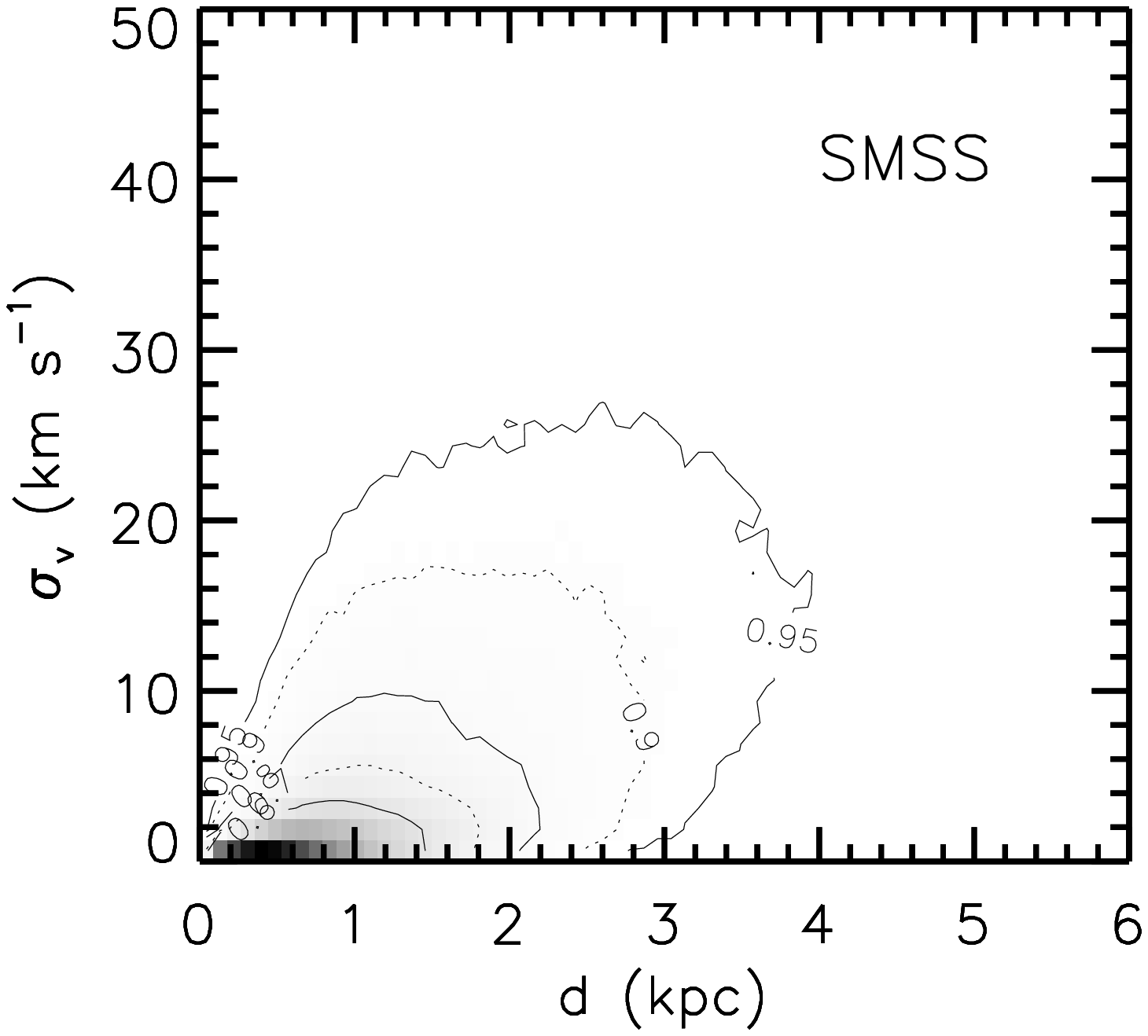}{0.36\textwidth}{\textbf{(f) Projected $\vphi$ uncertainty using SMSS distance.}}}
  \caption{Uncertainties associated with measurements of the distance (left panels) and projected $\vphi$ (right panels) of the sample stars used in this analysis. The top to bottom panels display measurements obtained from Gaia EDR3, photometric distance with its associated $\vphi$ derived from SDSS photometry, and measurements obtained from SMSS photometry, respectively. In the latter two cases, PS1 photometry is used in the parameter estimation, when available. The original uncertainty estimates of photometric distance and its associated $\vphi$ are rescaled by a factor of three (see text). The contour levels on each panel correspond to percentiles that encompass $60\%$, $70\%$, $80\%$, $90\%$, and $95\%$ of the stars.}
  \label{fig:error}
\end{figure*}

Figure~\ref{fig:error} shows the fractional uncertainties of distance and projected $\vphi$ of stars used in \S~\ref{sec:phase}. The top panels display Gaia parallaxes and $\vphi$ measurements computed from Gaia's proper motions based on Gaia parallaxes. In the middle and bottom panels, the same cases are shown from photometric approaches based purely on photometry from SDSS (middle panels) and SMSS (bottom panels) for all stars used in this work. Whenever possible, PS1 photometry is used, as in our main analysis. However, as found in the comparison with Gaia parallax in Figure~\ref{fig:comp2}, photometric distance uncertainties are likely overestimated by a factor of about three when photometry is used to estimate both distance and metallicity simultaneously. As $\vphi$ estimates are linearly dependent on distance, our original uncertainties in $\vphi$ are also overestimated by the same factor. The distribution in Figure~\ref{fig:error} shows these rescaled values.

As we move away from the Sun, both geometric and photometric distance measurements become less precise. However, Gaia parallaxes are affected more significantly, with uncertainties deteriorating to $15\%$ beyond a distance of $2.5$ kpc. In contrast, photometric distances based on SDSS exhibit a more gradual increase in uncertainties with distance for most stars. SMSS, with its limited survey depth, can provide useful distance estimates only for nearby stars. While there is a second clump of stars with larger distance uncertainties in both SDSS and SMSS samples, due to large errors near the main-sequence turn-off, its impact on the overall sample is negligible. Following the ridge line of the majority of stars in the top and middle left panels, one can see that the photometric approach based on SDSS provides better distances than Gaia beyond $1$ kpc.

To ensure the accuracy of our analysis in \S~\ref{sec:phase}, we select a sample of stars with good Gaia parallax measurements and corresponding $\vphi$ measurements. However, this sample is limited to a nearby volume within a distance of approximately $2$ kpc. We use this sample to study high proper-motion stars in Figures \ref{fig:vphi}--\ref{fig:vphi_gaia}, where accurate $\vphi$ measurements and precise sample cuts based on proper-motion measurements are necessary. Conversely, for a more extensive sample that requires a larger volume coverage, such as in estimating scale height and length in Figures \ref{fig:sheight}--\ref{fig:scale_smss}, we rely solely on photometric distance estimates and $\vphi$.

\begin{figure*}
  \centering
  \gridline{\fig{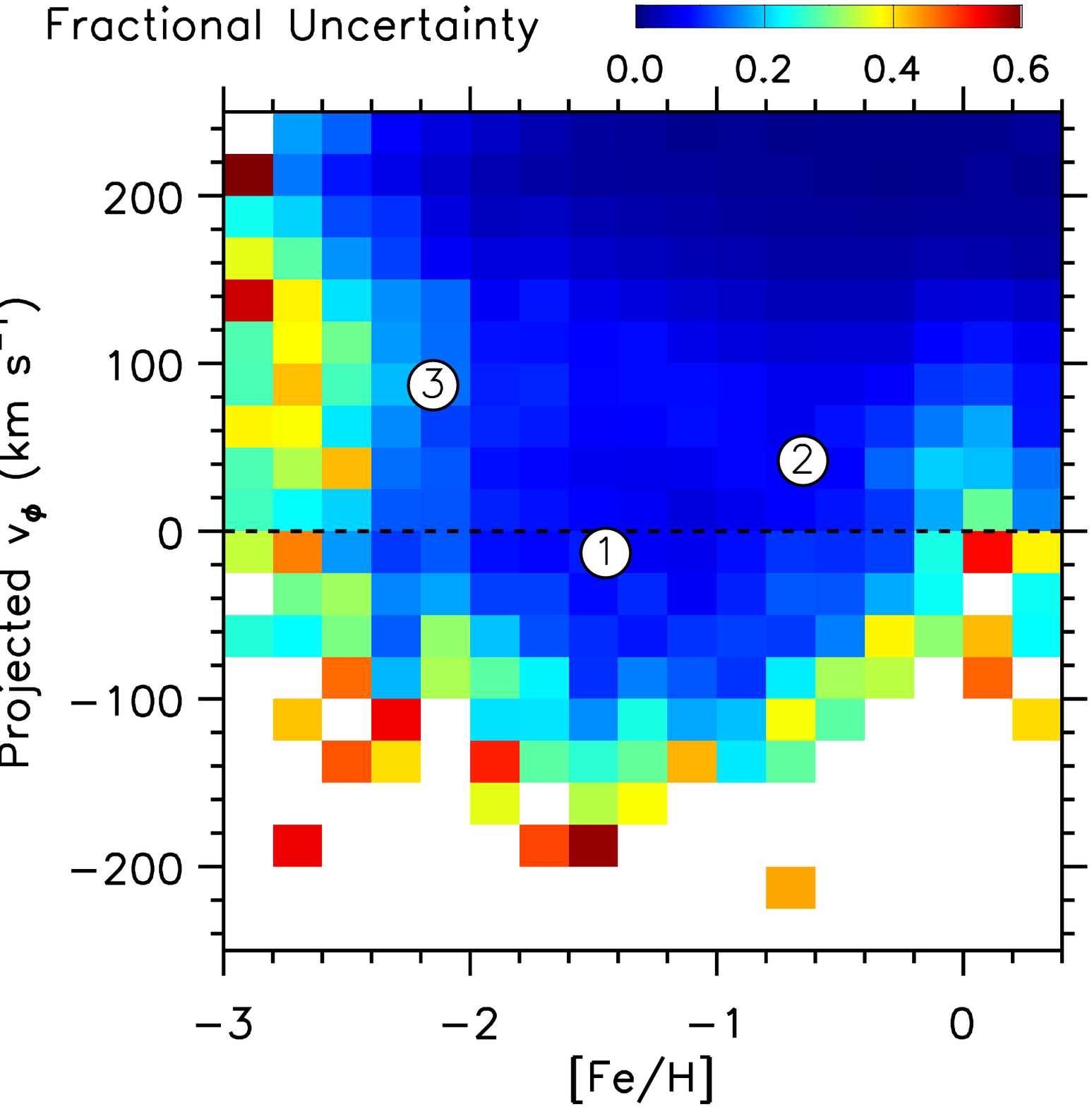}{0.4\textwidth}{\textbf{(a) Uncertainty in scale height in the Northern Galactic Hemispehre.}}
                \fig{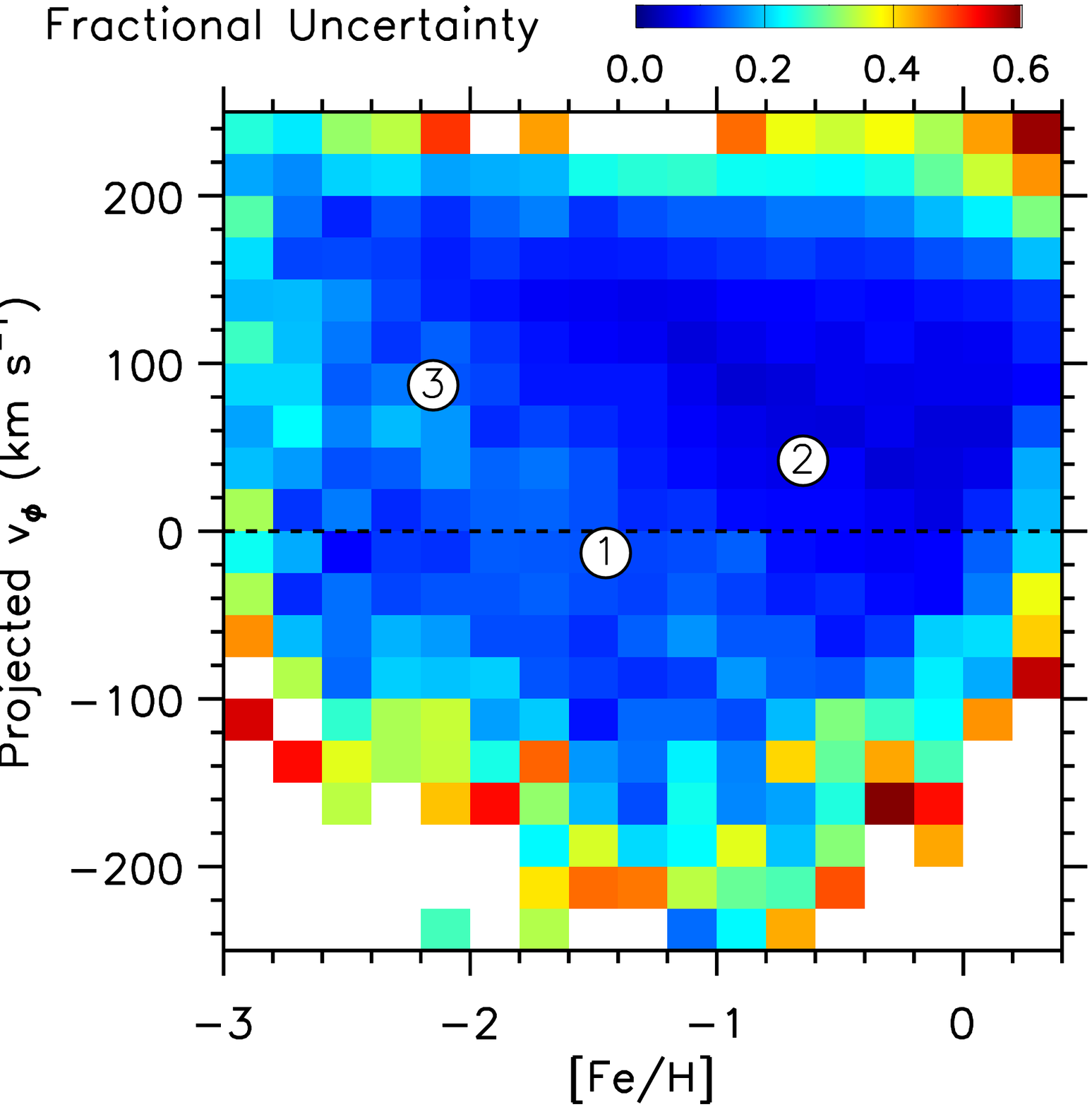}{0.4\textwidth}{\textbf{(c) Uncertainty in scale length in the Northern Galactic Hemispehre.}}}
  \gridline{\fig{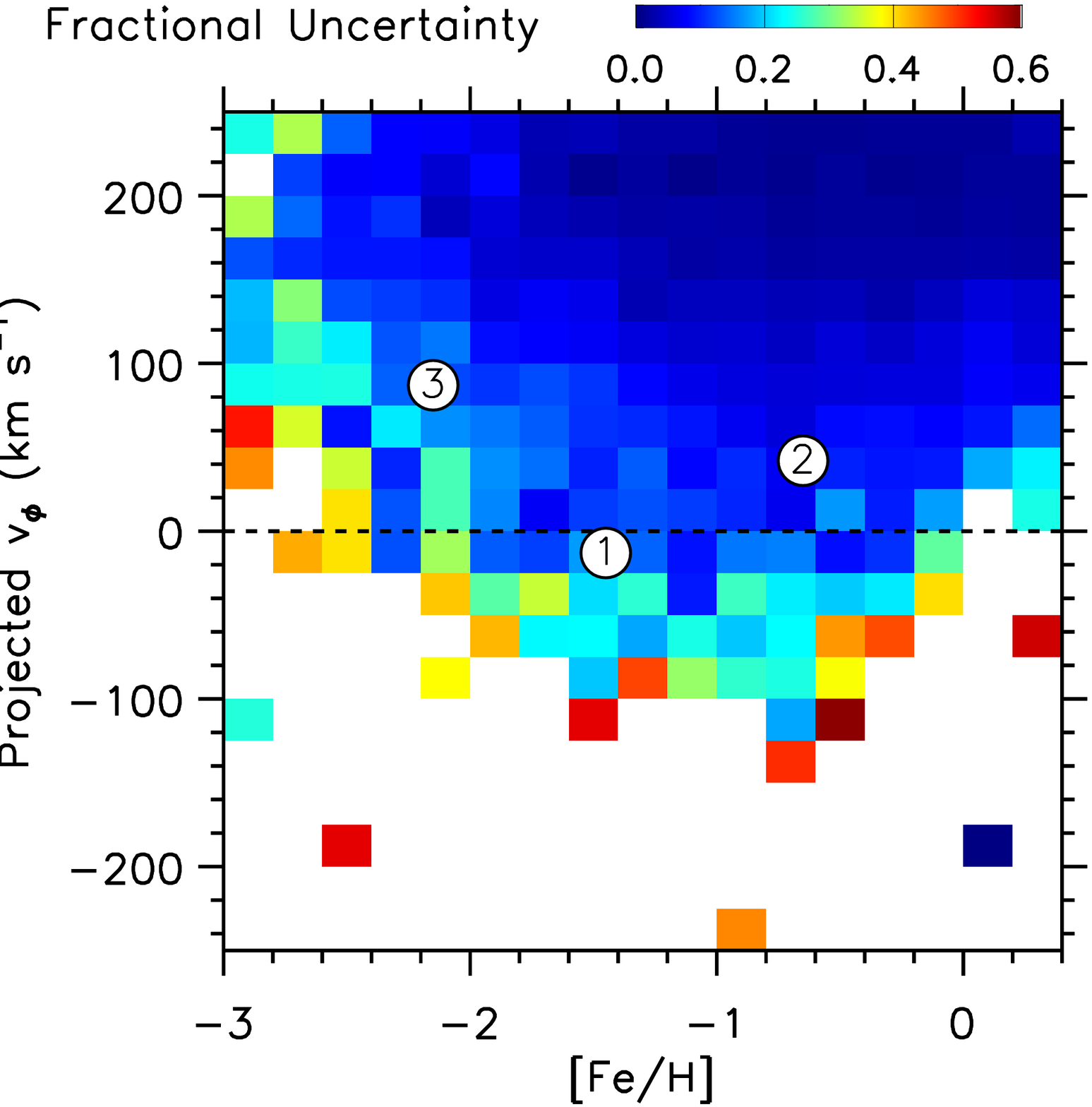}{0.4\textwidth}{\textbf{(b) Uncertainty in scale height in the Southern Galactic Hemispehre.}}
                \fig{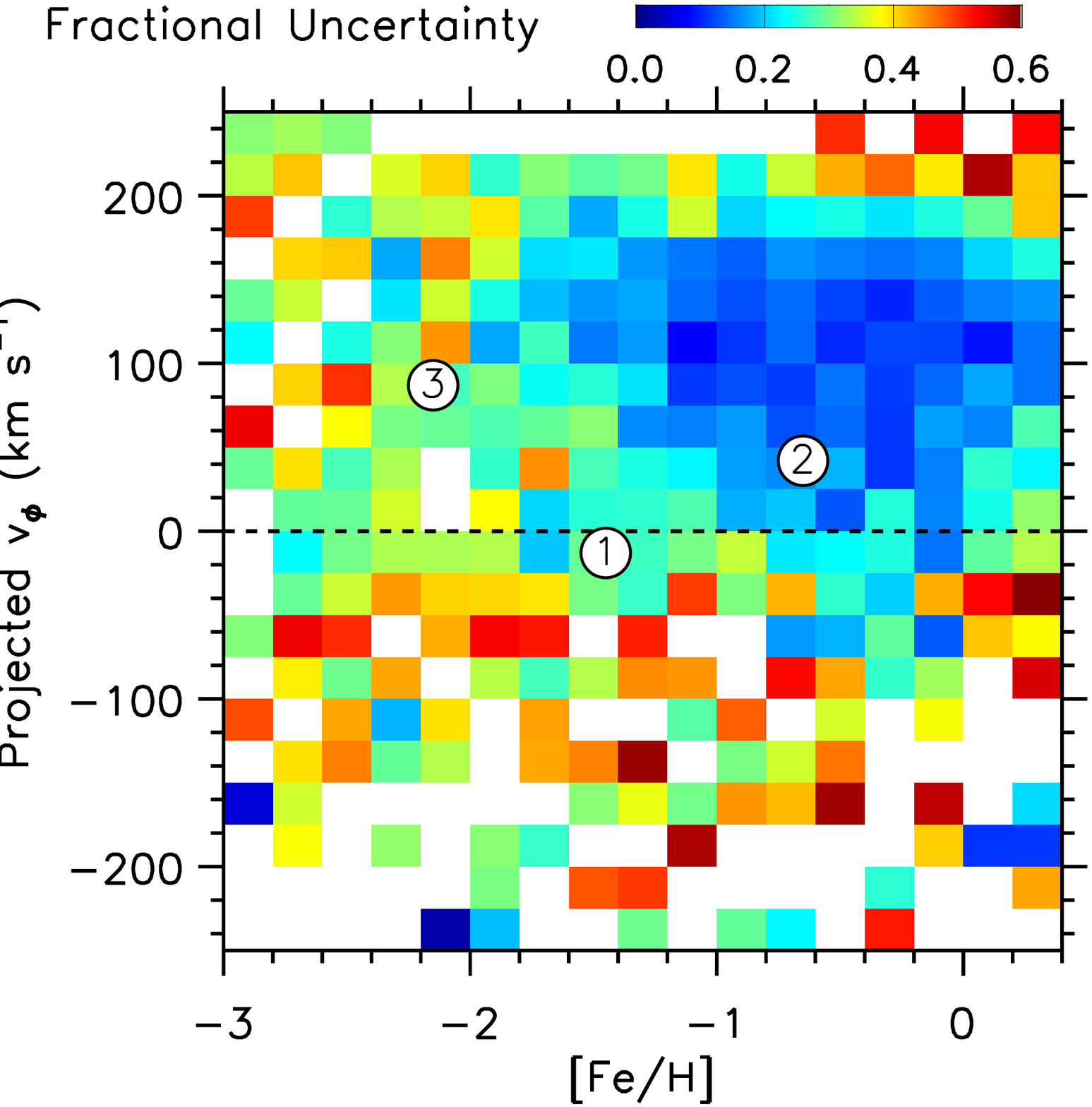}{0.4\textwidth}{\textbf{(d) Uncertainty in scale length in the Southern Galactic Hemispehre.}}}
  \caption{Fractional uncertainties in the measurement of scale height (left panels) and length (right panels) from the SDSS $\cap$ Gaia sample in Figures~\ref{fig:sheight} and \ref{fig:slength}. The top and bottom panels show the uncertainties from the Northern and Southern Galactic Hemispheres, respectively.}
  \label{fig:error2}
\end{figure*}

Figure~\ref{fig:error2} illustrates the distribution of uncertainties in our measurements of scale height and length, as shown in Figures~\ref{fig:sheight}--\ref{fig:slength}, based on SDSS photometry. These uncertainties are expressed as fractional values and are estimated by calculating the standard deviation of the best-fitting slope parameter in the logarithmic number density of stars with respect to either $R$ or $Z$ (\S~\ref{sec:scale}). The uncertainties in the Northern Galactic Hemisphere are relatively small, owing to the larger number of stars observed in SDSS. Additionally, we mark the approximate positions of the three major features observed in this work with circled numbers. These markings indicate that these features are not affected by accidental inaccuracies in certain pixels on the phase-space diagram.

\section{The Effects of Sampling and Uncertainties on Phase-Space Diagrams}\label{sec:phase_error}

\begin{figure*}
  \centering
  \gridline{\fig{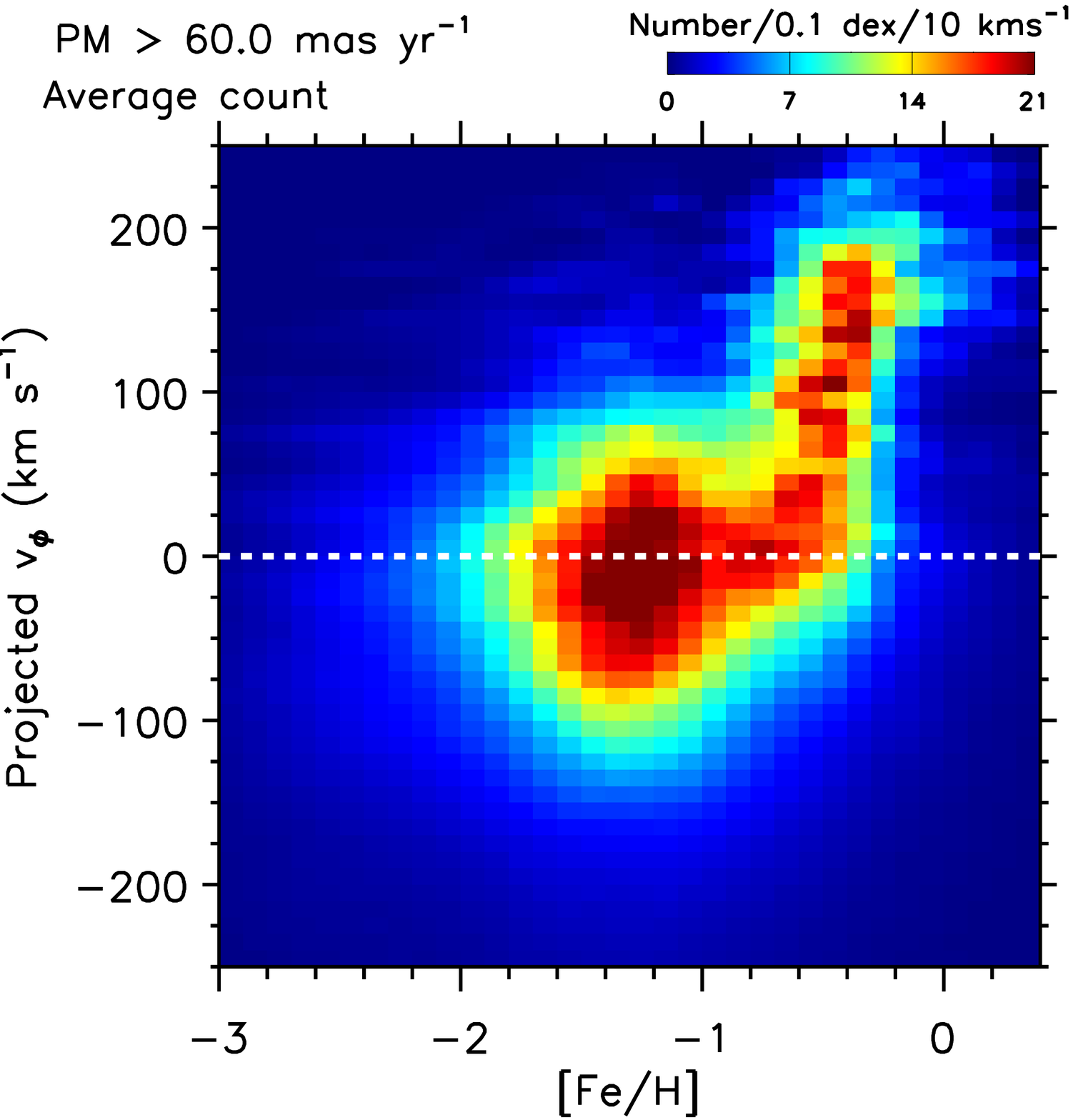}{0.4\textwidth}{\textbf{(a) All}}
                \fig{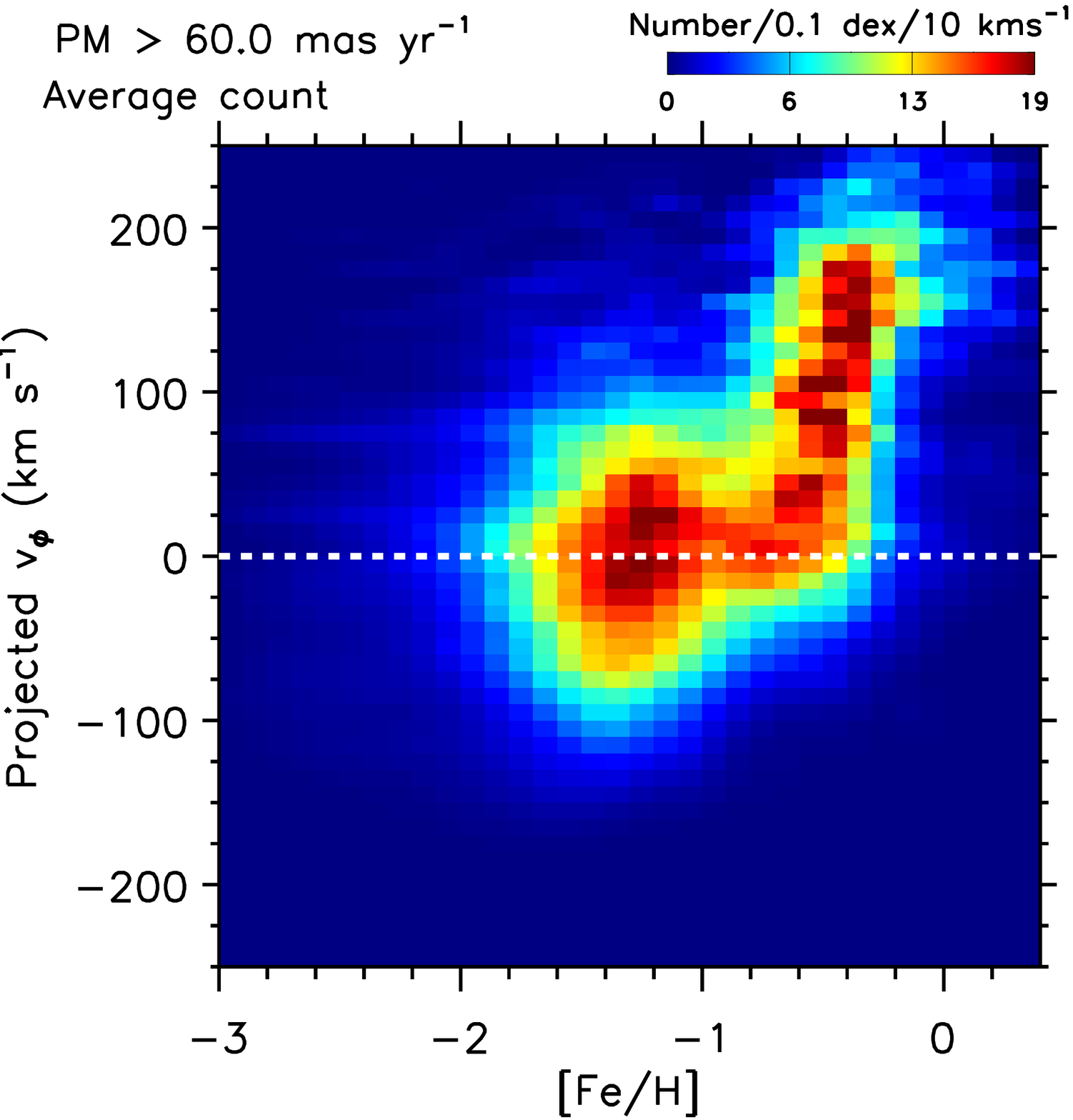}{0.4\textwidth}{\textbf{(b) $\sigma_{\rm [Fe/H]} < 0.3$~dex and $\sigma_v < 20\ \kms$}}}
  \gridline{\fig{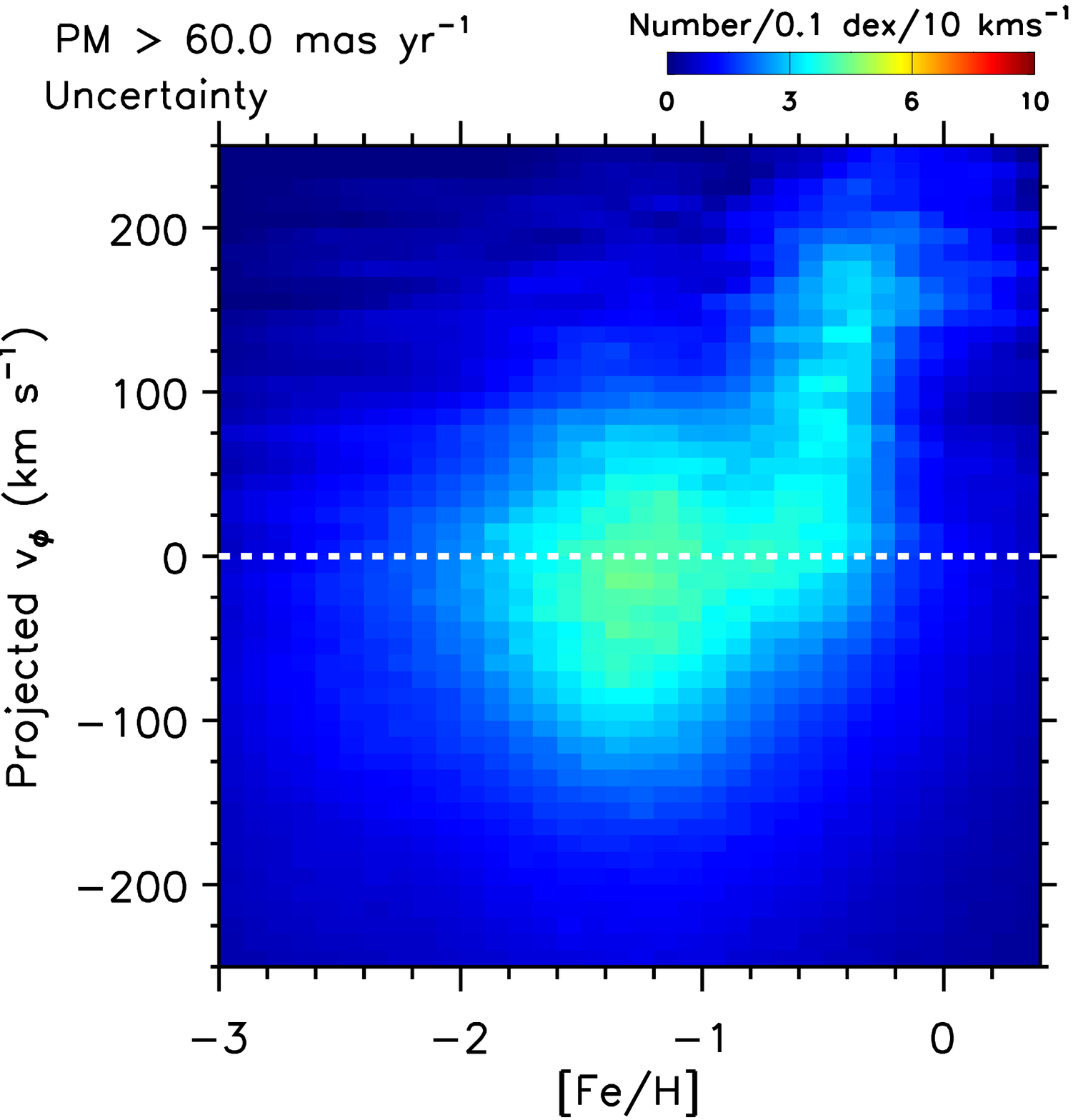}{0.4\textwidth}{\textbf{(c) All}}
                \fig{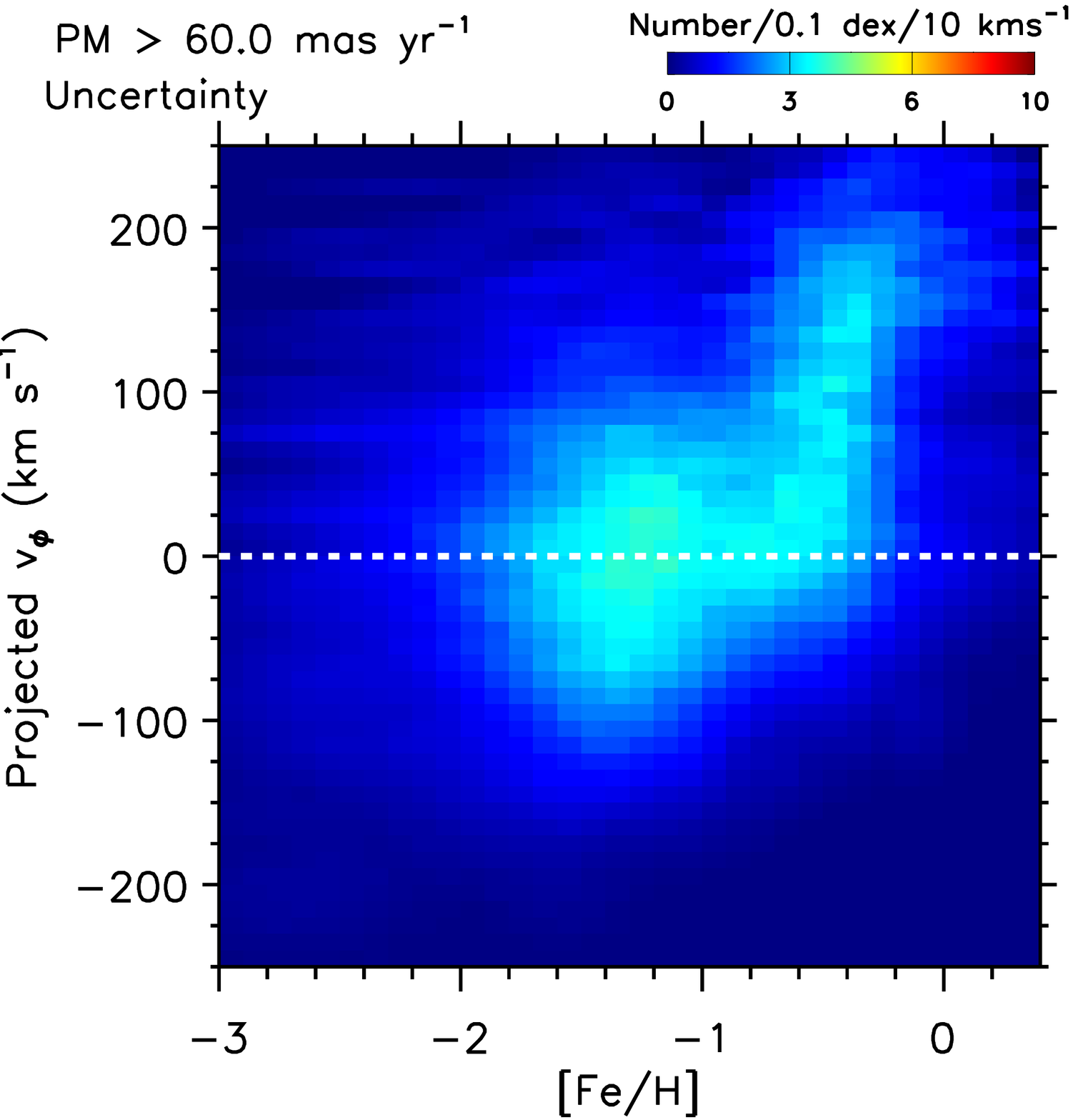}{0.4\textwidth}{\textbf{(d) $\sigma_{\rm [Fe/H]} < 0.3$~dex and $\sigma_v < 20\ \kms$}}}
  \caption{The average counts of stars belonging to the high proper-motion sample ($>60$~mas~yr$^{-1}$; top panels) and their uncertainty measurements (bottom panels) obtained from a Monte Carlo simulation (see text). The right panels display the results of applying restrictive cuts to the measurement uncertainty in [Fe/H] and $\vphi$.}
  \label{fig:gss_sample}
\end{figure*}

In Figure~\ref{fig:gss_sample}, panel (a) shows a phase-space diagram of high proper-motion stars ($>60$~mas~yr$^{-1}$) from the SDSS $\cap$ Gaia sample. This panel is equivalent to panel~(c) in Figure~\ref{fig:vphi}, but it is based on $1,000$ realizations in a Monte Carlo simulation. The plot shows the average raw count of stars without convolving the data with measurement uncertainties. The corresponding uncertainty distribution is shown in panel~(c) from the same suite of a simulation. Panels (b) and (d) show the results obtained when implementing restrictive cuts on the measurement uncertainty in [Fe/H] and $\vphi$. In these simulations, we use the original uncertainty measurements in [Fe/H] and $\vphi$, as Gaia parallaxes are used in the estimation of both quantities. Our results indicate that the significance and shape of the GSS remain nearly unchanged, even with strict sampling criteria. Even when applying the strictest condition, $\sigma_{\rm [Fe/H]} < 0.2$~dex and $\sigma_{v_\phi} < 10\ \kms$ (not shown), the structure remains nearly unaffected, although some stars are lost in the low metallicity tail of the GSS.

\begin{figure*}
  \centering
  \gridline{\fig{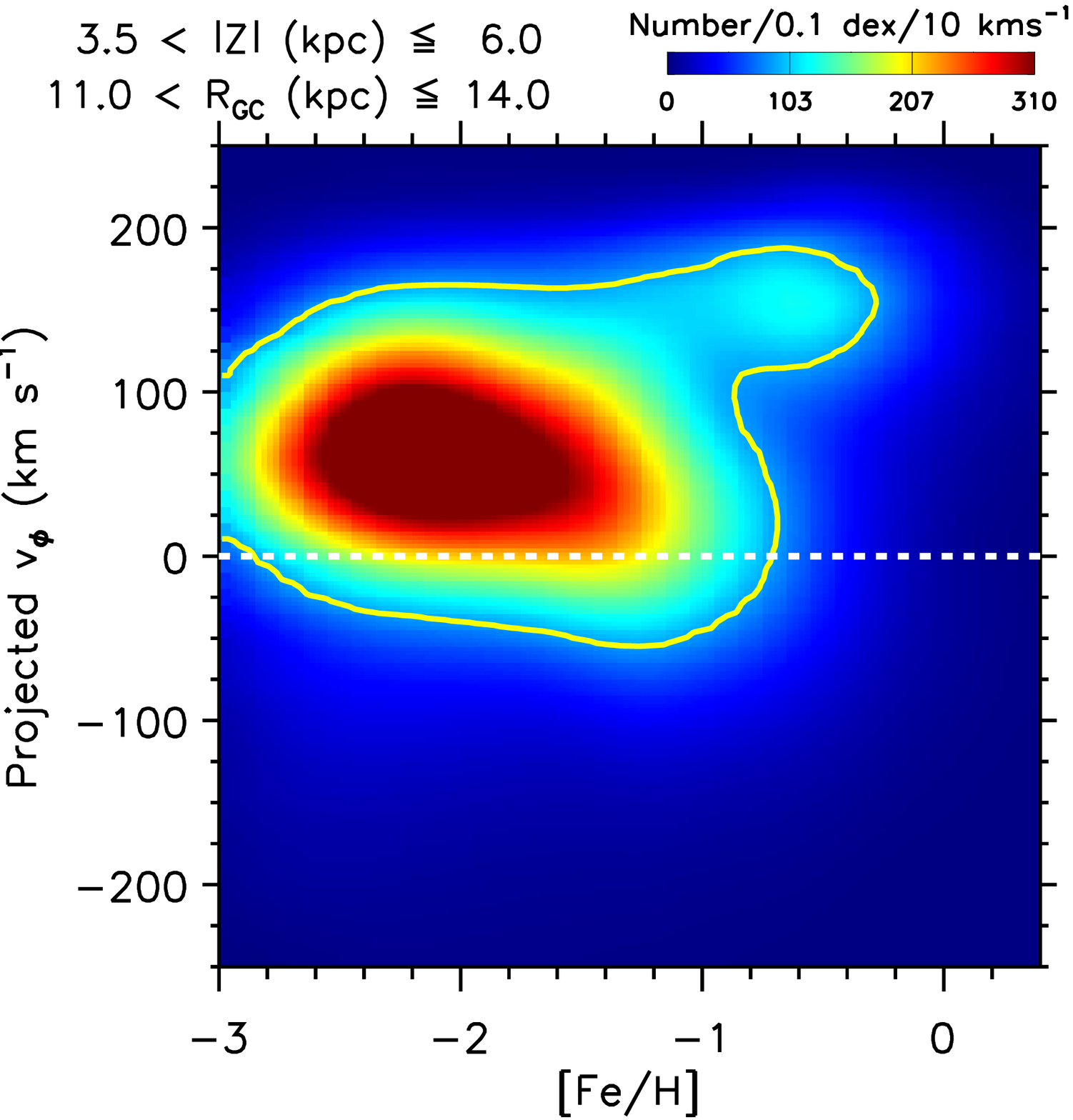}{0.4\textwidth}{\textbf{(a) Uncertainties reduced by a factor of $4$.}}
                \fig{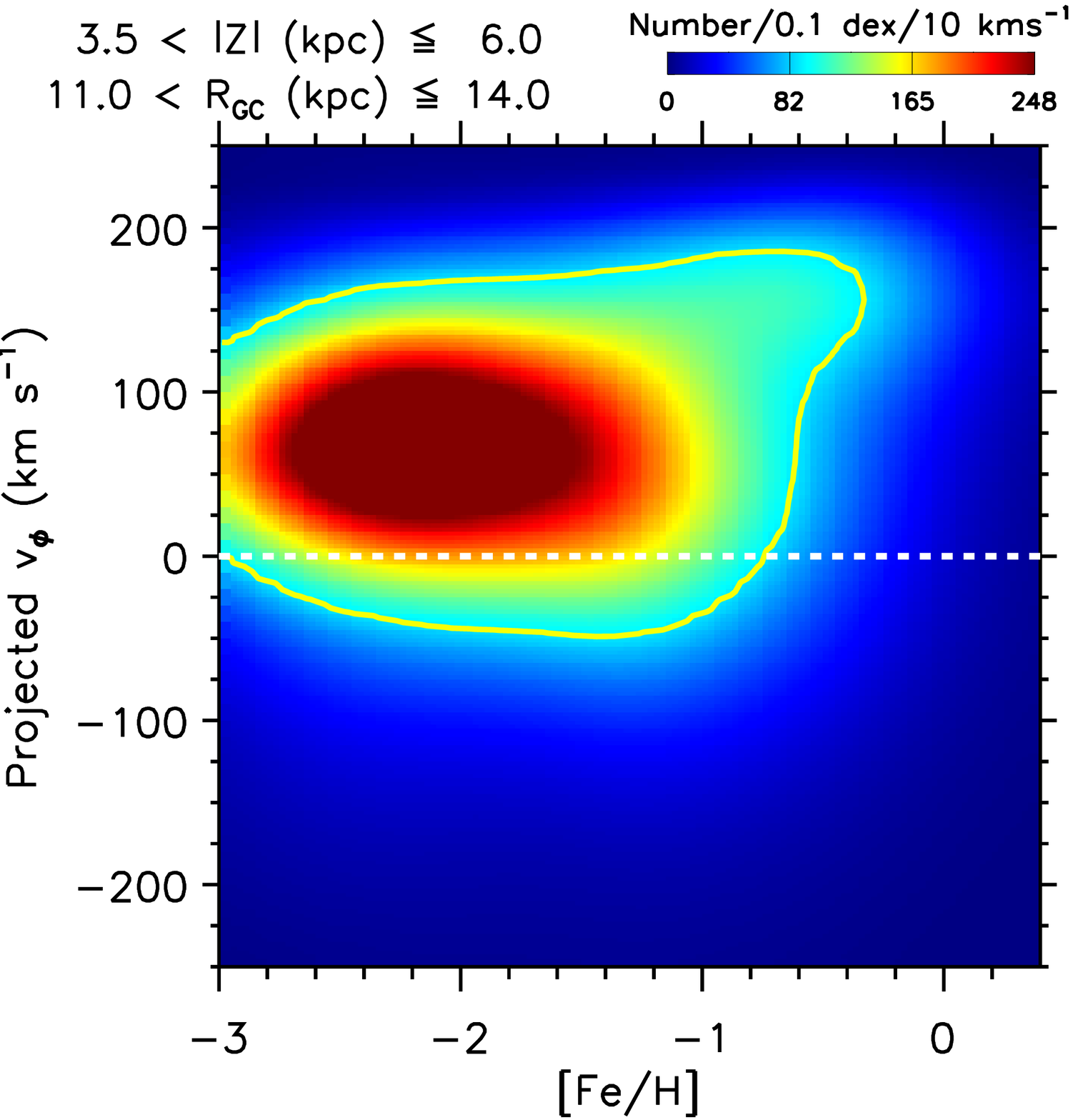}{0.4\textwidth}{\textbf{(b) Uncertainties reduced by a factor of $2$.}}}
  \caption{Same as Panel (b) in Figure~\ref{fig:phase}, but displaying a distribution of stars convolved using measurement uncertainties. The original measurement uncertainties are rescaled by a specified factor in both [Fe/H] and $\vphi$.}
  \label{fig:rescale_halo}
\end{figure*}

Figure~\ref{fig:rescale_halo} presents the same dataset shown in panel~(b) of Figure~\ref{fig:phase}. However, we convolve each count with a normalized Gaussian function, which has standard deviations determined by the uncertainty measurements in [Fe/H] and $\vphi$. It is worth noting that, when using a pure photometric solution for both [Fe/H] and distance, uncertainties may be overestimated by a factor of $2$--$3$ compared to photometric metallicity estimates based on Gaia parallaxes (see \S~\ref{sec:comparison}). To account for this, we rescale our original uncertainties by a factor of $4$ and $2$ in the left and right panels, respectively. This exercise demonstrates that adopting original uncertainties in the analysis fails to produce a clear separation between the thick disk and the halo. This offers additional evidence that our original uncertainty measurements on [Fe/H] and distance are overestimated, possibly because of correlations between photometric measurements in different passbands.

{}

\end{document}

%% file: tab1.tex
\begin{deluxetable*}{lclc}
\tablecaption{A Summary of Calibration Samples\label{tab:tab1}}
\tablehead{
   \colhead{Photometry} &
   \colhead{Bandpasses} &
   \colhead{Stellar sequences} &
   \colhead{Individual stars (spectroscopic sample)}
}
\startdata
SDSS & $ugriz$ & Gaia double, \citet{an:08} & SEGUE, GALAH \\
SMSS & $uvgriz$ & Gaia double & SEGUE, GALAH \\
PS1 & $grizy$ & Gaia double, \citet{bernard:14} & SEGUE, GALAH \\
APASS & $BV$ & Gaia double & SEGUE, GALAH \\
Stetson & $BVI_C$ & Stetson's standard photometry & \nodata \\
\enddata
\end{deluxetable*}

%% file: tab2.tex
\begin{deluxetable*}{lcccccc}
\tablecaption{Fundamental Parameters and Uncertainties of Stellar Sequences\label{tab:tab2}}
\tablehead{
   \colhead{Cluster/Sequence} &
   \colhead{[Fe/H]\tablenotemark{\scriptsize a}} &
   \colhead{$(m\, -\, M)_0$\tablenotemark{\scriptsize a}} &
   \colhead{$\ebv$\tablenotemark{\scriptsize a}} &
   \colhead{age\tablenotemark{\scriptsize a}} &
   \colhead{Min $M_r$\tablenotemark{\scriptsize b}} &
   \colhead{} \\
   \colhead{Name} &
   \colhead{(dex)} &
   \colhead{(mag)} &
   \colhead{(mag)} &
   \colhead{(Gyr)} &
   \colhead{(mag)} &
   \colhead{References\tablenotemark{\scriptsize c}}
}
\startdata
M15      & $-2.42\pm0.10$ & $15.25\pm0.15$ & $0.100\pm0.020$ & $13.0\pm2.6$ & $19.4$ & 1/1/1 \\
M92      & $-2.38\pm0.10$ & $14.64\pm0.15$ & $0.020\pm0.004$ & $13.0\pm2.6$ & $18.5$ & 1/2/1 \\
M13      & $-1.60\pm0.10$ & $14.38\pm0.15$ & $0.020\pm0.004$ & $13.0\pm2.6$ & $18.5$ & 1/2/1 \\
M3       & $-1.50\pm0.10$ & $15.02\pm0.15$ & $0.010\pm0.002$ & $13.0\pm2.6$ & $19.0$ & 1/1/1 \\
Gaia double (blue MS) & $-1.30\pm0.10$ & \nodata\tablenotemark{\scriptsize d} & \nodata\tablenotemark{\scriptsize d} & $13.0\pm2.6$ & $ 4.0$ & 3/./. \\
M5       & $-1.26\pm0.10$ & $14.46\pm0.15$ & $0.030\pm0.006$ & $12.0\pm2.4$ & $18.5$ & 1/2/1 \\
Gaia double (red MS) & $-0.40\pm0.10$ & \nodata\tablenotemark{\scriptsize d} & \nodata\tablenotemark{\scriptsize d} & $12.0\pm2.4$ & $ 4.0$ & 3/./. \\
M67      & $+0.00\pm0.01$ & $ 9.61\pm0.03$ & $0.041\pm0.004$ & $ 4.0\pm0.5$ & $13.0$ & 4/4/4 \\
NGC~6791 & $+0.37\pm0.07$ & $13.06\pm0.06$ & $0.120\pm0.020$ & $ 9.0\pm1.0$ & $17.6$ & 4/4/4 \\
\enddata
\tablenotetext{a}{Uniform uncertainties except for M67 and NGC~6791.}
\tablenotetext{b}{A minimum $M_r$ to select MS stars.}
\tablenotetext{c}{References for [Fe/H], $(m\, -\, M)_0$, and $\ebv$, respectively: (1) \citet{kraft:03}; (2) \citet{carretta:00}; (3) Paper~II; (4) \citet{an:19} and references therein.}
\tablenotetext{d}{Parallaxes in Gaia EDR3 are adopted for individual field stars, along with their foreground extinctions in \citet{schlegel:98}. See Paper~II.}
\end{deluxetable*}